\documentclass[12pt,preprint]{aastex}

\newcommand{\ang}{{\rm \AA}}

\newcommand{\iraf}{{\sl IRAF\/}}

\newcommand{\Lsun}{\mbox{$\rm L_\odot$}}

\newcommand{\SFRu}{\mbox{$\rm M_\odot\,yr^{-1}$}}

\newcommand{\ch}{\mbox{{\it Chandra\/}}}

\newcommand{\iron}{\mbox{IRAS\,19245-4140}}
\newcommand{\tol}{\mbox{Tol\,0440-381}}
\newcommand{\sbs}{\mbox{SBS\,0335-052}}

\newcommand{\ergcmsa} {\mbox{erg cm$^{-2}$ s$^{-1}$ \AA$^{-1}$}}
\newcommand{\cms} {\mbox{cm$^{-2}$}}

\newcommand{\fu}{\mbox{\it FUSE\/}}
\newcommand{\vv}{\mbox{VV\,114}}

\newcommand{\ovi}{\mbox{\ion{O}{6}}}
\newcommand{\oviw}{\mbox{\ion{O}{6}$\lambda1032$}}
\newcommand{\oviww}{\mbox{\ion{O}{6}$\lambda\lambda1032,1038$}}
\newcommand{\oi}{\mbox{\ion{O}{1}}}
\newcommand{\oiaw}{\mbox{\ion{O}{1}$\lambda989$}}
\newcommand{\oibw}{\mbox{\ion{O}{1}$\lambda1039$}}
\newcommand{\oicw}{\mbox{\ion{O}{1}$\lambda976$}}
\newcommand{\cii}{\mbox{\ion{C}{2}}}
\newcommand{\ciiw}{\mbox{\ion{C}{2}$\lambda1036$}}
\newcommand{\ciibw}{\mbox{\ion{C}{2}$\lambda1335$}}
\newcommand{\aliiw}{\mbox{\ion{Al}{2}$\lambda1671$}}
\newcommand{\siiiw}{\mbox{\ion{Si}{2}$\lambda1260$}}
\newcommand{\sthw}{\mbox{\ion{S}{3}$\lambda1012$}}
\newcommand{\naoww}{\mbox{\ion{Na}{1}$\lambda\lambda5890,5896$}}
\newcommand{\nai}{\mbox{\ion{Na}{1}}}
\newcommand{\naiww}{\mbox{\ion{Na}{1}$\lambda\lambda5890,5896$}}
\newcommand{\sth}{\mbox{\ion{S}{3}}}
\newcommand{\ciii}{\mbox{\ion{C}{3}}}
\newcommand{\ciiiw}{\mbox{\ion{C}{3}$\lambda977$}}

\newcommand{\niiiw}{\mbox{\ion{N}{3}$\lambda990$}}
\newcommand{\nii}{\mbox{\ion{N}{2}}}
\newcommand{\niiw}{\mbox{\ion{N}{2}$\lambda1084$}}
\newcommand{\hoo}{\mbox{Haro\,11}}
\newcommand{\ks}{{\rm{km\,s^{-1}}}}
\newcommand{\kms}{\mbox{$\rm{km\,s^{-1}}$}}

\newcommand{\lybw}{\mbox{$\rm{Ly}\-\beta\,\lambda1025$}}
\newcommand{\novi}{\mbox{$\rm N_{O\,VI}$}}
\newcommand{\htwo}{\mbox{$\rm{H_{2}}$}}

\newcommand{\izw}{\mbox{I\,Zw\,18}}

\newcommand{\flyabs}{\mbox{$\rm f_{esc,abs}$}}
\newcommand{\flyrel}{\mbox{$\rm f_{esc,rel}$}}

\shorttitle{\fu\,Observations of Starburst Galaxies}
\shortauthors{J. Grimes et al.}

\begin{document}

\title{Far- Ultraviolet Observations of Starburst Galaxies with \fu: Galactic Feedback in the Local Universe}

\author{J. P. Grimes\altaffilmark{1}, T. Heckman\altaffilmark{1},
A. Aloisi\altaffilmark{2}, D. Calzetti\altaffilmark{3}, 
C. Leitherer\altaffilmark{2},
C. L. Martin\altaffilmark{4}, G. Meurer\altaffilmark{1},
K. Sembach\altaffilmark{2}, D. Strickland\altaffilmark{1}}
\altaffiltext{1}{Center for Astrophysical Sciences, Johns Hopkins University,
    3400 N. Charles St, Baltimore, MD 21218; jgrimes@pha.jhu.edu, heckman@pha.jhu.edu,
    choopes@pha.jhu.edu,meurer@pha.jhu.edu,dks@pha.jhu.edu
    }
\altaffiltext{2}{Space Telescope Science Institute, 3700 San Martin Drive,
   Baltimore, MD, 21218; aloisi@stsci.edu,leitherer@stsci.edu,sembach@stsci.edu}
\altaffiltext{3}{Department of Astronomy, University of Massachusetts, 710 N. Pleasant St, 
	Amherst, MA 01003}
\altaffiltext{4}{Department of Physics, University of California, Santa Barbara, CA 93106
	cmartin@physics.ucsb.edu}
\begin{abstract}

We have analyzed \fu\, (905-1187 \AA) spectra of a sample of 
16 local starburst galaxies.
These galaxies cover almost three orders of magnitude in
star-formation rates and over two orders of magnitude in stellar mass.
Absorption features from the stars and interstellar medium are observed
in all the spectra. The strongest interstellar absorption features are generally
blue-shifted by $\sim$ 50 to 300 km s$^{-1}$, 
implying the almost ubiquitous presence of starburst-driven galactic winds in this sample.
The outflow velocites increase with both the star formation rate and the star formation rate
per unit stellar mass, consistent with a galactic wind driven by the population of massive stars.
We find outflowing coronal-phase gas (T $\sim 10^{5.5}$ K) detected via the \ovi\ absorption-line in nearly
every galaxy. 
The \ovi\ absorption-line
profile is optically-thin, is generally weak near the galaxy systemic velocity,
and has a higher mean outflow velocity than seen in the lower ionization lines.
The relationship between the line width and column density for the \ovi\ absorbing gas
is in good agreement with expectations for radiatively cooling and outflowing gas. Such gas will be created in the interaction of the
hot out-rushing wind seen in X-ray emission and cool dense ambient material.
\ovi\, emission is not generally detected in our sample, suggesting that radiative cooling by
the coronal gas is not dynamically significant in draining energy from galactic winds.
We find that the measured
outflow velocities in the HI and HII phases of the interstellar gas in a given galaxy
increase with the strength (equivalent width) of the absorption
feature and not with the ionization potential of the species. The strong lines often have profiles
consisting of a broad and optically-thick component centered near the galaxy systemic velocity
and weaker but highly blue-shifted absorption.
This suggests that
the outflowing gas with high velocity has a lower column density than the more quiescent
gas, and can only be readily detected in the strongest absorption lines.
From direct observations below the Lyman edge in the galaxy rest-frame, we find no evidence of Lyman continuum 
radiation escaping from any of the
galaxies in the sample.  Moreover, the small escape fraction of light
in the center of the strong \cii\ absorption feature confirms the high opacity
below the Lyman limit in the neutral ISM. The absolute fraction of escaping Lyman continuum photons is typically $<$1\%. 
This sample provides a unique window on the global properties
of local star-forming galaxies as observed in the far-UV
and also provides a useful comparison sample 
for understanding spectra of high redshift star-forming galaxies.

\end{abstract}

\keywords{ galaxies: starburst  ---
             galaxies: halos ---
             UV: galaxies ---
             galaxies: individual (Haro11)
             galaxies: individual (VV114)
             galaxies: individual (NGC1140)
             galaxies: individual (SBS0335-052)
             galaxies: individual (TOL0440-381)
             galaxies: individual (NGC1705)
             galaxies: individual (NGC1741)
             galaxies: individual (IZW18)
             galaxies: individual (NGC3310)
             galaxies: individual (Haro3)
             galaxies: individual (NGC3690)
             galaxies: individual (NGC4214)
             galaxies: individual (MRK54)
             galaxies: individual (IRAS19245-4140)
             galaxies: individual (NGC7673)
             galaxies: individual (NGC7714)}

\section{Introduction}

While $\Lambda$CDM  simulations
of a universe dominated by dark matter and
dark energy have led to a robust understanding
of the large scale structure of the universe,
many questions remain when studying
smaller scales where 
baryonic physics become important \citep{rob05,som99,ben03}. 
Thus, understanding the interplay
between a galaxy's components (e.g. gas, stars, radiation, 
black holes) and its environment is an important next step
for cosmology studies.

In particular, feedback on the surrounding gas-phase baryons provided
by massive stars is believed to play a major role
in the formation and evolution of galaxies. Starburst driven
galactic winds are the most obvious manifestation of these processes.
These ``superwinds'' appear to be omnipresent
in galaxies with star formation rates (SFR)
per unit area $\Sigma_\ast \gtrsim 
10^{-1}\,\rm{M_\odot\,yr^{-1}\,kpc^{-2}}$ \citep{heck03}.
Locally, these starbursts alone account for
roughly 20\% of the high-mass star formation and 
10\% of the radiant energy production 
\citep{heck05}.  Powered by supernovae and
stellar winds these outflows
drive gas, dust, energy, and metals out of galaxies 
and possibly into the inter-galactic medium \citep[IGM,][]{heck98,veill05}.  
The higher overall star-formation rates at
$z\sim1$ and above \citep{bunk04,hop06}
imply that star-formation driven winds
are even more important in the early universe.
Such winds at high redshift almost
certainly had a profound
impact on the energetics, ionization, and chemical composition
of the IGM \citep{ag05}. 

Lyman break galaxies  \citep[LBGs,][]{steid99} are the most widely
studied population of high redshift
star forming galaxies as they can be efficiently
selected using the Lyman-break dropout technique.
Rest frame UV spectroscopy of LBGs have shown that
starburst driven galactic winds appear to be a
ubiquitous property of LBGs \citep{shap03}. 
During the cosmological important
epoch of $z\sim2-4$ LBGs constitute a 
significant (and possibly dominant) fraction 
of the population of star forming galaxies  \citep{peac00}.

Studies of the outflows in star forming galaxies are complicated by their
multiphase nature \citep{strick04b}. 
The mechanical energy from the supernovae and stellar winds 
create an over-pressurized cavity of hot gas at  $\sim10^8$ K \citep{chev85}. 
The hot gas expands
and cools adiabatically as it
accelerates into lower pressure regions.  
Shock fronts form at the interface between the hot outflowing gas
and the ambient material \citep{strick00}.  Some of this ambient material
is subsequently swept up into the outflow itself.  Unfortunately, 
directly studying the hot wind fluid is only possible in a
few local galaxies \citep[e.g.][]{strick07} due to the low emissivity of the hot gas
and observational constraints in the hard X-ray band.  
Observations
then of superwinds focus on either material entrained 
in the wind fluid 
(e.g. neutral, ionized, molecular gas and dust) or created through
the interface between the hot wind and the ambient 
material \citep{marc05}.

The coronal 
gas ($T \sim 10^5 - 10^6$ K) observed in galactic
winds is of particular importance.
This phase is  intimately connected to the mechanical/thermal energy that drives the 
outflows and places important limits on the cooling efficiency of the outflow.
The coronal phase is probed through observations of the \oviww\, 
doublet found in the rest frame FUV.  Unfortunately, observations of this
wavelength regime in LBGs are generally limited by  
its location deep within the Lyman $\alpha$ forest.
Our knowledge of the properties of the 
winds in LBGs is generally then limited to what can be inferred from 
the interstellar UV absorption lines longward of Lyman $\alpha$. 

Studies of local star-forming galaxies in this wavelength regime
have been limited by UV absorption in the Earth's atmosphere.  
Hopkins ultraviolet telescope (HUT) observations of star forming galaxies
analyzed by \citet{leith02} probed the 912-1800 \AA\, regime but with very low spectral resolution.
\citet{heck98} studied a large sample of 45 starburst galaxies 
using the {\sl International Ultraviolet Explorer} (IUE)
with similarly low resolution and wavelength coverage only longward of about 1150 \ang.  
HST UV ($>1215$ \AA) spectral studies have also been undertaken \citep[e.g.][]{vaz05,schwa06}
although they focus primarily on specific star clusters given the 
small spectroscopic apertures of {\sl HST}.  

The situation changed however with the launch
of the {\it Far-ultraviolet Spectrographic Explorer} (\fu) 
in 1999.  \fu\, provides high resolution spectroscopy
with an aperture that is well matched to observing 
global FUV properties of local star-forming galaxies.
Two previous works using overlapping data sets have used \fu\ to study
dust attenuation in the FUV \citep{buat02} and
stellar populations \citep{pell07}.
However, 
while several previous works have studied specific galaxies
with \fu\,\citep[e.g.][]{heck01,al03,hoop03,hoop04,grim05,berg06,grim06},
a study of the global properties of outflows in a representative sample of star-forming 
galaxies has not previously been undertaken.  

In this paper we will focus on the absorption lines
produced by the interstellar medium (ISM) and observed in the \fu\, wavebands.  We will first introduce
our sample selection method and analysis techniques in Section \ref{s:fusesampledata}.
Next we will discuss the outflow velocity distributions of the
various ISM absorption features in Section \ref{s:velvel}.
Section \ref{s:ovi} focuses on the 
different physical origin of the coronal gas as traced by  \ovi.
We then compare the outflow velocities and strengths of the
absorption features to the physical properties of the galaxies in
Sections \ref{s:velsfr} and \ref{s:eqw} respectively.
Section \ref{s:lyman} discusses the possible escape of Lyman continuum 
emission in local
starburst galaxies.

\section{\fu~Observations\label{s:fusesampledata}}

The \fu\,satellite is 
a high resolution spectrometer with sensitivity in
the 905 - 1187 \ang\, range.  
Four co-aligned  prime focus 
telescopes, each with its own Rowland spectrograph,
comprise the four \fu\, channels.  Each channel has its own
focal plane assembly with four  apertures.  
Only two science apertures are relevant to this work, the
low-resolution aperture (LWRS; $30\arcsec \times 30\arcsec$)
and the medium-resolution aperture (MDRS: $4\arcsec \times 20\arcsec$).
Spectral resolution (full width at half maximum) ranges from about $17 \ks$
for a point source to about $100 \ks$ for a uniform source filling the LWRS.
Two channels have an Al+LiF optical coating while the other two
are SiC providing sensitivity for the $\sim$990 - 1187 \ang\, and
$\sim$905 - 1070 \ang\, ranges respectively.  The spectra are imaged onto
two micro-channel plate (MCP) detectors (1 \& 2) each with two sides (A \& B).  
The four channels and two sides to each MCP produce eight individual spectra
for every \fu\, observation.  Telescope acquisition and guiding is controlled 
by a fine error sensor which is
aligned with the LiF 1 channel for the observations in this paper.

\subsection{Sample Selection}

One of the primary motivations for this work is to study the properties 
of  \ovi\, absorption in star-forming galaxies.
The \ovi\, feature is actually a doublet at 1032 and 1038 \ang.
In our galaxies the feature is optically thin, and the $\lambda$1038 line will be half
as strong as the $\lambda$1032 line. Since the weaker $\lambda$1038 line is severely blended
with \ciiw\, and \oibw, we do not attempt to measure it in our spectra.
Therefore, our sample selection is driven by the desire to
pick star-forming galaxies in which we {\it could} detect \oviw.
While \oviw\, is generally a prominent feature in these galaxies, it can
be very broad and is found in a complicated region of the
spectra.   The wavelength region around \oviw\,
is particularly intricate due to the presence of several other strong features 
most noteably \lybw\, and \ciiw.  More problematically, Milky Way (MW) \ovi\, and \cii\,
are strong absorption features that may also fall on the observed redshifted extra-galactic \oviw\, feature.
Therefore, we cut galaxies with systemic
velocities $\rm{v_{sys}}<600\,\ks$ and $1000<\rm{v_{sys}}<1500\, \ks$
as extragalactic \ovi\, would then be blended with the strong
Milky Way \ovi\, and \cii\, features respectively.  The two highest redshift galaxies in the 
original sample were also dropped as the \ovi\,was shifted onto the gap between the 
LiF channels at $\sim$1090\ang.  
We also require 
S/N $> 4$ in 0.078 \ang\, bins in the LiF 1A galaxy spectrum.  The
LiF 1A spectrum is the highest S/N spectrum
of the eight spectra that are obtained during every \fu\, exposure.
This reduced our sample from the 54 star-forming
galaxies we found in the archive to only 15 objects.  We have however made an exception for 
one galaxy, NGC 4214.  One of the side-effects of dropping galaxies
with $\rm{v_{sys}}<600\,\ks$ is that we dropped a majority of the lower SFR galaxies found in the archive.  
In the case of NGC 4214, it is a particularly high quality dataset (S/N=14.6) with sufficiently narrow 
absorption features so that they are cleanly separated from the MW features.  We therefore include
NGC 4214 in our sample for a total of 16 galaxies.  Table \ref{t:fuselumins} lists
the galaxies in our sample along with many of their physical properties.  

\begin{figure}
\centering
\leavevmode
\includegraphics[width=6in]{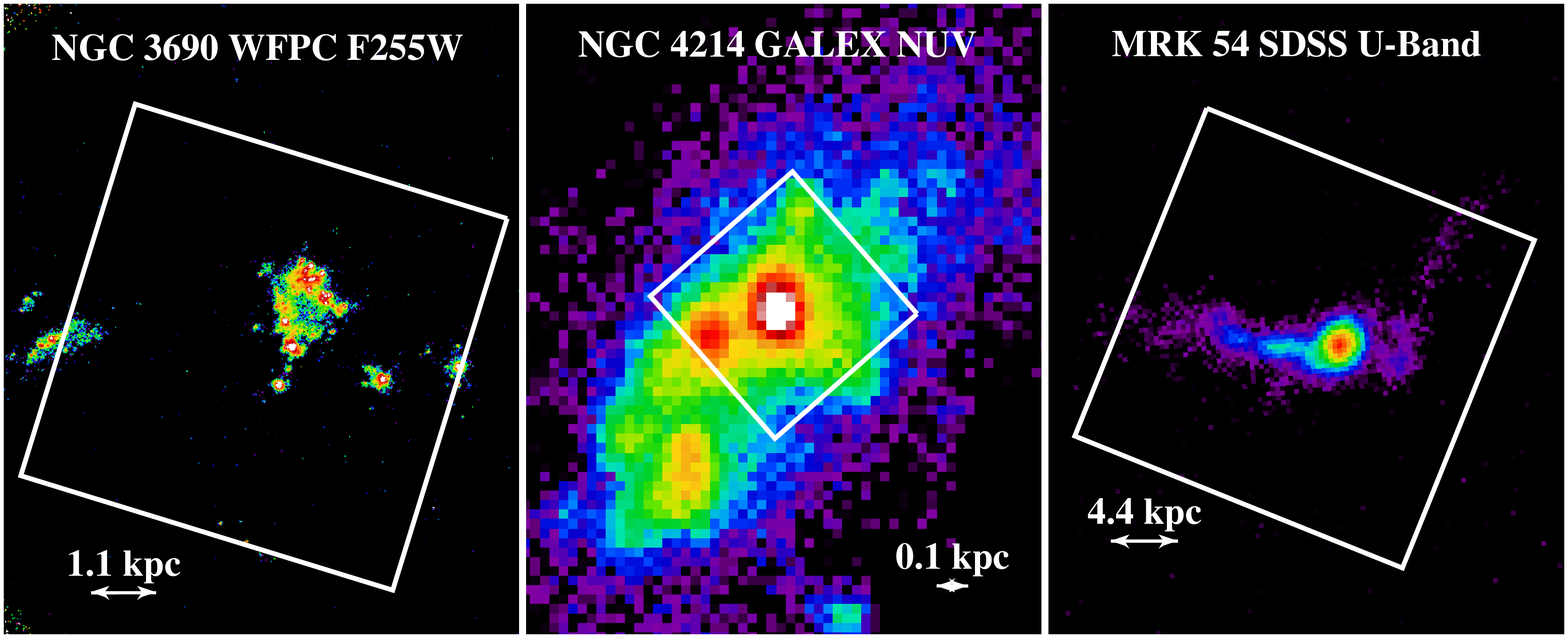}
\includegraphics[width=6in]{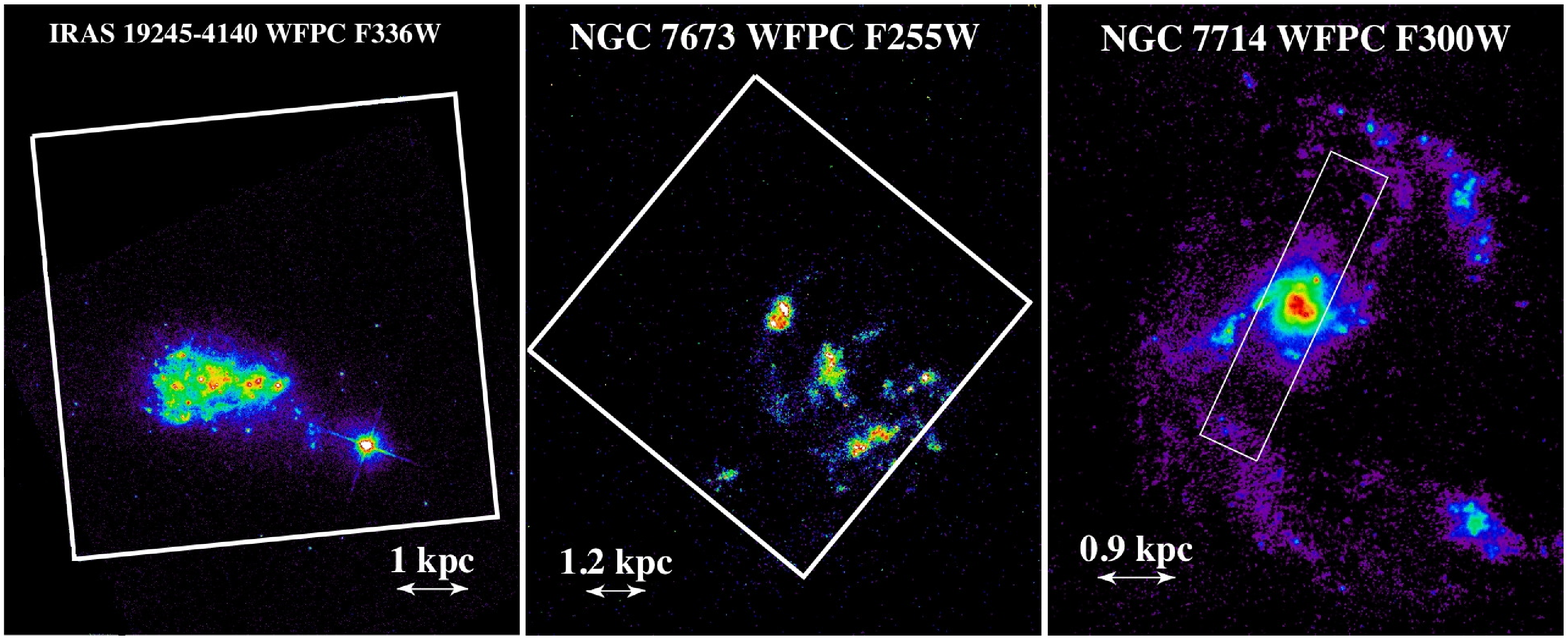}
\caption{
UV images with \fu\,aperture overlay for NGC\,3690, NGC\,4214, Mrk\,54,
 \iron, NGC\,7673, and NGC\,7714.  The \fu\, LWRS aperture has been overlaid at the nominal
position and orientation for the \fu\,observations.   The images
of NGC\,4214 ({\sl GALEX } NUV) and Mrk\,54 (SDSS U-band) are at significantly lower
spatial resolutions than observed in the other panels.
\label{f:images2}}
\end{figure}

The \fu\, apertures for these observations are generally pointed directly
at the location of the most intense SF in every galaxy.
As the FUV emission of most of the starbursts in the sample
is strongly peaked near the galaxy center,
the LWRS aperture includes the majority of the FUV emission
for each galaxy.  We are therefore probing the FUV properties
of the entire galaxy, similar to observations of high redshift LBGs.
The lone MDRS pointing, NGC\,7714 has a compact nuclear starburst
so that the majority of FUV emission of the galaxy is enclosed even in the smaller
aperture.  We have overlaid the \fu\,aperture and orientation on FUV or UV images for
15 of the galaxies in this sample in Figures \ref{f:images1} and \ref{f:images2}.  \tol\,
has been excluded as a UV image of that object is not available.  
All of the images in this sample are continuum UV images although the majority are at significantly
 longer wavelengths than our \fu\, spectroscopy.  Most of the images are {\sl HST} WFPC 
 using either the  F255W ($\lambda_c=2599$\,\AA), F300W ($\lambda_c=2987$\,\AA), or F336W 
 ($\lambda_c=3344$\,\AA) filters.  All but three of the images are from high spatial resolution
 {\it HST} instruments with the exception of the low resolution images of Mrk\,54 
 (SDSS U-Band), NGC\,4214 ({\it GALEX} NUV), and Haro\,3 ({\it GALEX} NUV).  
 A logarithmic image scaling has been used to highlight the various features
 of these galaxies although it does somewhat misrepresent the relative strength
 of the central features.  Specifically, only NGC\,3310 has significant extended emission
 outside of the central regions of the \fu\,aperture.
 
 The selected galaxies in our sample represent a wide range in galaxy types
 and properties.  The galaxies span over two orders of magnitude in 
 stellar mass and nearly three orders of magnitude in SFR.  We have included several dwarf galaxies (e.g. \izw, NGC\,4214,
 NGC\,1705), spirals (NGC\,3310, NGC\,7714), compact ultraviolet luminous galaxies 
 (UVLG; \vv, \hoo, Mrk\,54), and highly IR-luminous mergers (NGC\,3690, \vv).  
 The UVLGs are of particular interest due to their 
 strong similarities to LBGs in terms of their masses, UV luminosities, and
 UV surface brightnesses \citep{heck05,hoop06,over08}.  
 

\subsection{Data Reduction}

We have chosen to use a consistent data
reduction method throughout the sample aimed at maximizing S/N.  
This method involves several extra analysis steps 
but provides superior data quality compared
to the standard data reduction method for the targets in our sample.
While some of the observations have high
S/N ratios in the LiF channels, the lower effective area SiC 
channels are almost always relatively low S/N.  Our study focuses principally 
on the LiF channel data
but we use the SiC data for measuring the very strong \ciiiw\,absorption line and
investigating possible escaping Lyman continuum emission.

\begin{figure}
\centering
\leavevmode
\includegraphics[width=4in]{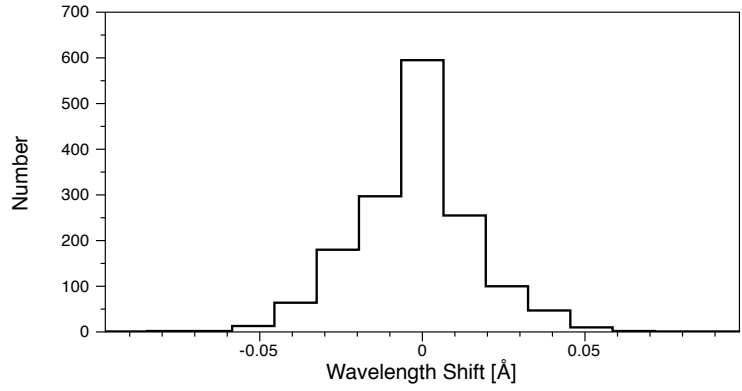}
\caption{
Histogram of all of the individual wavelength offsets used to correct for 
zero point shifts in the wavelength calibrations.  The vast majority
of the corrections are less than 0.02 \AA\, (6 km s$^{-1}$).  The overlaid arrows
are 5\arcsec in length.
\label{f:wlshifts}}
\end{figure}

The first step was to run all  raw exposure data through the CalFUSE 3.1.8 
data pipeline to produce intermediate data (IDF) files.
We then extracted all eight flux calibrated spectra for every \fu\, exposure.  
See \citet{dix07} for a complete discussion of the calibrations and corrections applied 
to the data.  Mirror misalignment and thermally-induced
rotations of the spectrograph gratings cause
small zero point shifts in the wavelength calibration.  
Therefore we cross calibrate the flux calibrated exposures
to determine the individual zero-point wavelength
shifts.  Given that most of our observations
are relatively low S/N spectra of extended sources
it is difficult to accurately determine many of these wavelength offsets.
We subsequently have chosen conservative offsets.
A histogram of the offsets used for all of the individual exposures in
shown in  Figure \ref{f:wlshifts}.  The majority of the offsets are 
$< 0.013$ \ang\, ($< 4\, \ks$) which is significantly less than the wavelength inaccuracies
caused by localized distortions of the detector (0.025 \ang\, $\sim 8\, \ks$).  

After determining the wavelength offset for each individual spectra we retroactively apply
the correction to the individual IDF files.  
We also examine the count rates of the individual exposures
and throw out the exposures with significant discrepancies.
These variations are usually caused by misalignment of a mirror
on the target.  The IDF files for each exposure are then combined
to create one total IDF file for each channel.  
We then re-extract a combined flux calibrated spectrum which
also has the zero-point wavelength corrections applied.  More
importantly, however, the  longer, combined IDF files 
allow the CalFUSE pipeline to do an improved background subtraction.
In particular, the pipeline fits the various backgrounds (e.g. detector, airglow, and geo-coronal
scattered light).  For the shorter, individual exposures it simply scales the background
files by exposure time when extracting flux calibrated spectra.  
As there can be significant background variations, exposure time
scaling can introduce significant background subtraction errors.
The backgrounds
dominate the signal for many of these spectra so that errors in the background subtraction
introduces systematic offsets in the flux scale.  Even when using
background fitting, these systematic offsets can still be seen in many
of the low S/N SiC spectra.

In order to minimize the detector background, we have extracted our
spectra using the smaller, point source detector regions \citep{dix07} for 
all objects in our sample with the exception of NGC\,1705, NGC\,3310, and NGC\,4214.
A visual examination of the 2-D detector images shows that these three
galaxies have source counts falling outside of the point source aperture.  
These are the three highest flux galaxies within our sample.
Figure \ref{f:images1} suggests
that at least for NGC\,3310, the galaxy has considerable extent within the
\fu\, aperture. By using the smaller point 
source region for the remaining 14 galaxies we include
the same amount of source flux but minimize the included background.

For galaxies which contain  a significant fraction of night only data, 
we have thrown out the daytime data.  Significant airglow features contaminate
the day time spectra and are best avoided whenever possible.  Table \ref{t:fusesample}
lists the \fu\, observational data for all of the galaxies included in our sample.

\subsection{Data}

\begin{figure}
\centering
\leavevmode
\includegraphics[width=5in,angle=90]{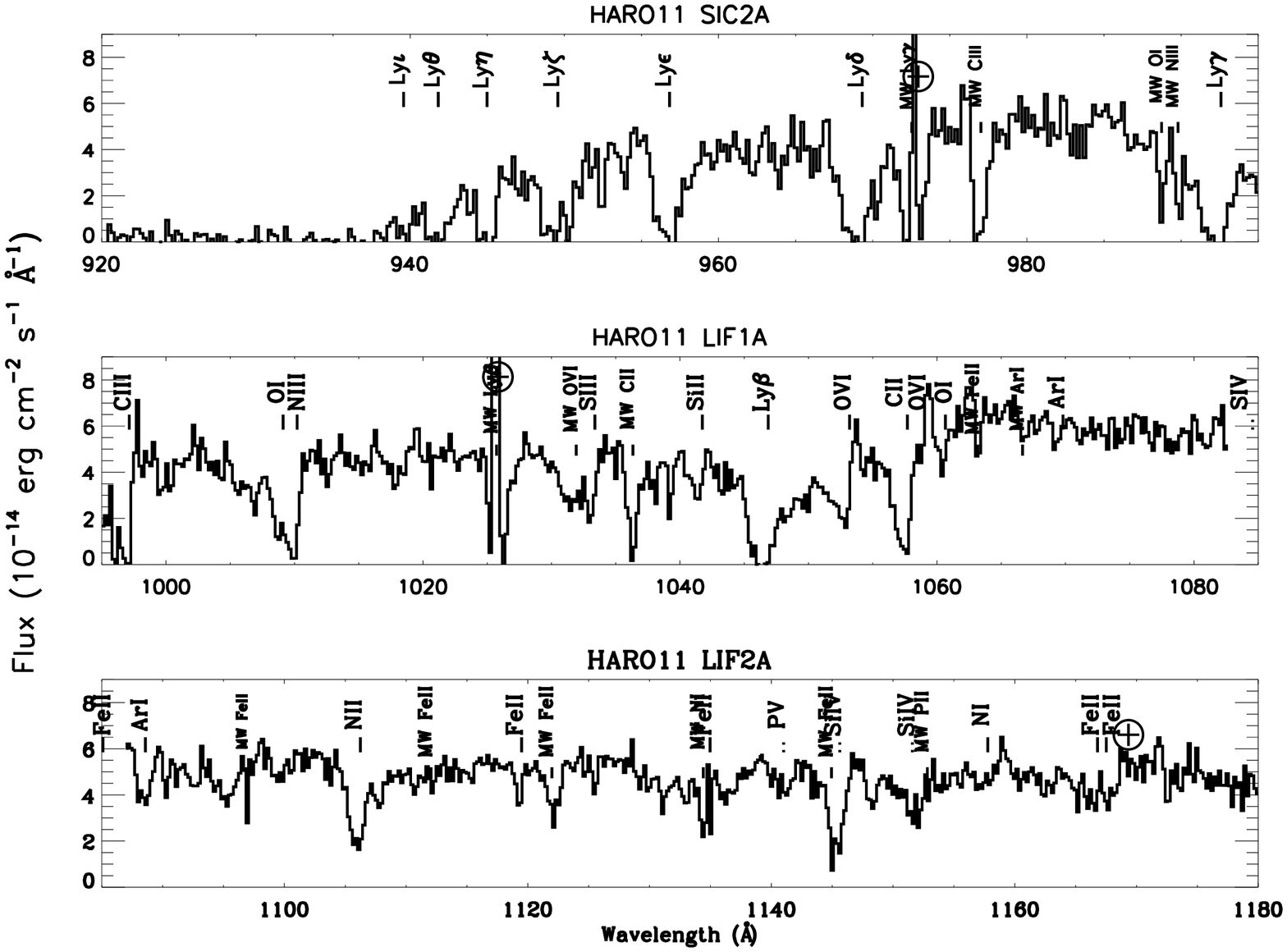}
\caption{Spectra of Haro\,11 from the {\it FUSE} SiC 2A, LiF 1A, and LiF 2A channels.
Locations of prominent ISM, stellar photospheric (dashed lines), Milky Way, and airglow ($\earth$) features have 
been identified.
\label{f:Haro11spec}}
\end{figure}

\begin{figure}
\centering
\leavevmode
\includegraphics[width=5in,angle=90]{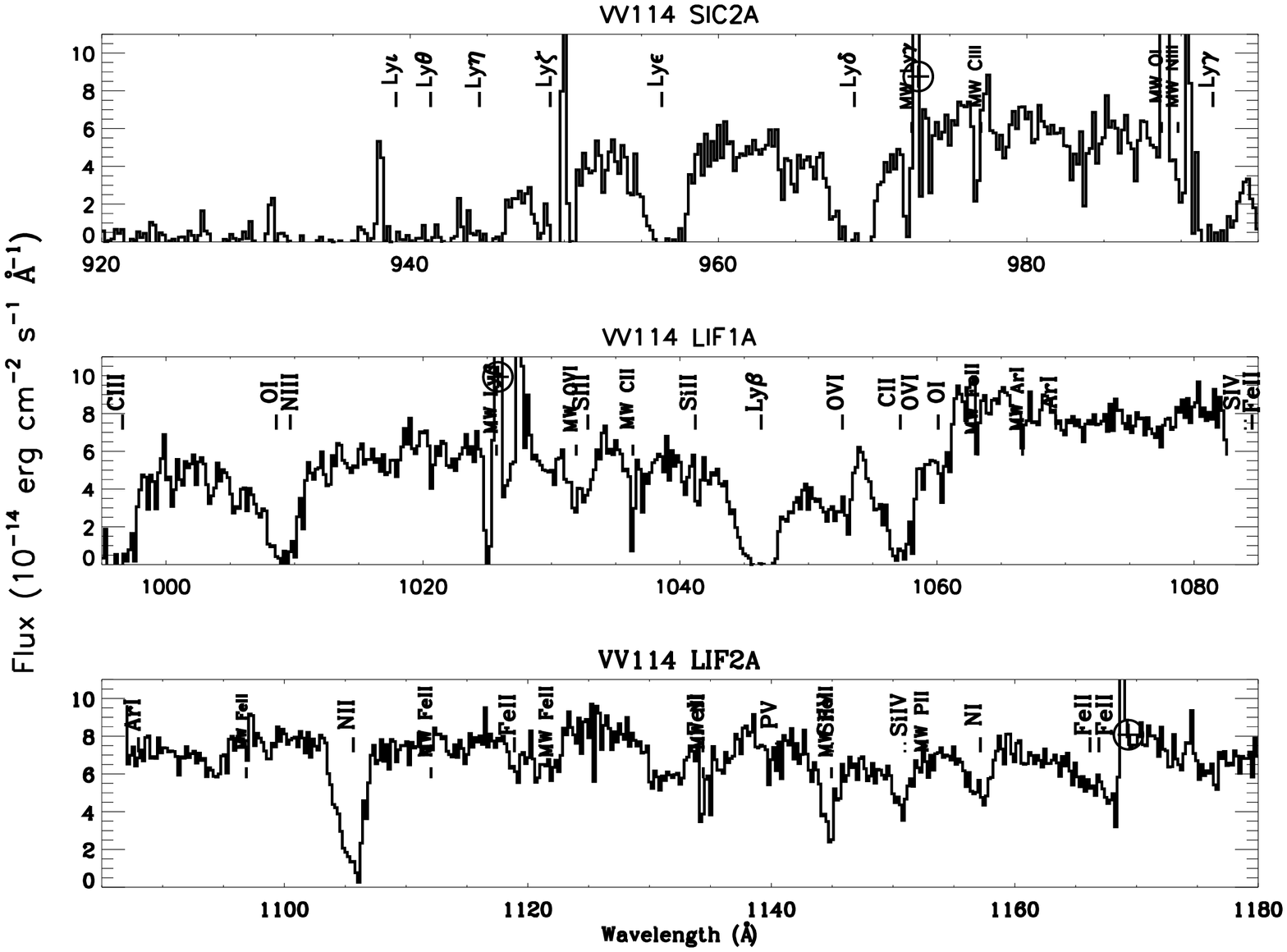}
\caption{Spectra of VV\,114 from the {\it FUSE} SiC 2A, LiF 1A, and LiF 2A channels.
Locations of prominent ISM, stellar photospheric (dashed lines), Milky Way, and airglow ($\earth$) features have 
been identified.
\label{f:VV114spec}}
\end{figure}

\begin{figure}
\centering
\leavevmode
\includegraphics[width=5in,angle=90]{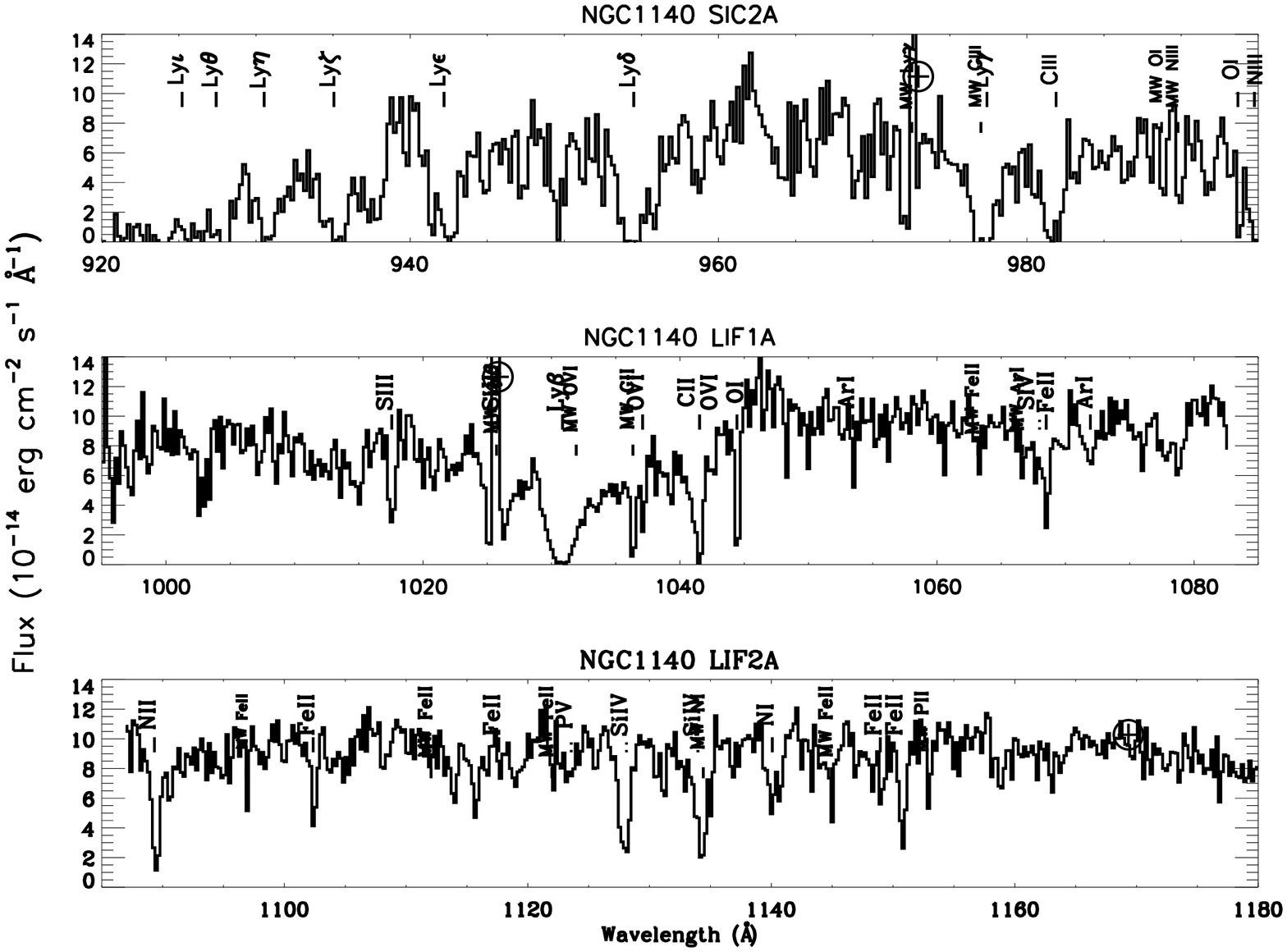}
\caption{Spectra of NGC1140 from the {\it FUSE} SiC 2A, LiF 1A, and LiF 2A channels.
Locations of prominent ISM, stellar photospheric (dashed lines), Milky Way, and airglow ($\earth$) features have 
been identified.
\label{f:NGC1140spec}}
\end{figure}

\begin{figure}
\centering
\leavevmode
\includegraphics[width=5in,angle=90]{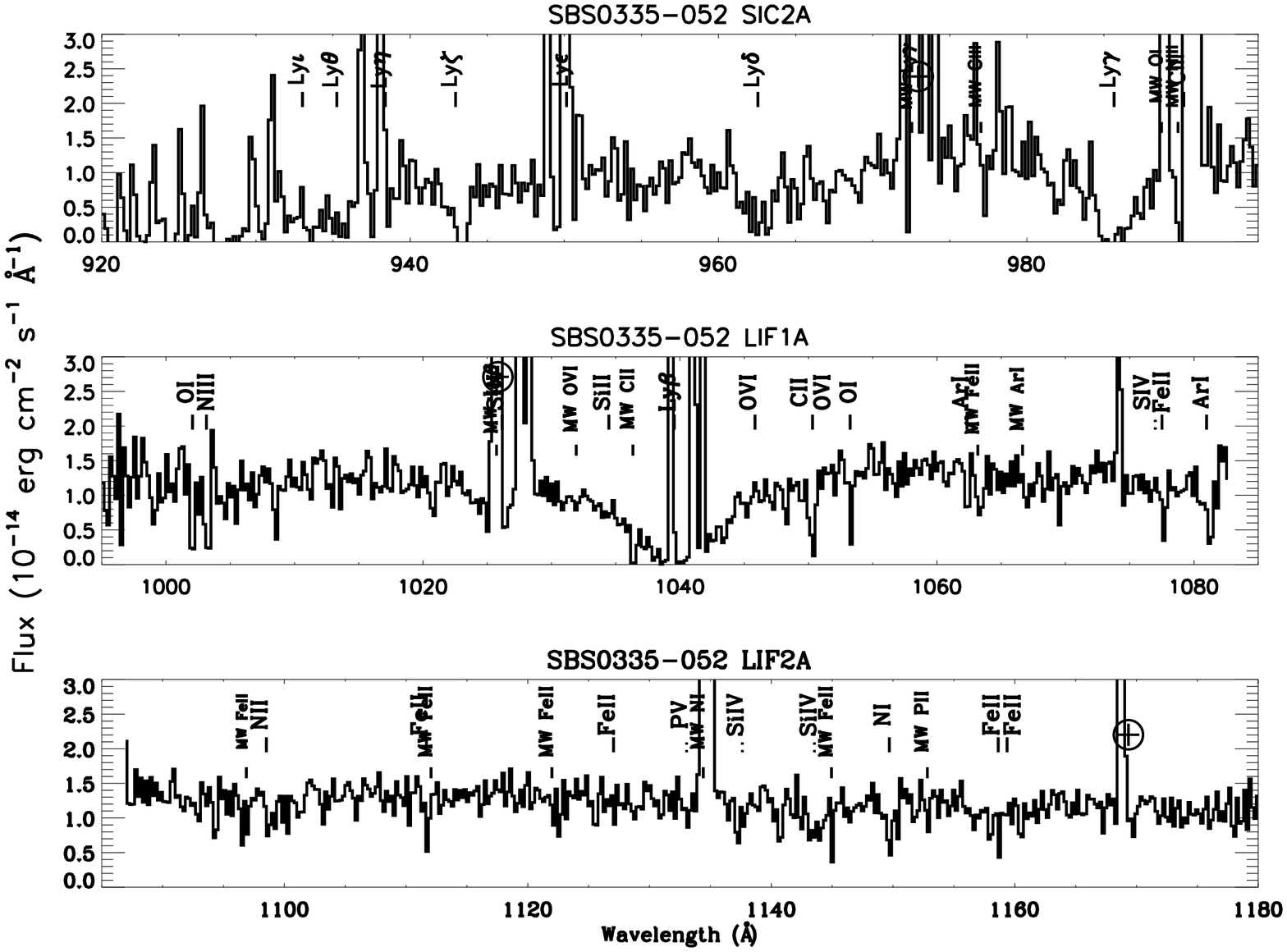}
\caption{Spectra of SBS\,0335-052 from the {\it FUSE} SiC 2A, LiF 1A, and LiF 2A channels.
Locations of prominent ISM, stellar photospheric (dashed lines), Milky Way, and airglow ($\earth$) features have 
been identified.
\label{f:SBS0335-052spec}}
\end{figure}

\begin{figure}
\centering
\leavevmode
\includegraphics[width=5in,angle=90]{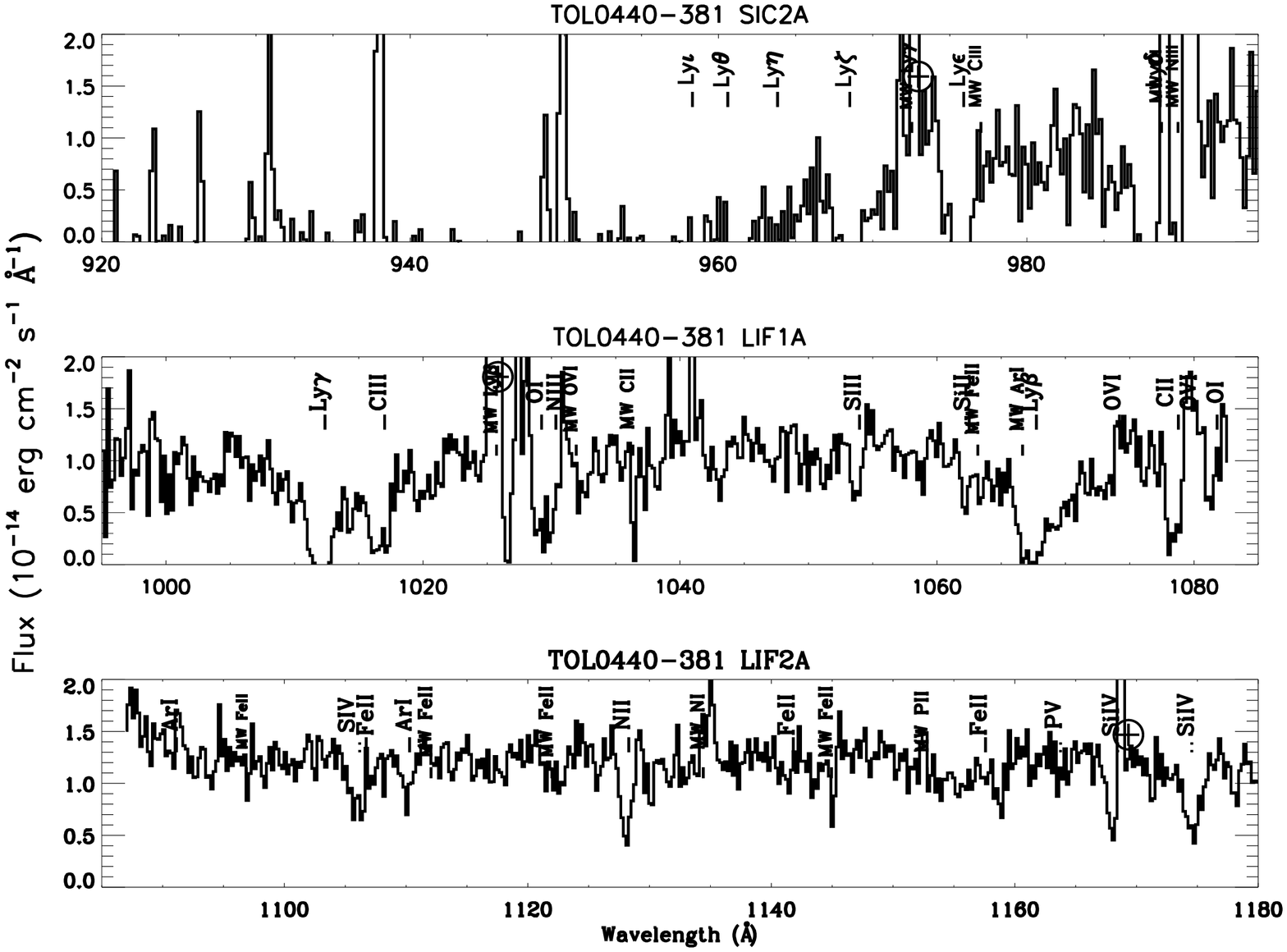}
\caption{Spectra of Tol\,0440-381 from the {\it FUSE} SiC 2A, LiF 1A, and LiF 2A channels.
Locations of prominent ISM, stellar photospheric (dashed lines), Milky Way, and airglow ($\earth$) features have 
been identified.
\label{f:Tol0440-381spec}}
\end{figure}

\begin{figure}
\centering
\leavevmode
\includegraphics[width=5in,angle=90]{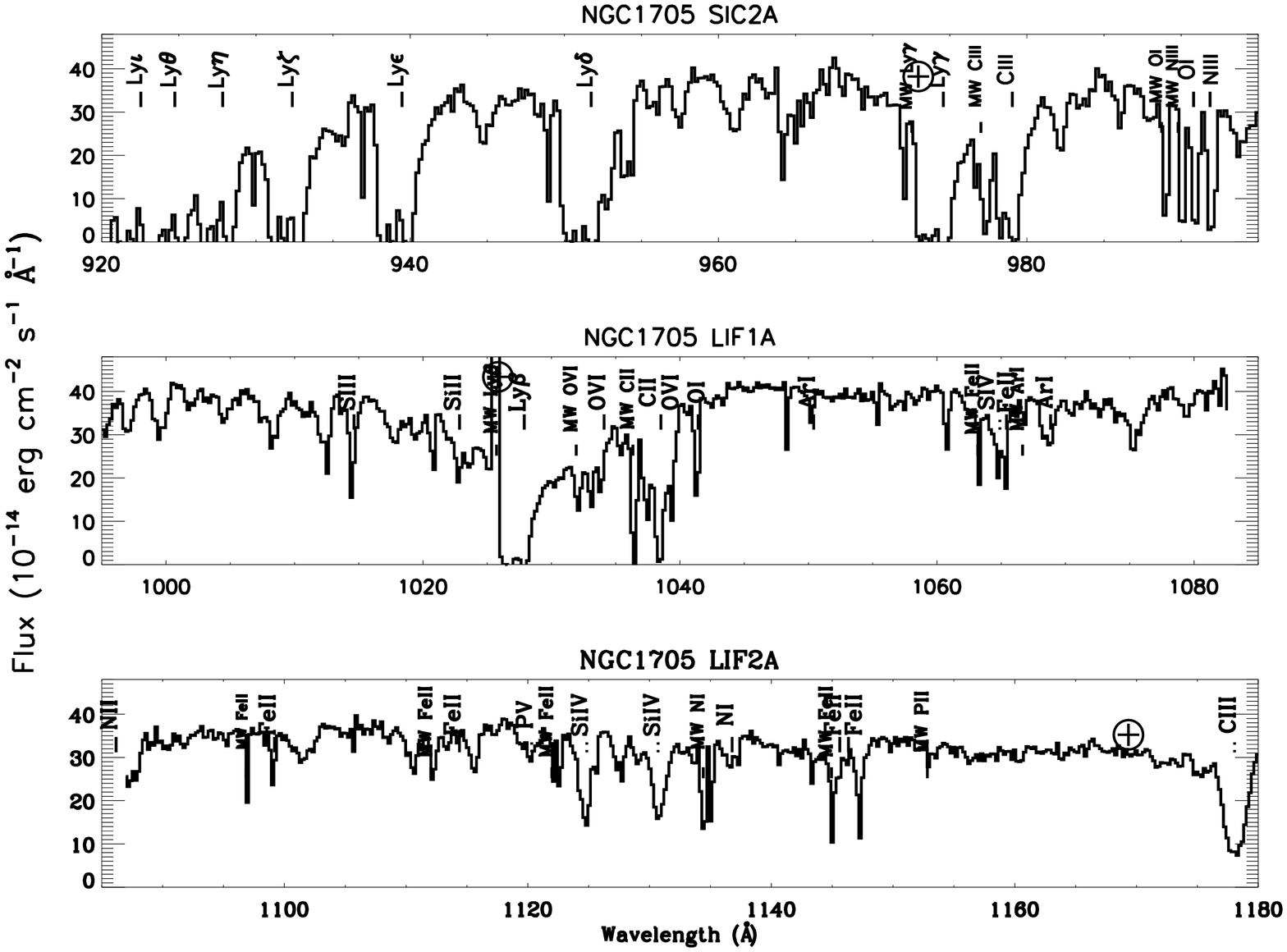}
\caption{Spectra of NGC\,1705 from the {\it FUSE} SiC 2A, LiF 1A, and LiF 2A channels.
Locations of prominent ISM, stellar photospheric (dashed lines), Milky Way, and airglow ($\earth$) features have 
been identified.
\label{f:NGC1705spec}}
\end{figure}

\begin{figure}
\centering
\leavevmode
\includegraphics[width=5in,angle=90]{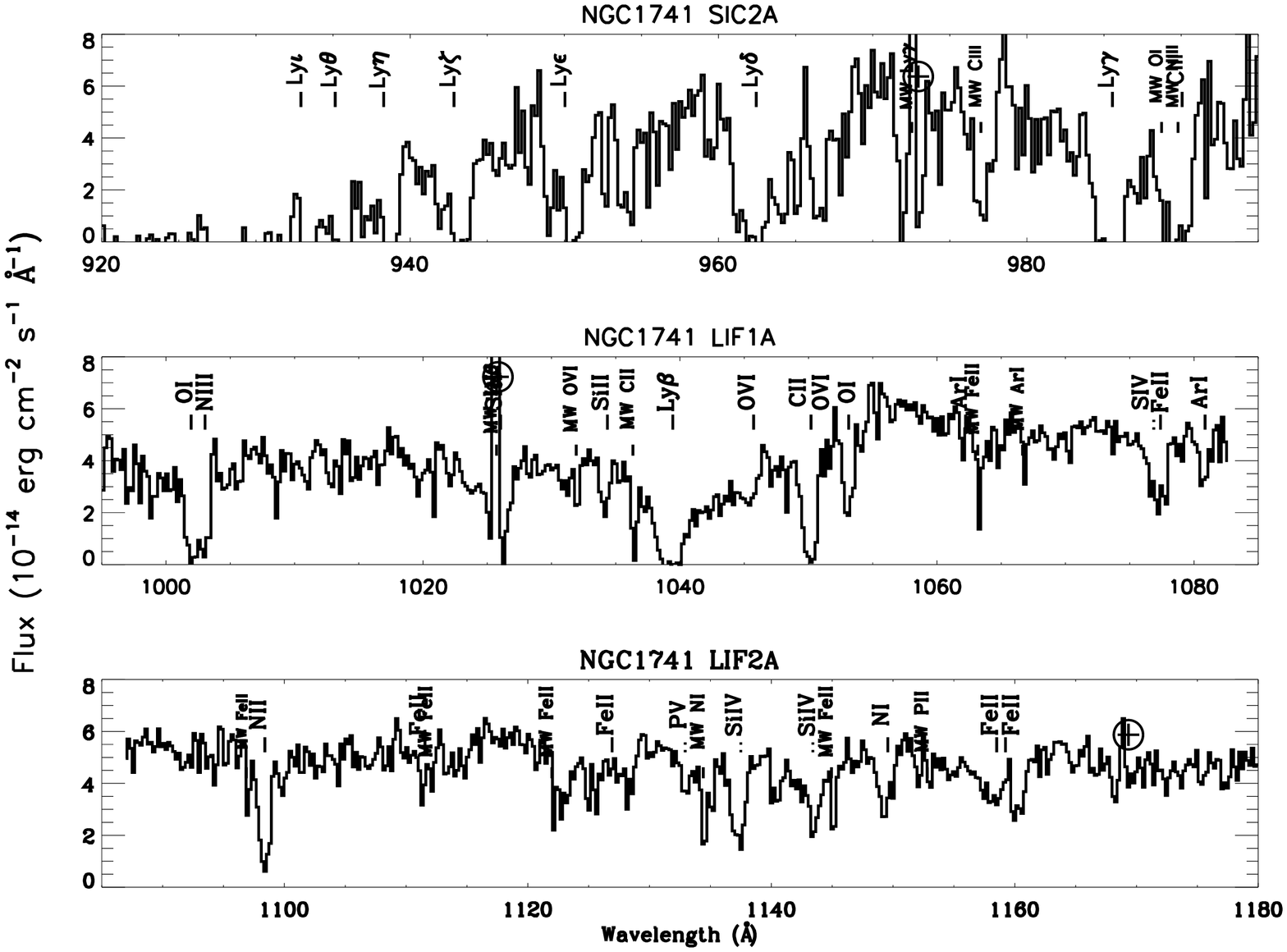}
\caption{Spectra of NGC\,1741 from the {\it FUSE} SiC 2A, LiF 1A, and LiF 2A channels.
Locations of prominent ISM, stellar photospheric (dashed lines), Milky Way, and airglow ($\earth$) features have 
been identified.
\label{f:NGC1741spec}}
\end{figure}

\begin{figure}
\centering
\leavevmode
\includegraphics[width=5in,angle=90]{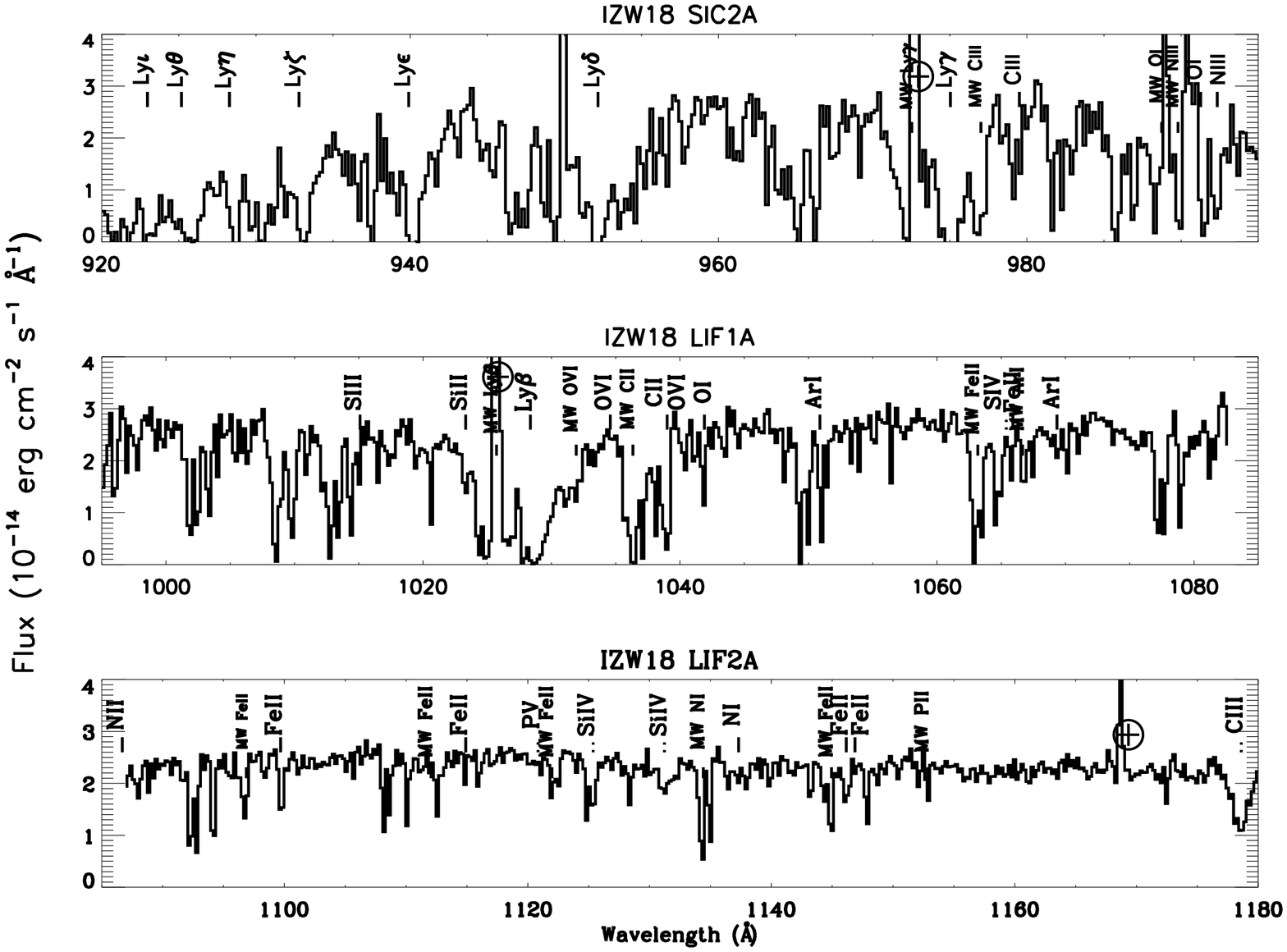}
\caption{Spectra of IZW\,18 from the {\it FUSE} SiC 2A, LiF 1A, and LiF 2A channels.
Locations of prominent ISM, stellar photospheric (dashed lines), Milky Way, and airglow ($\earth$) features have 
been identified.
\label{f:IZW18spec}}
\end{figure}

\begin{figure}
\centering
\leavevmode
\includegraphics[width=5in,angle=90]{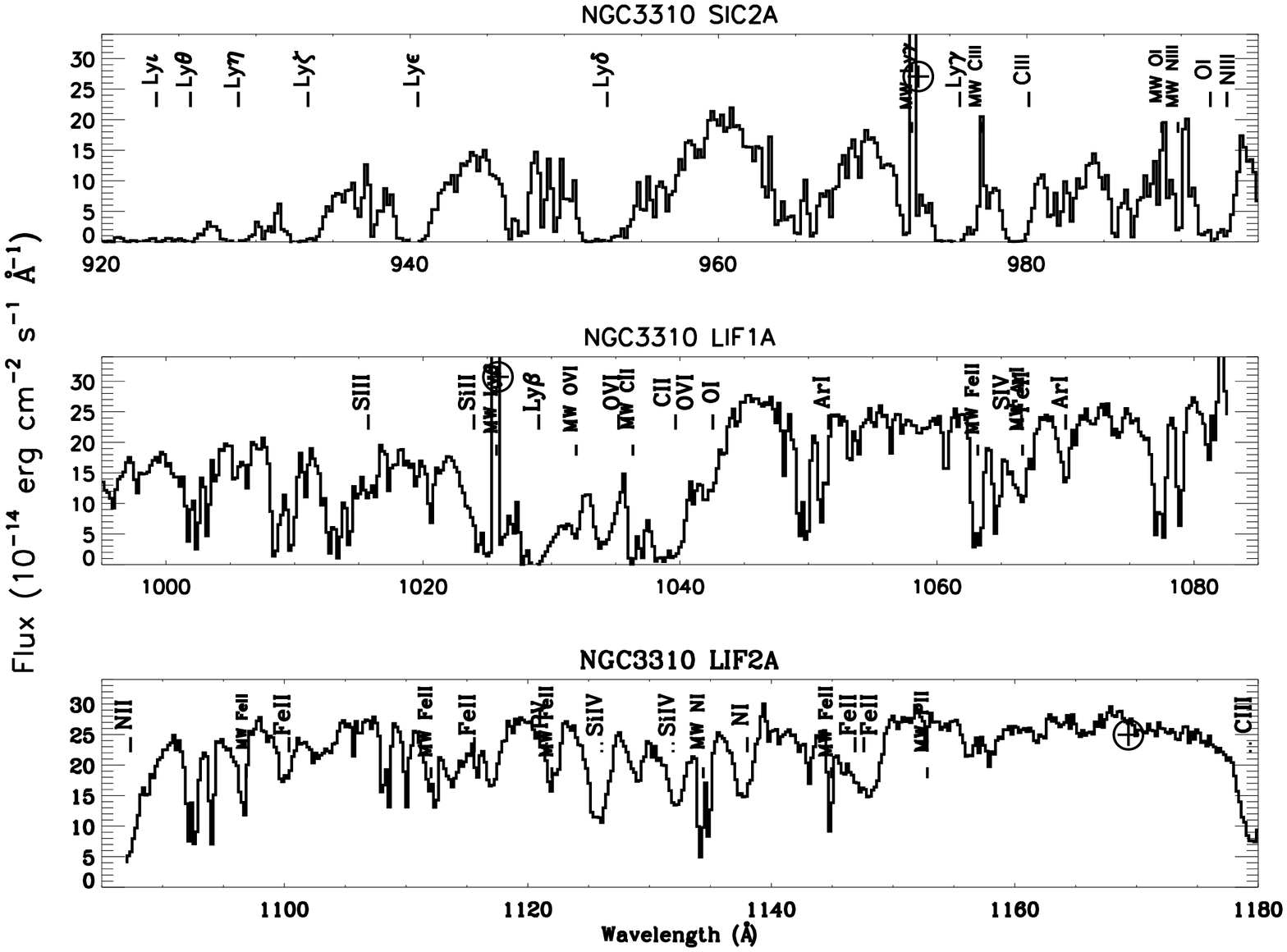}
\caption{Spectra of NGC\,3310 from the {\it FUSE} SiC 2A, LiF 1A, and LiF 2A channels.
Locations of prominent ISM, stellar photospheric (dashed lines), Milky Way, and airglow ($\earth$) features have 
been identified.
\label{f:NGC3310spec}}
\end{figure}

\begin{figure}
\centering
\leavevmode
\includegraphics[width=5in,angle=90]{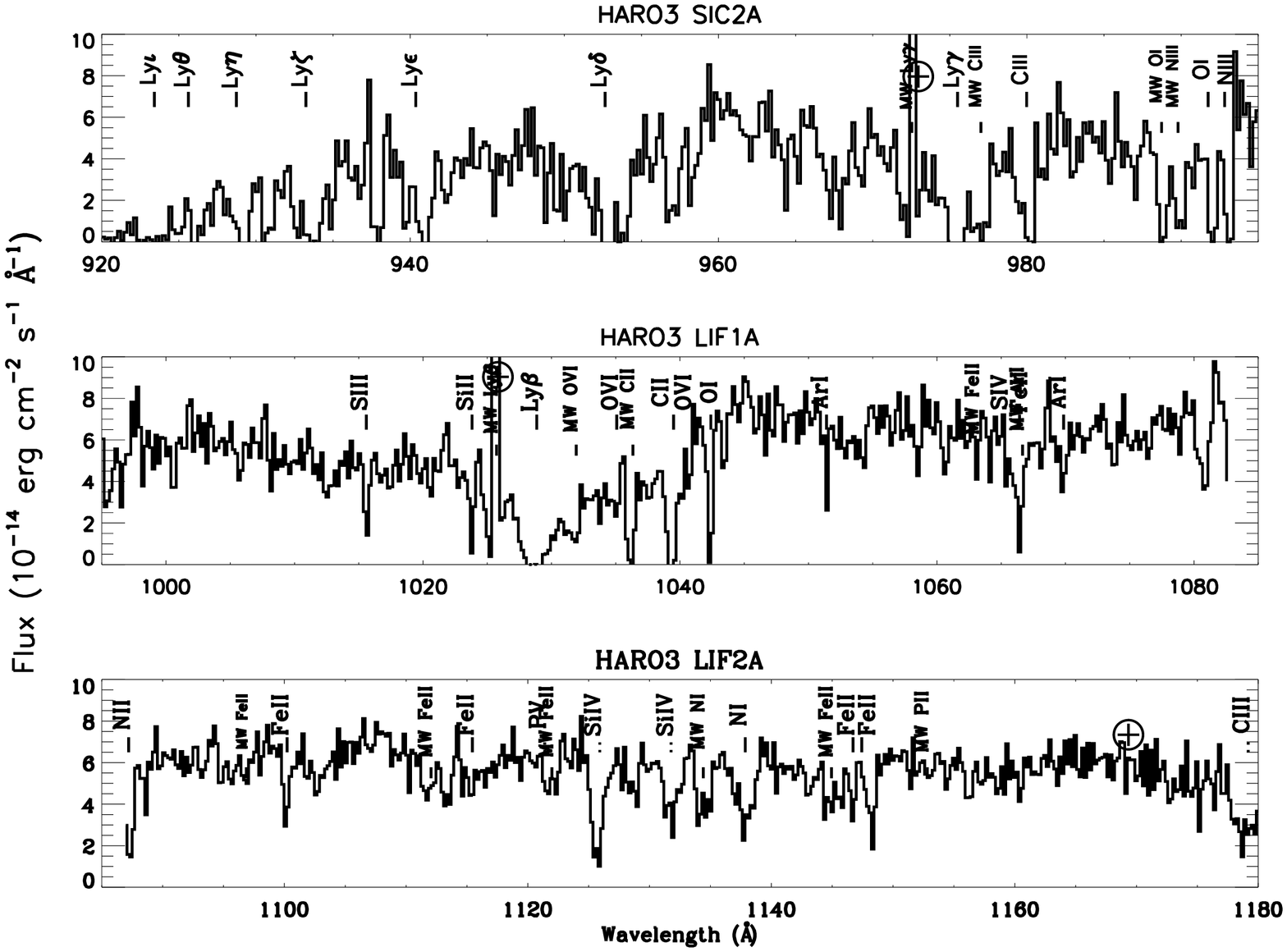}
\caption{Spectra of Haro\,3 from the {\it FUSE} SiC 2A, LiF 1A, and LiF 2A channels.
Locations of prominent ISM, stellar photospheric (dashed lines), Milky Way, and airglow ($\earth$) features have 
been identified.
\label{f:Haro3spec}}
\end{figure}

\begin{figure}
\centering
\leavevmode
\includegraphics[width=5in,angle=90]{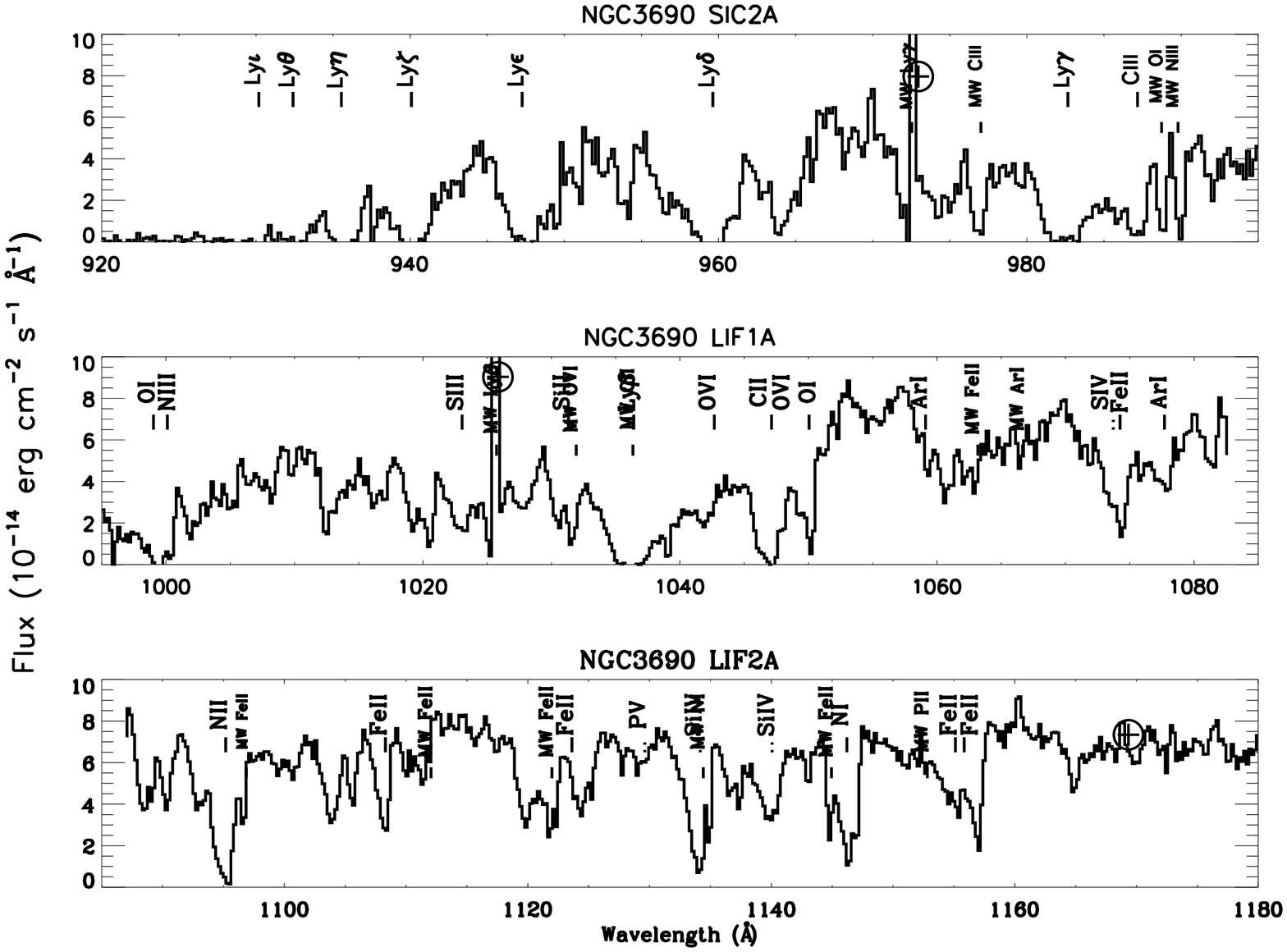}
\caption{Spectra of NGC\,3690 from the {\it FUSE} SiC 2A, LiF 1A, and LiF 2A channels.
Locations of prominent ISM, stellar photospheric (dashed lines), Milky Way, and airglow ($\earth$) features have 
been identified.
\label{f:NGC3690spec}}
\end{figure}

\begin{figure}
\centering
\leavevmode
\includegraphics[width=5in,angle=90]{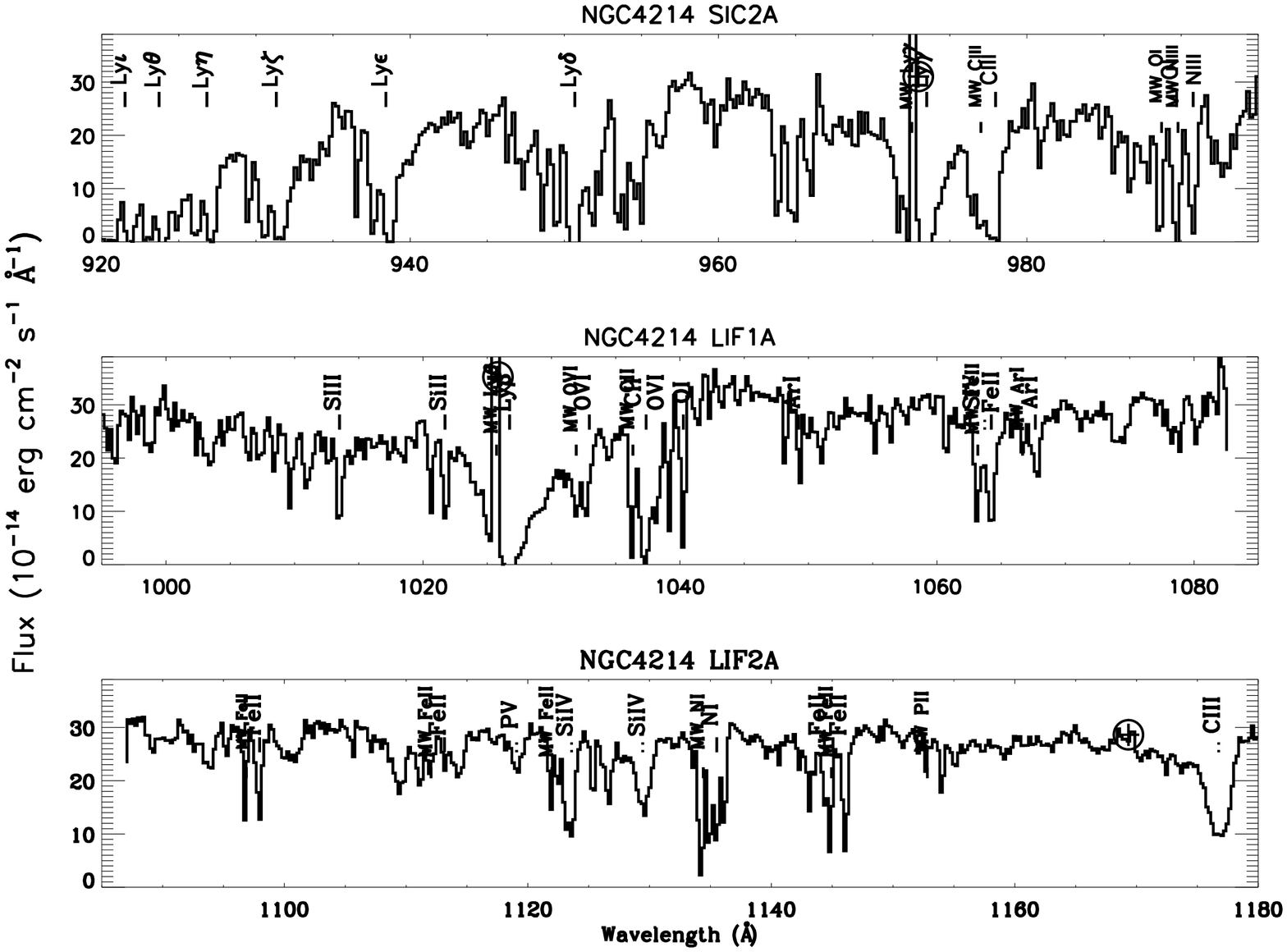}
\caption{Spectra of NGC\,4214 from the {\it FUSE} SiC 2A, LiF 1A, and LiF 2A channels.
Locations of prominent ISM, stellar photospheric (dashed lines), Milky Way, and airglow ($\earth$) features have 
been identified.
\label{f:NGC4214spec}}
\end{figure}

\begin{figure}
\centering
\leavevmode
\includegraphics[width=5in,angle=90]{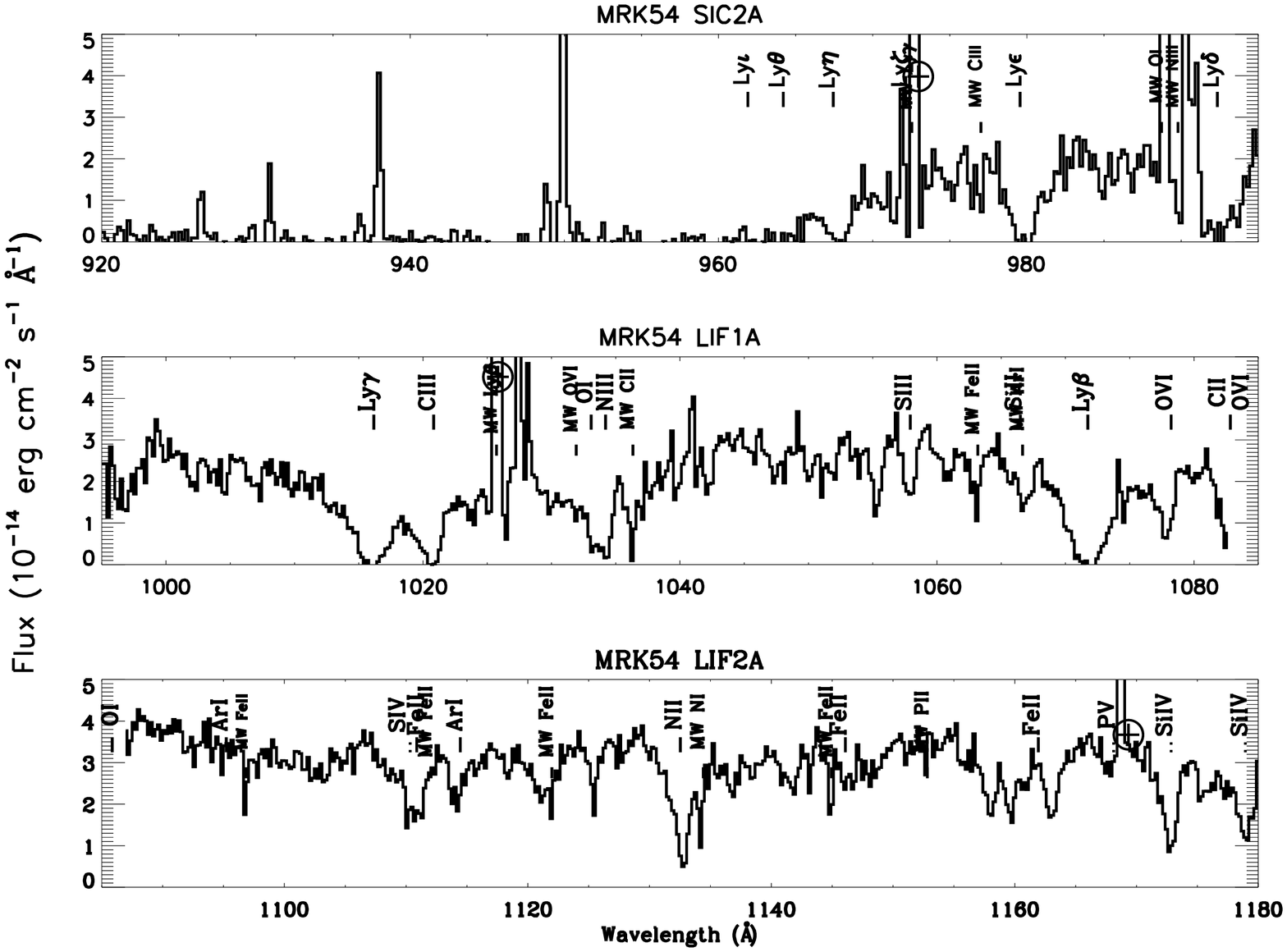}
\caption{Spectra of Mrk\,54 from the {\it FUSE} SiC 2A, LiF 1A, and LiF 2A channels.
Locations of prominent ISM, stellar photospheric (dashed lines), Milky Way, and airglow ($\earth$) features have 
been identified.
\label{f:Mrk54spec}}
\end{figure}

\begin{figure}
\centering
\leavevmode
\includegraphics[width=5in,angle=90]{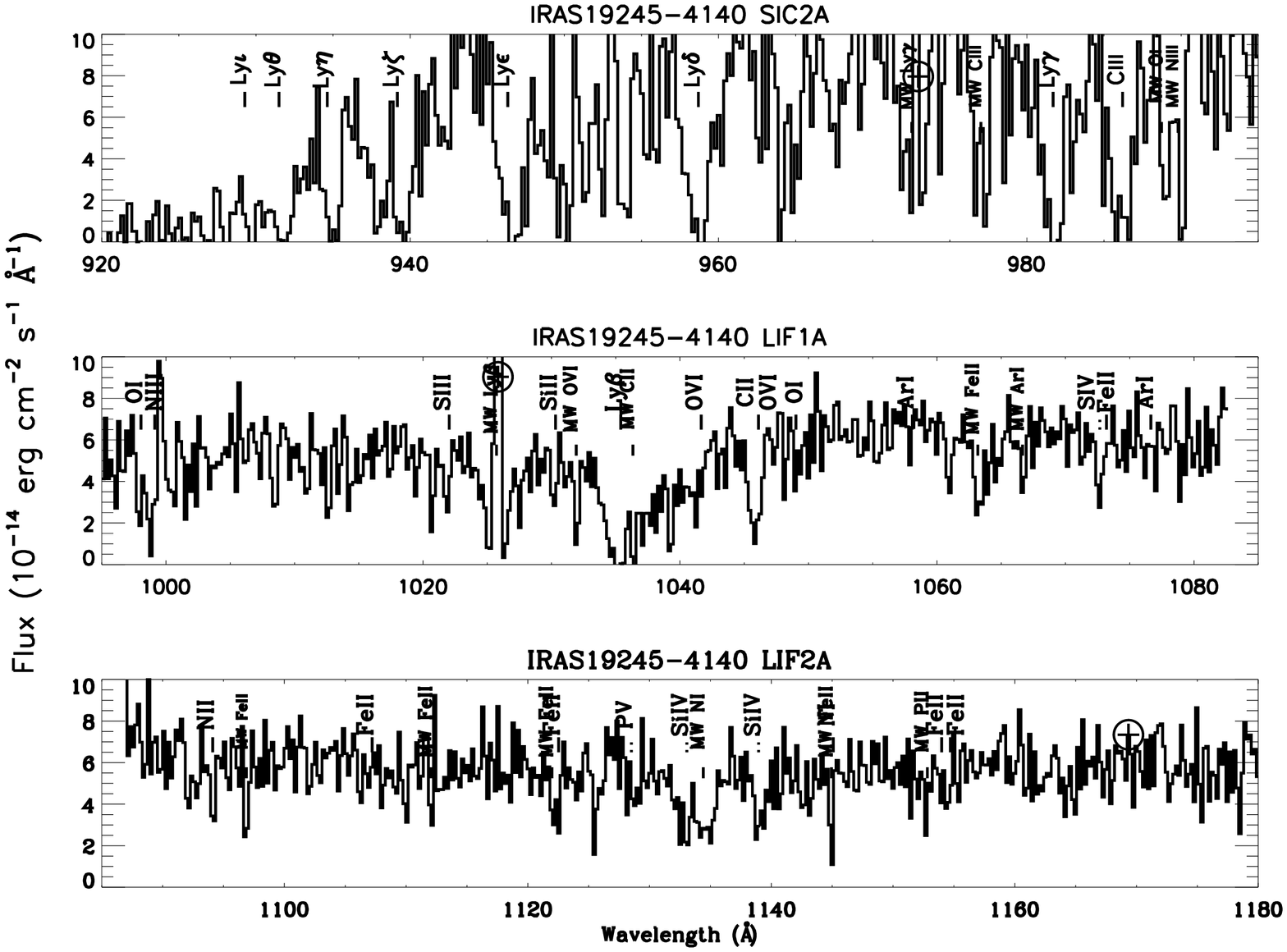}
\caption{Spectra of IRAS\,19245-4140 from the {\it FUSE} SiC 2A, LiF 1A, and LiF 2A channels.
Locations of prominent ISM, stellar photospheric (dashed lines), Milky Way, and airglow ($\earth$) features have 
been identified.
\label{f:IRAS19245-4140spec}}
\end{figure}

\begin{figure}
\centering
\leavevmode
\includegraphics[width=5in,angle=90]{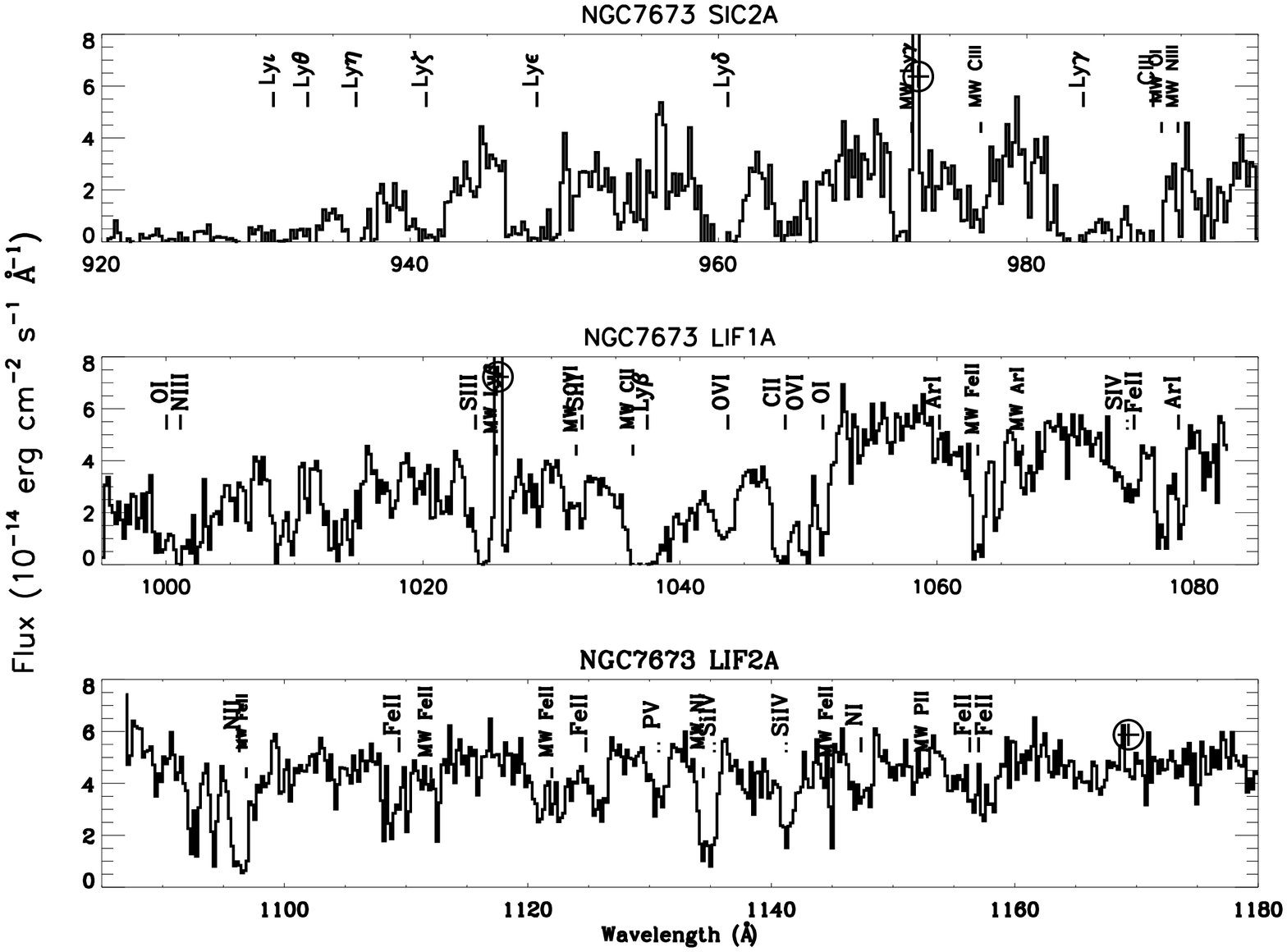}
\caption{Spectra of NGC\,7673 from the {\it FUSE} SiC 2A, LiF 1A, and LiF 2A channels.
Locations of prominent ISM, stellar photospheric (dashed lines), Milky Way, and airglow ($\earth$) features have 
been identified.
\label{f:NGC7673spec}}
\end{figure}

\begin{figure}
\centering
\leavevmode
\includegraphics[width=5in,angle=90]{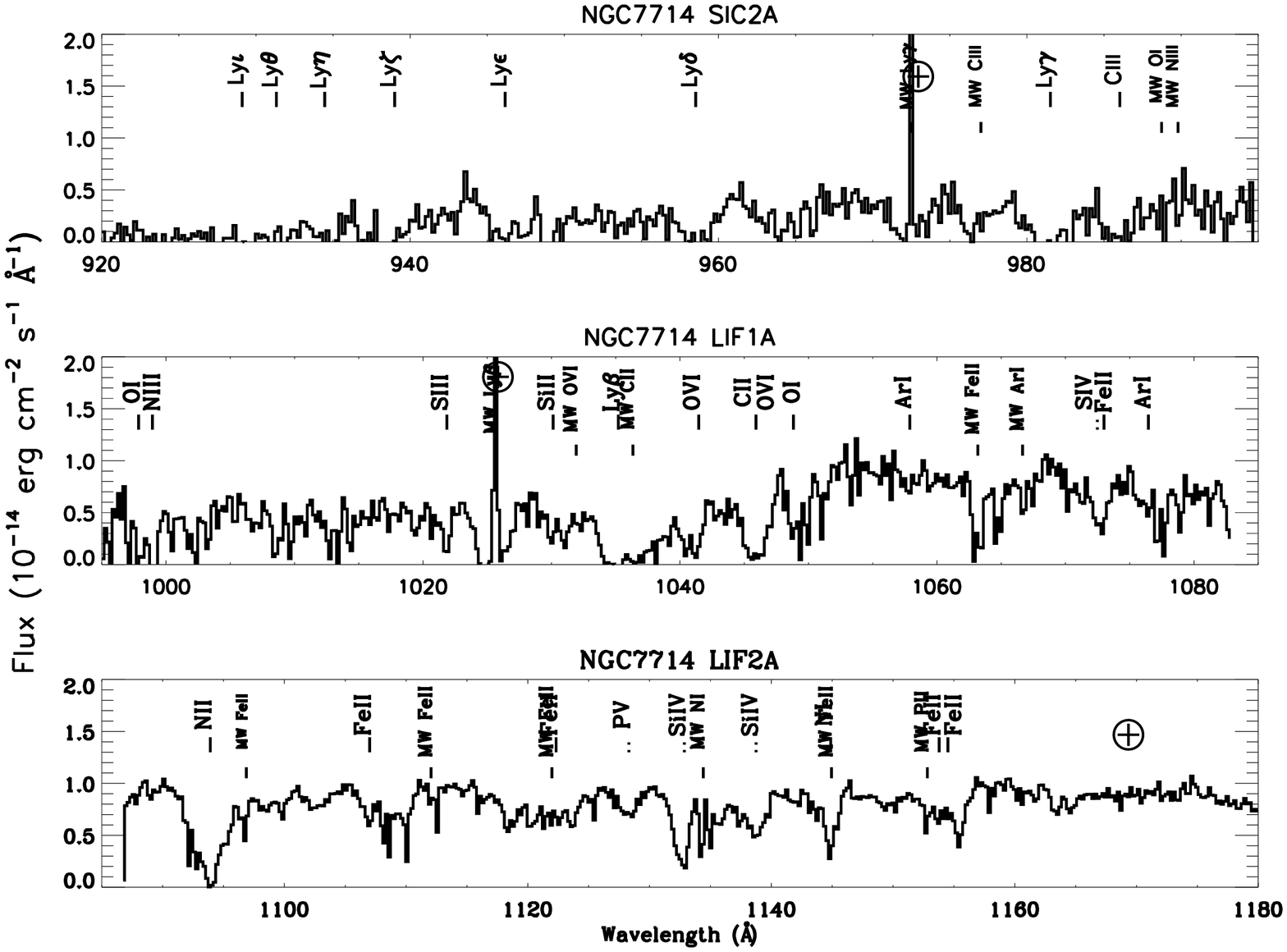}
\caption{Spectra of NGC\,7714 from the {\it FUSE} SiC 2A, LiF 1A, and LiF 2A channels.
Locations of prominent ISM, stellar photospheric (dashed lines), Milky Way, and airglow ($\earth$) features have 
been identified.
\label{f:NGC7714spec}}
\end{figure}

The SiC 2A, LiF 1A, and LiF 2A spectra for every galaxy
in our sample are shown in Figures \ref{f:Haro11spec} - \ref{f:NGC7714spec}. 
These channels cover most of the wavelength
regime between 920 - 1180 \ang.  We use LiF 2B for these panels instead of LiF 1B
as it is rarely affected by the ``worm''.  The ``worm'' is depression in the observed flux
that can span over 50 \ang\,in length \citep{dix07}.
Shadowing by a grid of quantum efficiency wires where the secondary focus falls above the detectors and close to the location of the wires causes these observed drops in flux.
This feature is seen in almost all of the
LiF 1B spectra in our sample but does not affect any of the other channels.

 Most of the common MW, galactic, and airglow features are
 marked on these figures.  The galactic absorption lines probe the stellar 
 population, interstellar medium (ISM), and outflowing gas.  
 \htwo, usually of MW origin, is present in many of the galaxies
 and is a significant hurdle in identifying and fitting many of the 
 weaker ISM absorption features.

\subsubsection{Gaussian Fits\label{s:gaussian}}

We have fit absorption lines to most of the most 
prominent features using the \iraf\, \citep{tod96}
 tool {\it specfit} \citep{kriss94}.  We have not attempted
 to fit every possible feature but have instead focused on
 the strongest, separable lines that we
 positively identify (this can be particularly difficult in
 galaxies contaminated by MW \htwo).
 Each line was fit using a freely varying powerlaw for 
 the continuum and a symmetric gaussian absorption line.  
 When absorption lines 
 are  blended (e.g. \ion{O}{1}~$\lambda$989 and \ion{N}{3}~$\lambda$990) we 
 fit them simultaneously. Due to the differences in the redshifts of the
galaxies (and hence differing degrees of blending or contamination by 
lines due to the Milky Way), it was not possible to consistently define
the spectral windows over which the adjacent continuum was fit for any given
absorption feature.

We note that the compact UV sizes of most of these galaxies implies
that the velocity resolution in our spectra willi typically  be roughly $20 \ks$ (FWHM).
The worst resolution will be for NGC 3310, which based on its angular size
we estimate will be about $60 \ks$. In all cases, the observed absorption-lines
are well-resolved. 

 For the majority of absorption lines we have used the LiF 
channels to make two independent measurements of the equivalent width, velocity, and 
full-width half-maximum (FWHM).  For lines with only one
measurement, the line either fell in the gap between LiF 
channels or was in the region (1125-1160\AA) 
affected by the worm on the LiF 1B channel.  
We have generally ignored the SiC channels above 1000 \AA\,
due to their lower S/N.  The results of our fits are listed in Tables 
\ref{t:HARO11dat} - \ref{t:NGC7714da}.

The listed statistical errors are $1\sigma$ and are
calculated from the minimization error matrix which is 
re-scaled by the reduced $\chi^2$ value 
\citep{kriss94}.  Systematic errors are probably not
large in the FWHM measurement but may 
be important to the equivalent width and centroid velocity.
The equivalent width determination is largely a problem for
saturated lines in low S/N spectra.  Background over-subtraction
is a common problem for these spectra so it is difficult
to determine the continuum level and flux zero point. 
Systematic wavelength calibration errors exist and are particularly
hard to quantify for our extended sources using the LWRS aperture.  
\citet{dix07} suggest that they are order 10 \kms\, which seems
consistent with the observed variations in our results.  
A comparison of the derived parameters for 
features with data on two LiF channels shows broad agreement,
generally well within the $1\sigma$ error bars as  expected.
In these cases, we have used the error
weighted value to create a single measurement for the analysis that follows.

\begin{figure}
\centering
\leavevmode
\includegraphics[width=4in]{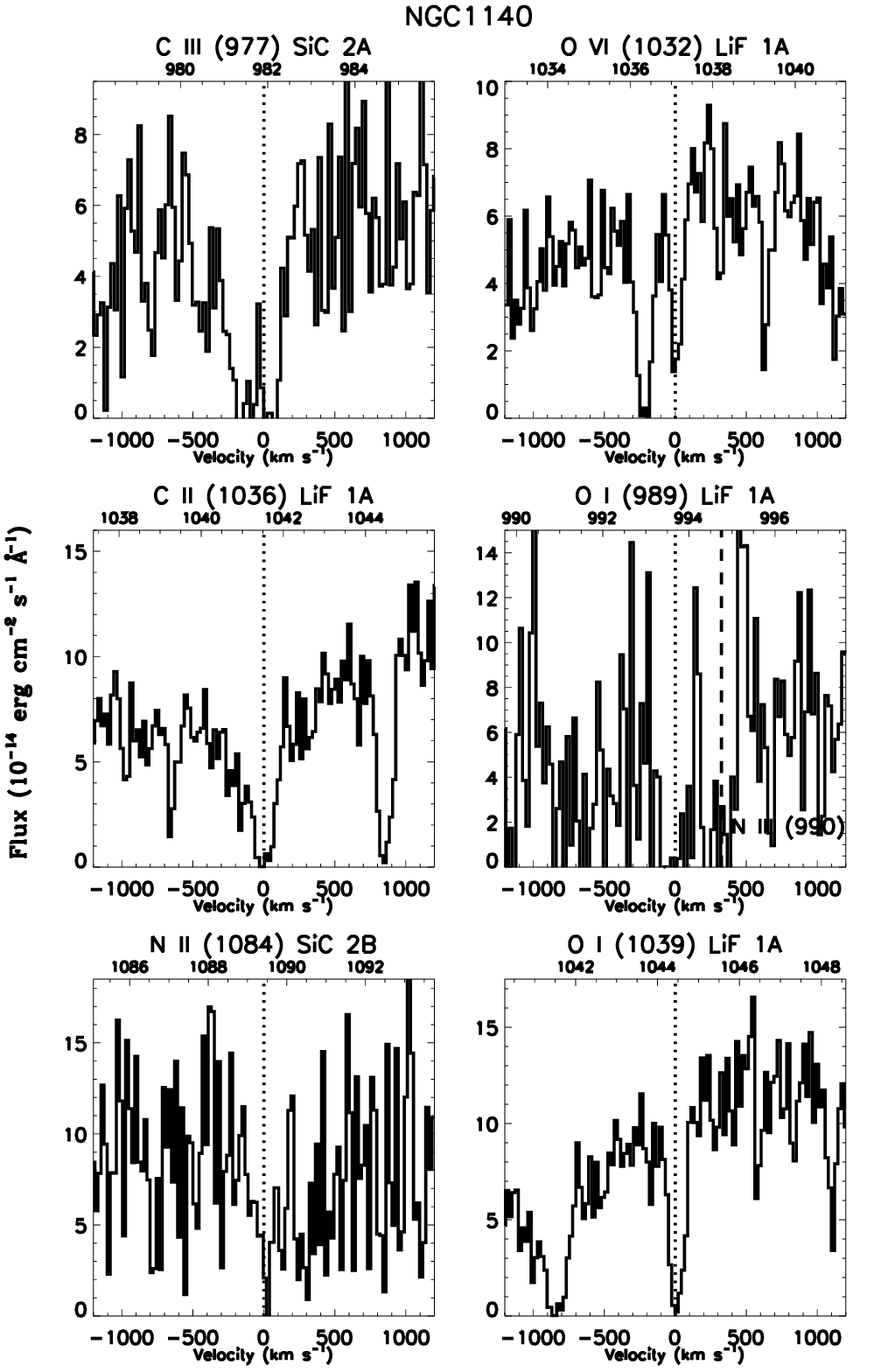}
\caption{Prominent spectral absorption lines plotted for NGC1140.
Note: \ion{O}{1}$\lambda$989 is blended with \ion{N}{3}$\lambda$990. 
\label{f:NGC1140lines}}
\end{figure}

\begin{figure}
\centering
\leavevmode
\includegraphics[width=4in]{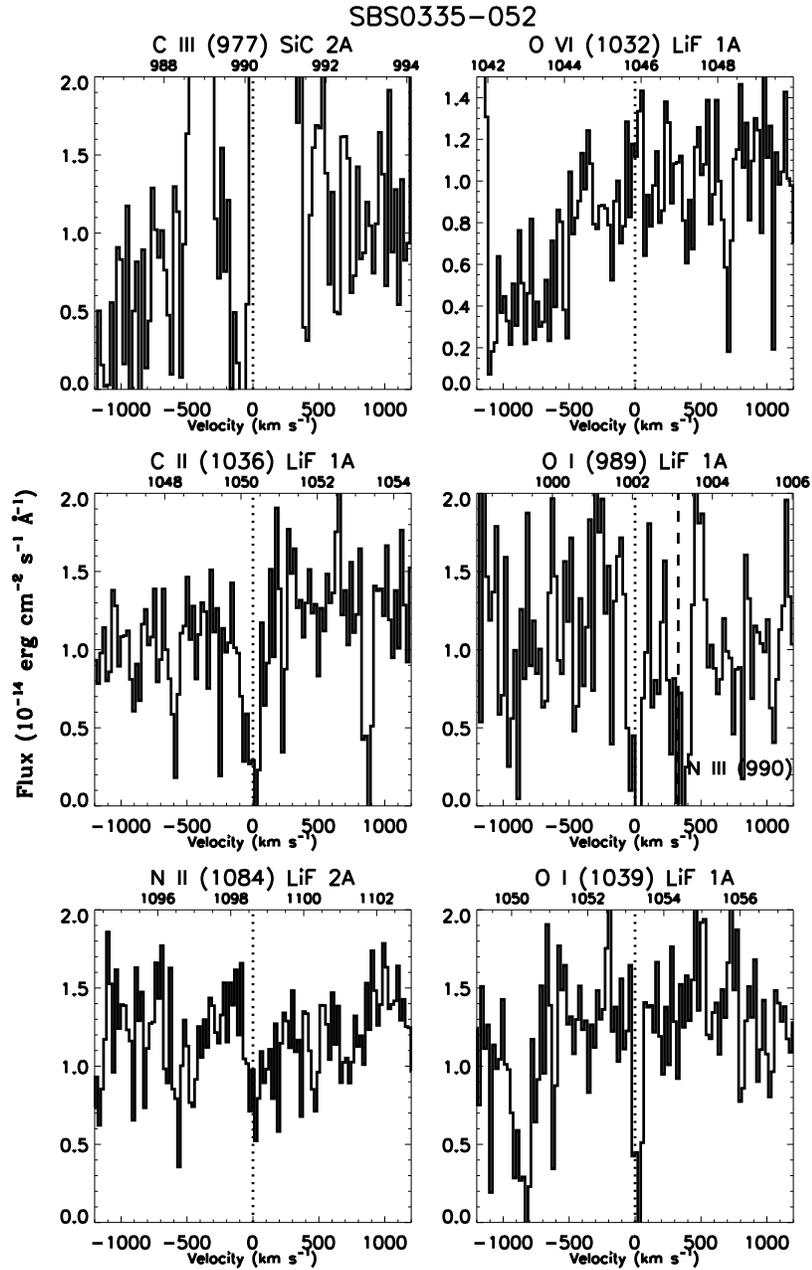}
\caption{Prominent spectral absorption lines plotted for SBS0335-052.
Note: \ion{O}{1}$\lambda$989 is blended with \ion{N}{3}$\lambda$990. 
\label{f:SBS0335-052lines}}
\end{figure}

\begin{figure}
\centering
\leavevmode
\includegraphics[width=4in]{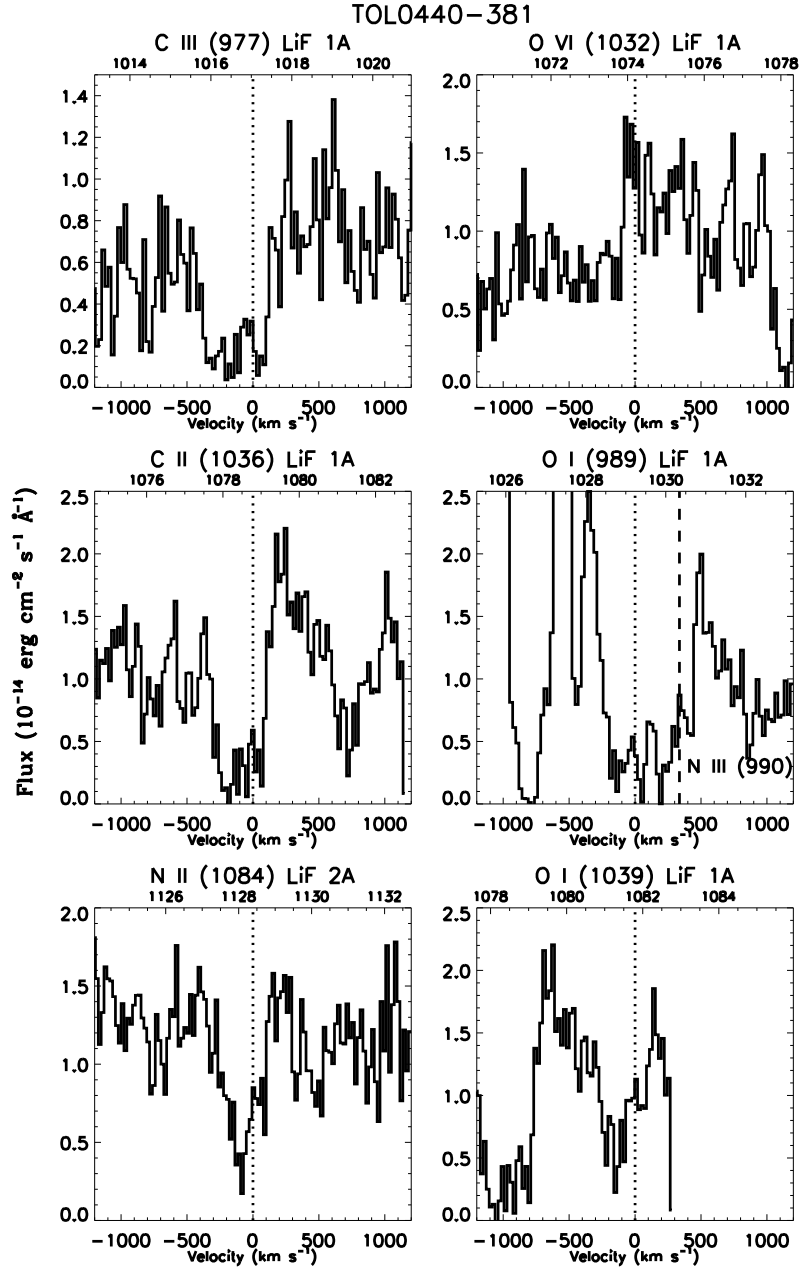}
\caption{Prominent spectral absorption lines plotted for Tol0440-381.
Note: \ion{O}{1}$\lambda$989 is blended with \ion{N}{3}$\lambda$990. 
\label{f:Tol0440-381lines}}
\end{figure}

\begin{figure}
\centering
\leavevmode
\includegraphics[width=4in]{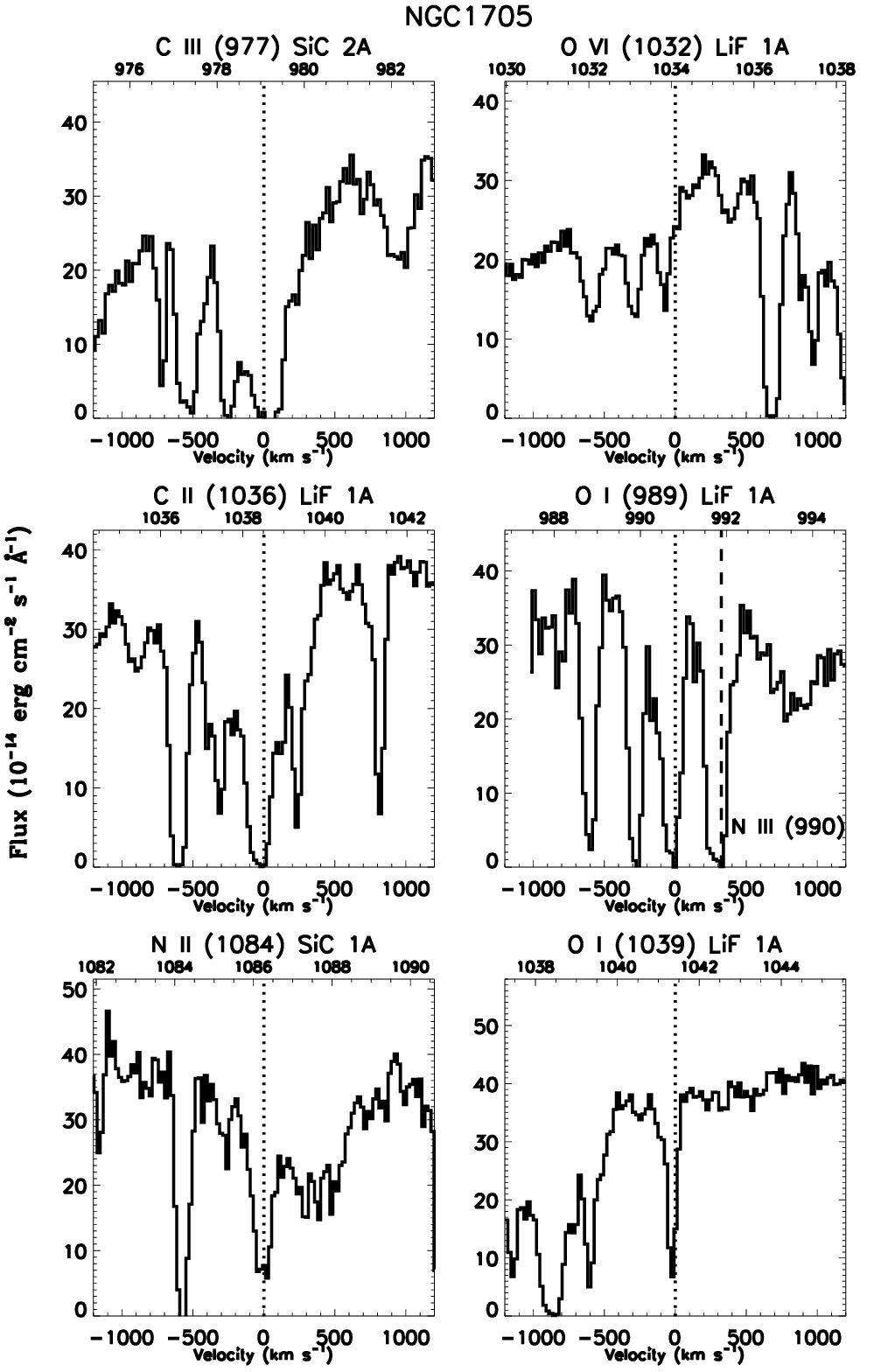}
\caption{Prominent spectral absorption lines plotted for NGC1705.
Note: \ion{O}{1}$\lambda$989 is blended with \ion{N}{3}$\lambda$990. 
\label{f:NGC1705lines}}
\end{figure}

\begin{figure}
\centering
\leavevmode
\includegraphics[width=4in]{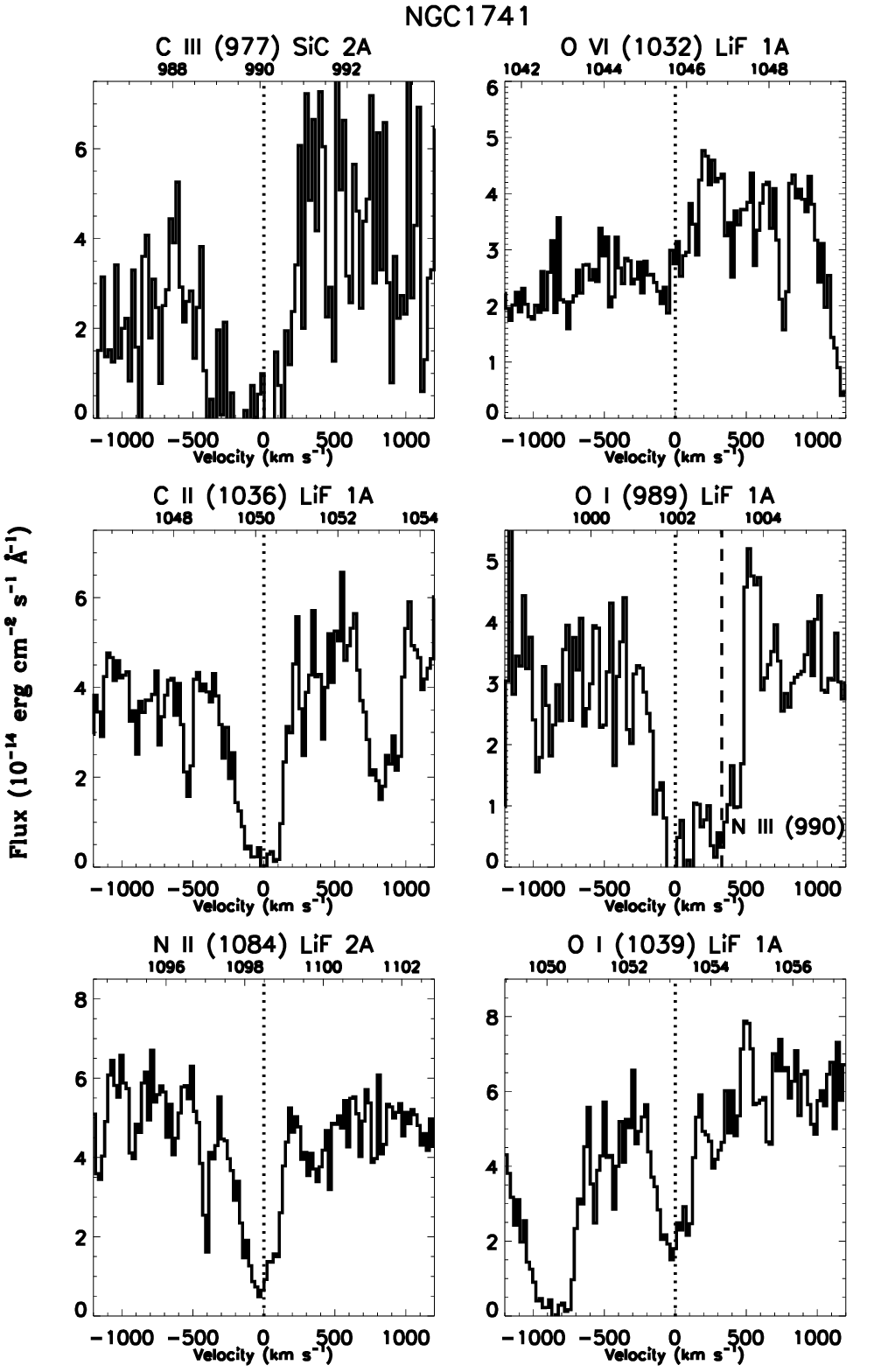}
\caption{Prominent spectral absorption lines plotted for NGC1741.
Note: \ion{O}{1}$\lambda$989 is blended with \ion{N}{3}$\lambda$990. 
\label{f:NGC1741lines}}
\end{figure}

\begin{figure}
\centering
\leavevmode
\includegraphics[width=4in]{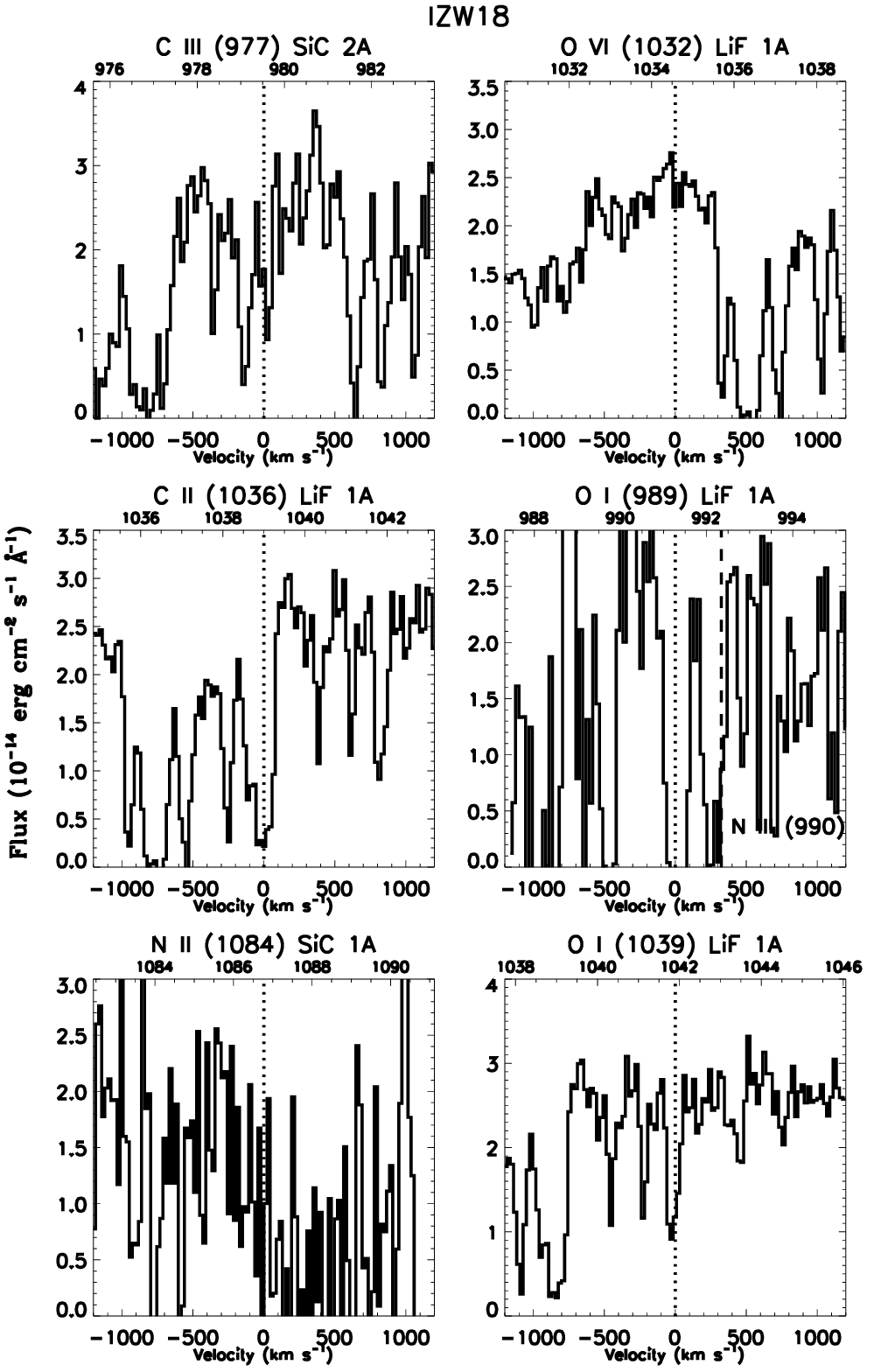}
\caption{Prominent spectral absorption lines plotted for IZW18.
Note: \ion{O}{1}$\lambda$989 is blended with \ion{N}{3}$\lambda$990. 
\label{f:IZW18lines}}
\end{figure}

\begin{figure}
\centering
\leavevmode
\includegraphics[width=4in]{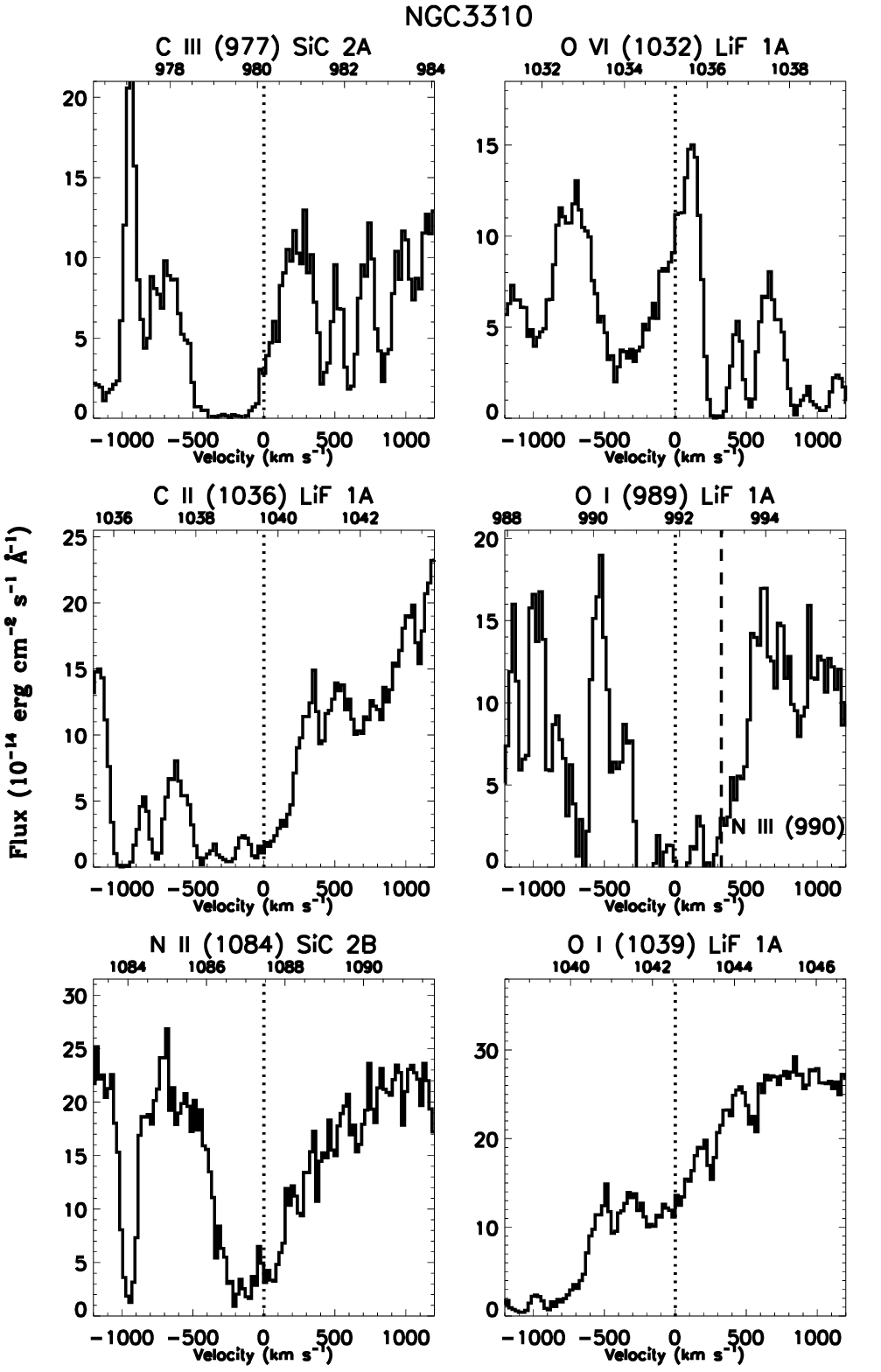}
\caption{Prominent spectral absorption lines plotted for NGC3310.
Note: \ion{O}{1}$\lambda$989 is blended with \ion{N}{3}$\lambda$990. 
\label{f:NGC3310lines}}
\end{figure}

\begin{figure}
\centering
\leavevmode
\includegraphics[width=4in]{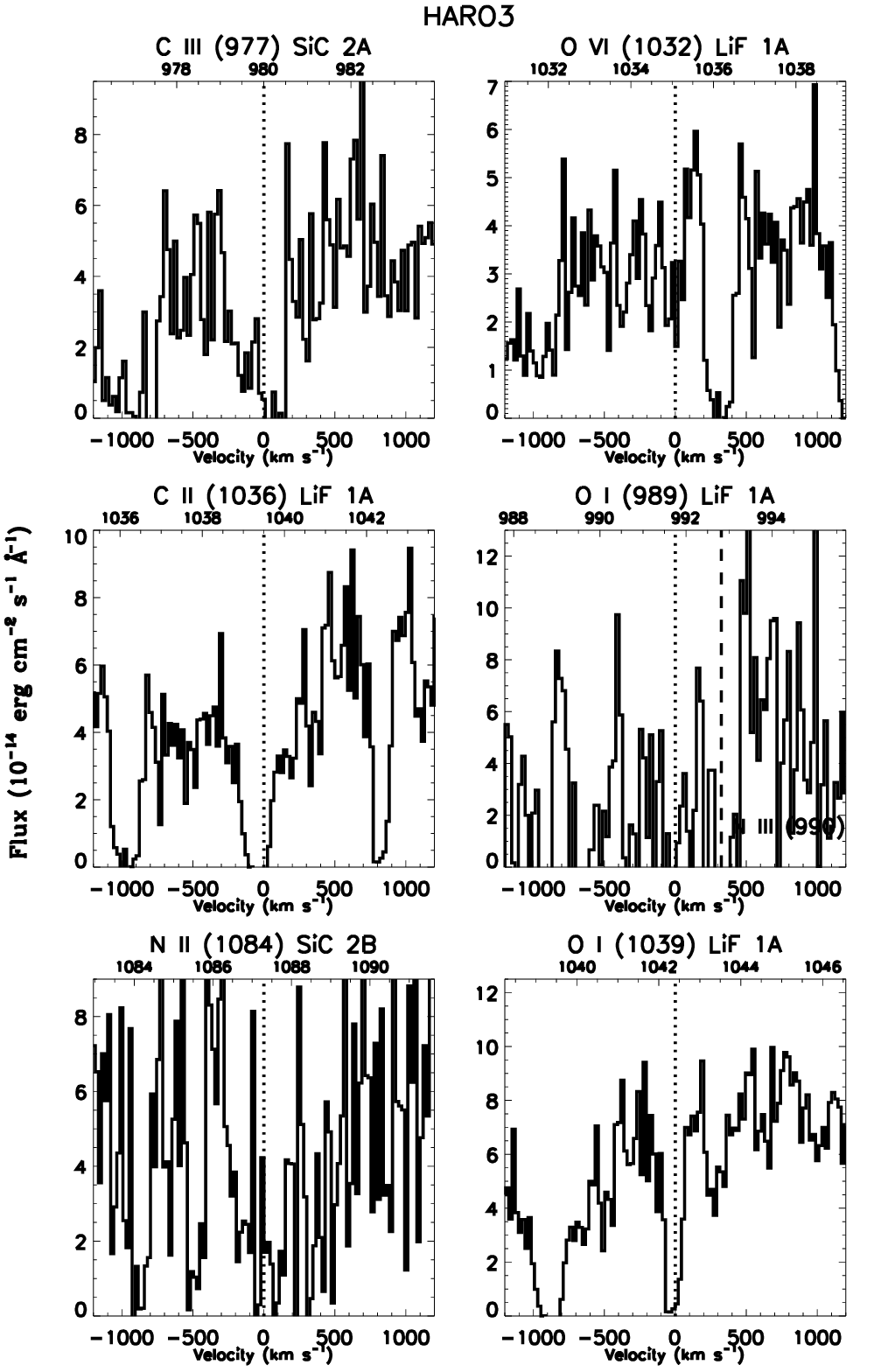}
\caption{Prominent spectral absorption lines plotted for Haro3.
Note: \ion{O}{1}$\lambda$989 is blended with \ion{N}{3}$\lambda$990. 
\label{f:Haro3lines}}
\end{figure}

\begin{figure}
\centering
\leavevmode
\includegraphics[width=4in]{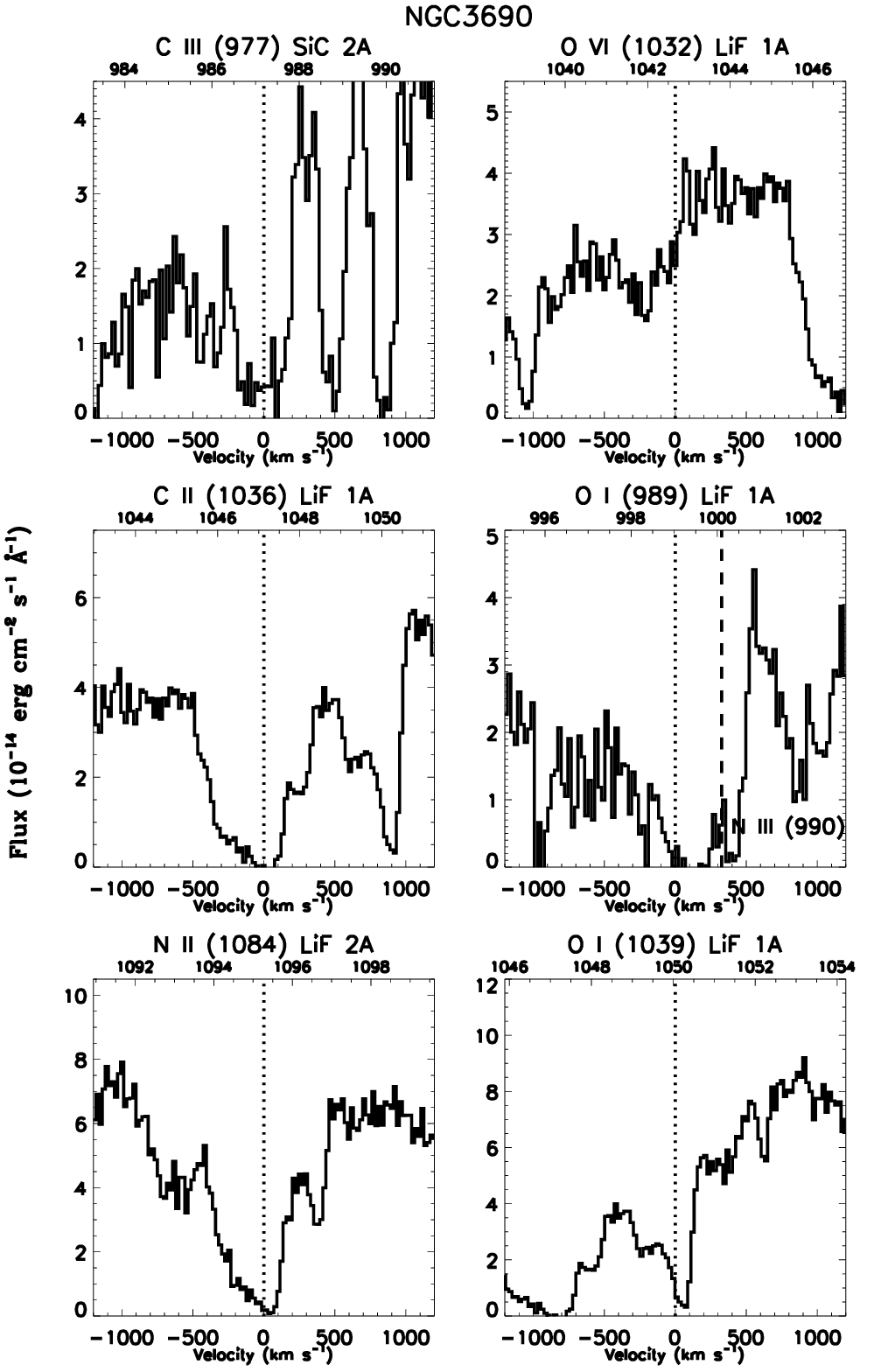}
\caption{Prominent spectral absorption lines plotted for NGC3690.
Note: \ion{O}{1}$\lambda$989 is blended with \ion{N}{3}$\lambda$990. 
\label{f:NGC3690lines}}
\end{figure}

\begin{figure}
\centering
\leavevmode
\includegraphics[width=4in]{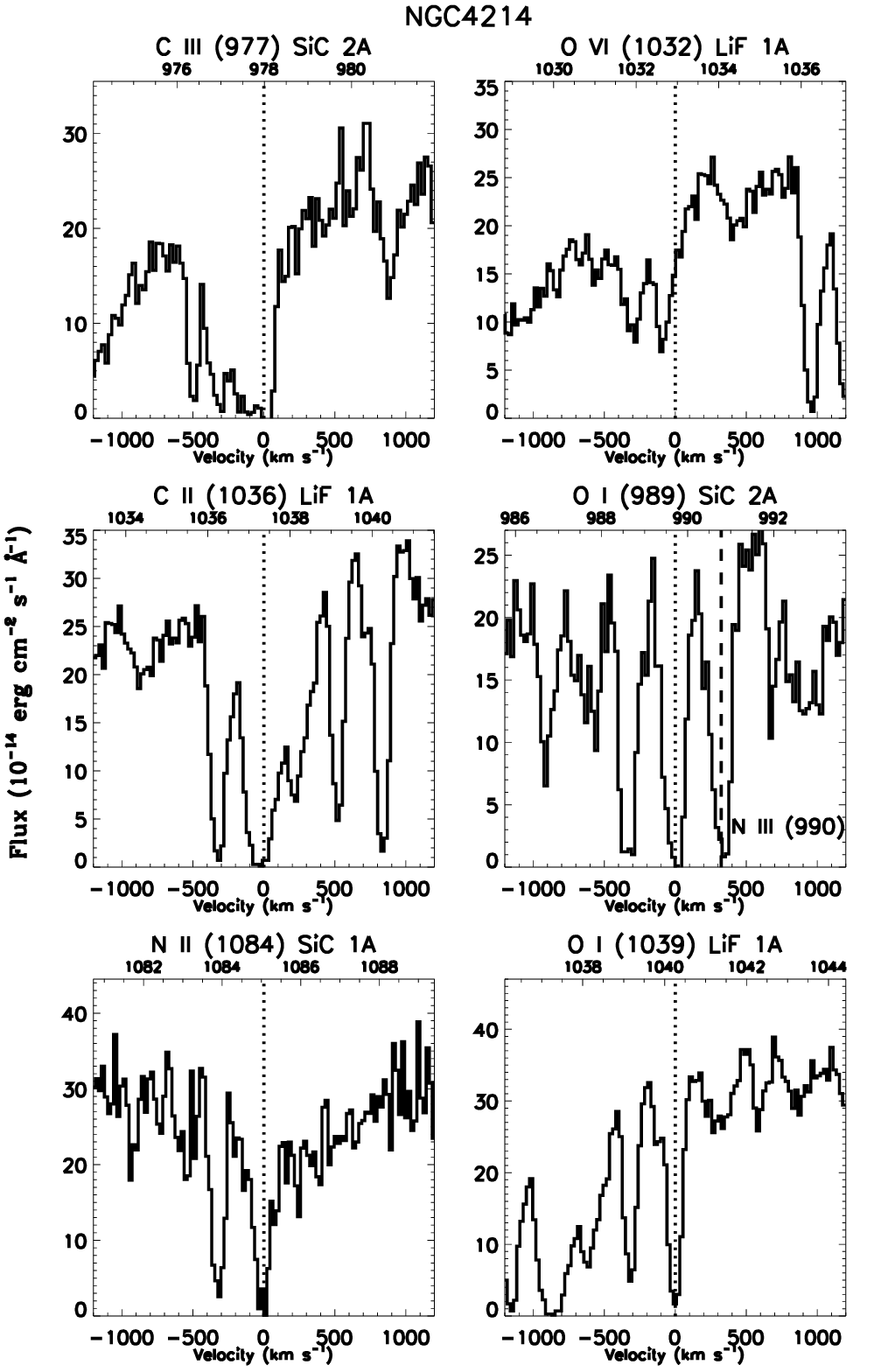}
\caption{Prominent spectral absorption lines plotted for NGC4214.
Note: \ion{O}{1}$\lambda$989 is blended with \ion{N}{3}$\lambda$990. 
\label{f:NGC4214lines}}
\end{figure}

\begin{figure}
\centering
\leavevmode
\includegraphics[width=4in]{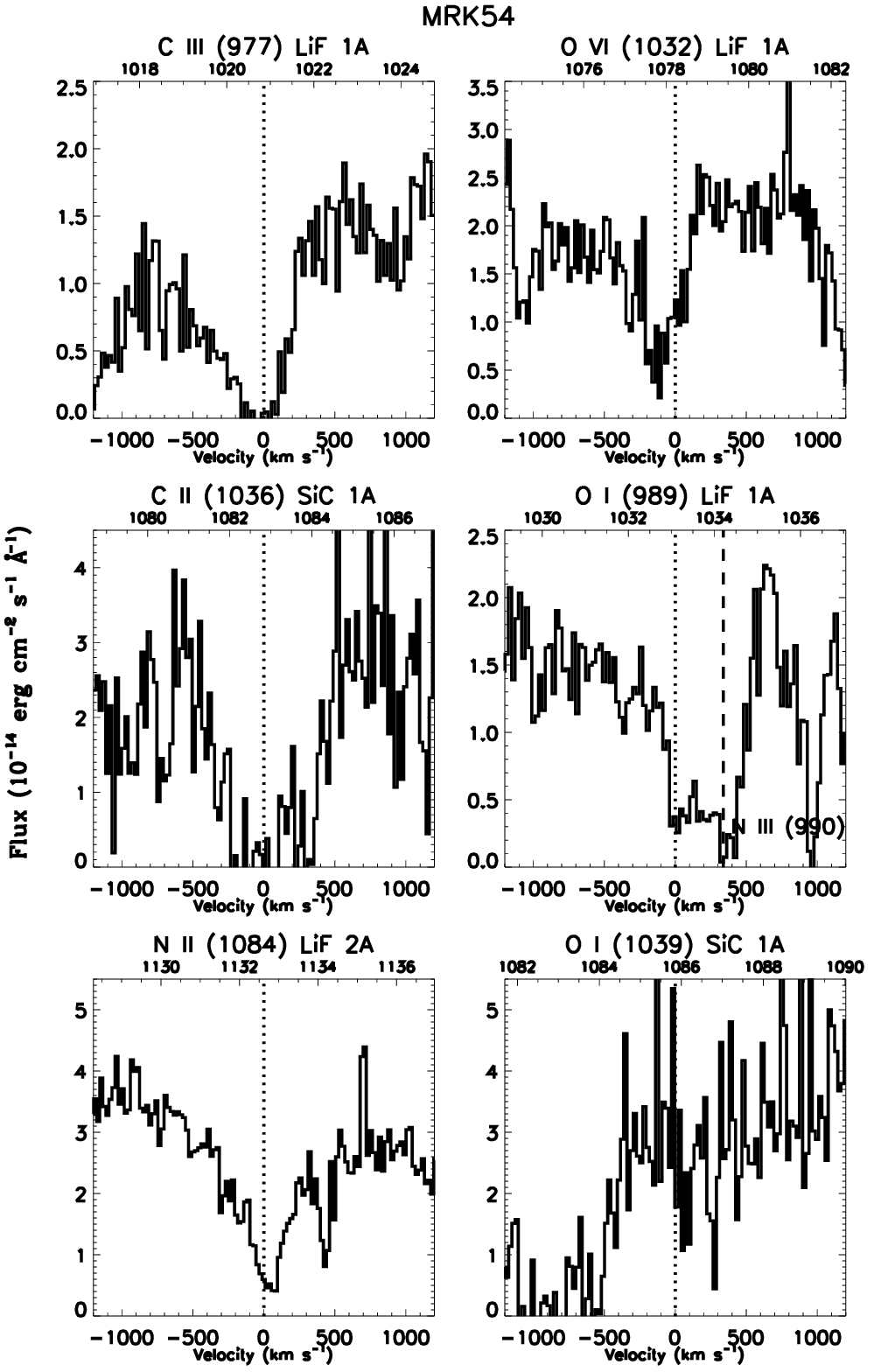}
\caption{Prominent spectral absorption lines plotted for Mrk54.
Note: \ion{O}{1}$\lambda$989 is blended with \ion{N}{3}$\lambda$990. 
\label{f:Mrk54lines}}
\end{figure}

\begin{figure}
\centering
\leavevmode
\includegraphics[width=4in]{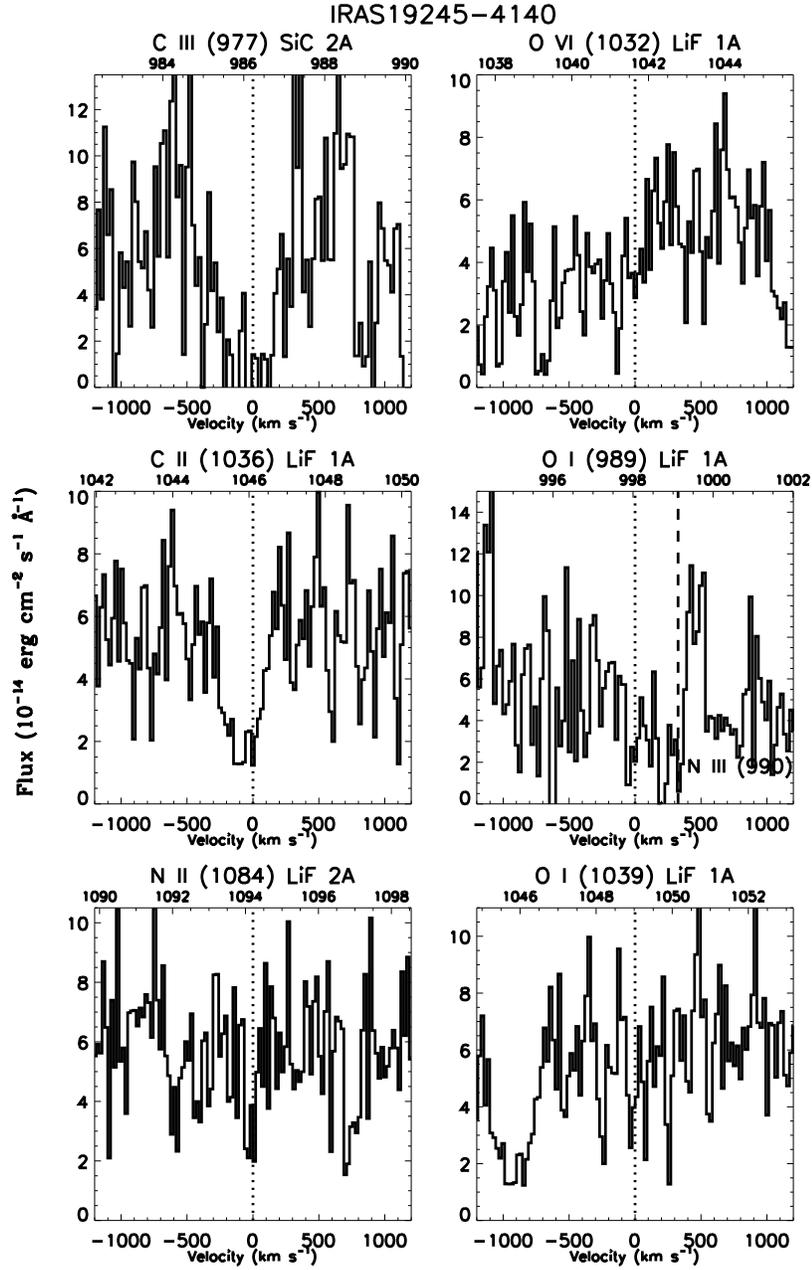}
\caption{Prominent spectral absorption lines plotted for IRAS19245-4140.
Note: \ion{O}{1}$\lambda$989 is blended with \ion{N}{3}$\lambda$990. 
\label{f:IRAS19245-4140lines}}
\end{figure}

\begin{figure}
\centering
\leavevmode
\includegraphics[width=4in]{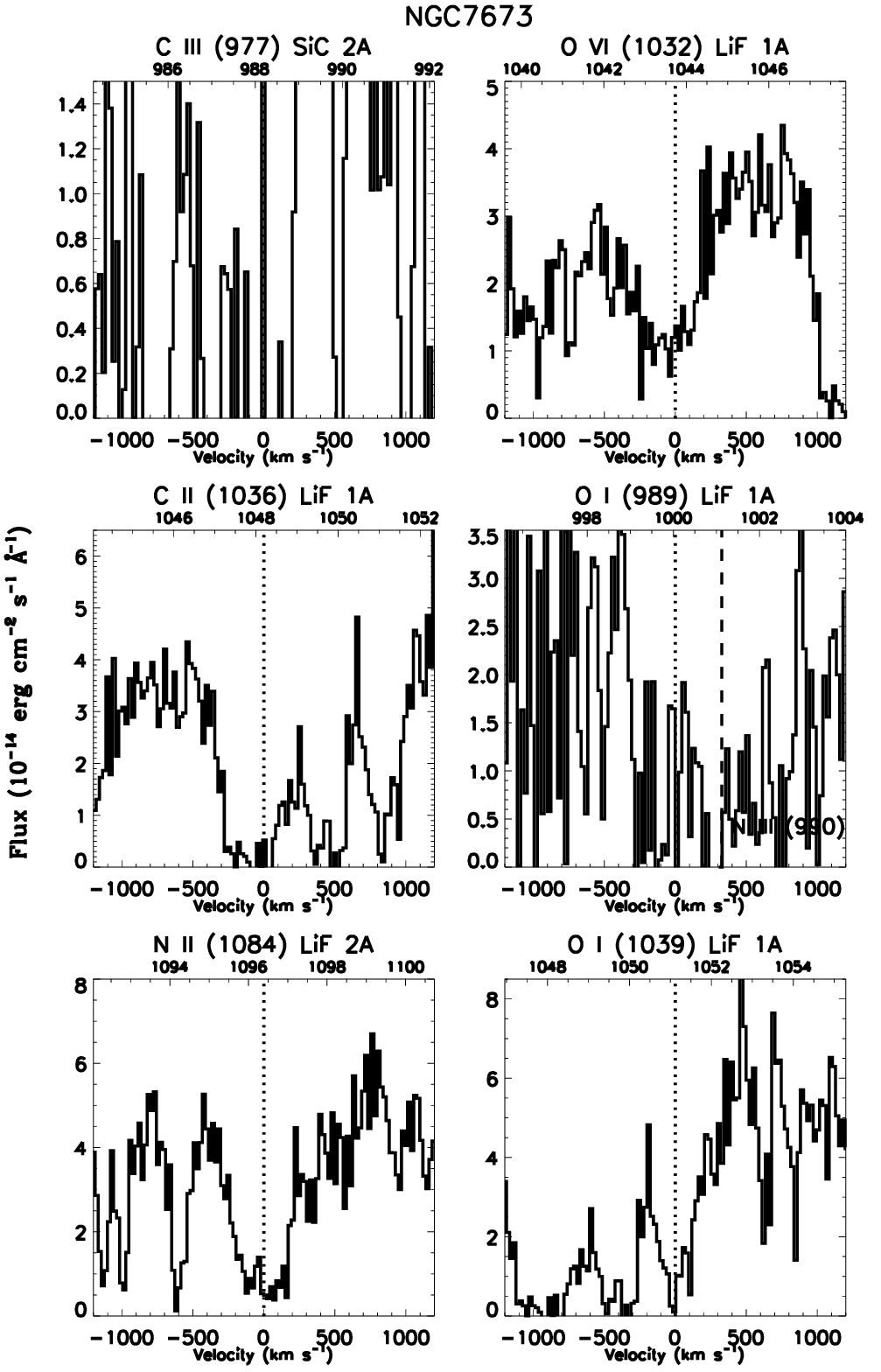}
\caption{Prominent spectral absorption lines plotted for NGC7673.
Note: \ion{O}{1}$\lambda$989 is blended with \ion{N}{3}$\lambda$990. 
\label{f:NGC7673lines}}
\end{figure}

\begin{figure}
\centering
\leavevmode
\includegraphics[width=4in]{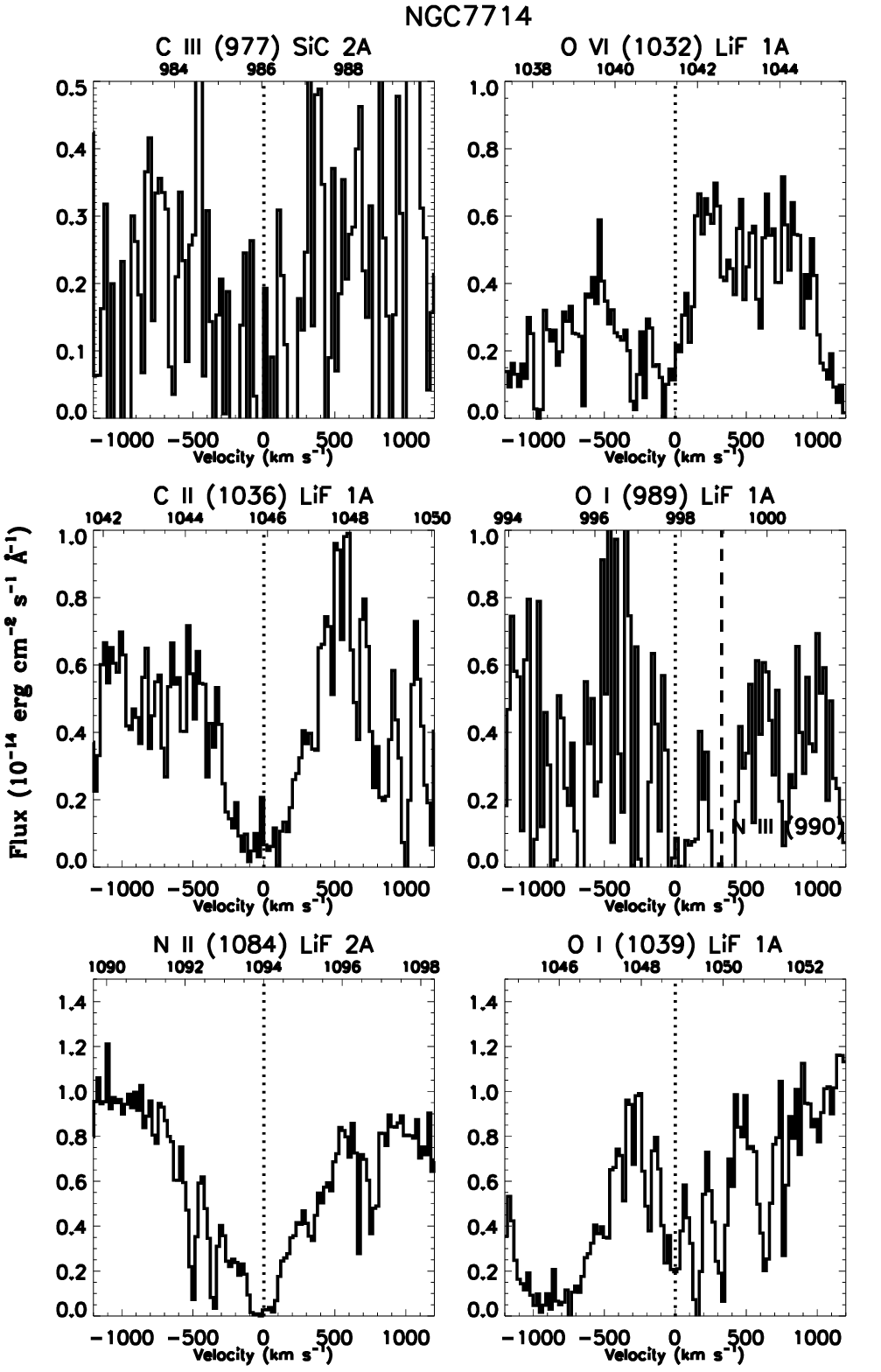}
\caption{Prominent spectral absorption lines plotted for NGC7714.
Note: \ion{O}{1}$\lambda$989 is blended with \ion{N}{3}$\lambda$990. 
\label{f:NGC7714lines}}
\end{figure}

Close up plots of six strong absorption features for all
16 galaxies are shown in Figures \ref{f:Haro11lines} -
 \ref{f:NGC7714lines}.   The plots focus
on the ISM absorption lines observed in almost all of the
spectra, specifically \oviw, \ciiw, \ciiiw, \oiaw, \niiw, and \oibw.  
We have displayed these spectra even when the
absorption feature was not observed.  MW ISM contamination, 
\htwo, airglow, instrument coverage gaps, and signal quality play an
important role in which features are actually detected.  Line blending is also 
important most particularly for \oiaw\, as it falls right next to the  
strong \niiiw \, feature.

\begin{figure}
\centering
\leavevmode
\includegraphics[width=3in,angle=90]{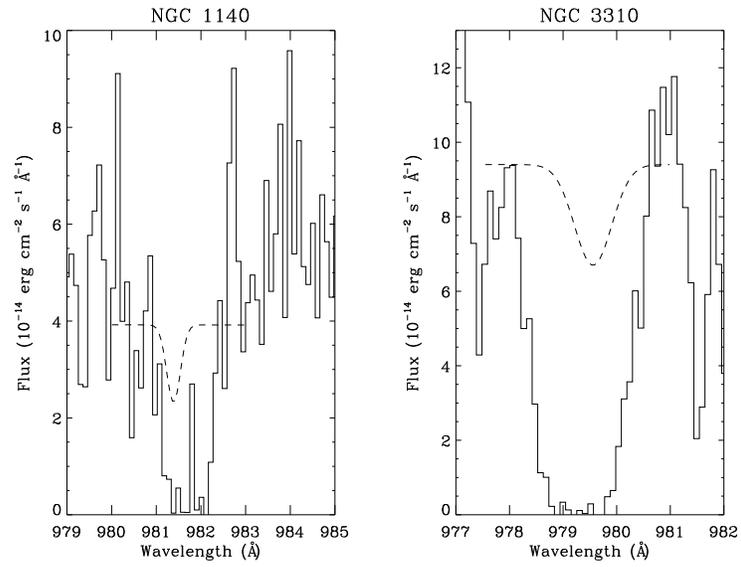}
\caption{
Overlay of \oicw\, on \ciiiw\, for NGC 1140 and NGC 3310.  
The \oi\, feature is significantly weaker than the \ciii\, profile
and does not significantly affect our measured parameters for
the \ciiiw\, line.
\label{f:oi976}}
\end{figure}

Another potentially problematic blending is 
 \ciiiw\, with \oicw.  However,
 we have checked the significance of the contamination
 by modeling the \oicw\, line using our gaussian fits to \oibw.  From 
 \citet{mort91} we expect \oicw\, to be a factor of 2.7 weaker 
 than \oibw\, for an optically thin feature.  In Figure \ref{f:oi976} we have overlayed
 an \oicw\,absorption feature with the same velocity and FWHM
 as the \oibw\, profile but we have lowered the equivalent width by 
 a factor of 2.7 for NGC 1140 and NGC 3310.   As shown in these plots, 
 this feature is significantly weaker
 than the observed \ciiiw\, profile.
 Fitting these profile with the \oicw\, contribution
 does not change the derived equivalent widths,
 centroid velocities, and FWHMs in a 
 significant manner.  The largest percentage change is
 seen in the FWHMs which lower by
 12 and 11 \kms\, for NGC 1140 and NGC 3310 respectively.
 This is only 25\% of the $1\sigma$ statistical error.  
We therefore conclude that the weak \oicw\, is not significantly
affecting our fits to the \ciiiw\, profile.

The strength of the absorption lines varies widely across the sample.  
As discussed in our previous work \citet{grim06}
and \citet{grim07}, most of the strong  absorption line profiles
are non-gaussian.  The absorption profiles
frequently show significant structure suggesting
that in many cases these are an
aggregate profile representing
multiple absorbers spread throughout
the galaxy's ISM and starburst driven outflow.  
Many of the profiles are asymmetric, with stronger absorption on the 
blueward side of the line core.  This is consistent with absorbing gas
entrained in and accelerated by an outflowing wind fluid.

In our previous work we have successfully modeled many of
the stronger lines with two gaussian component fits.  Physically, for both
\hoo\, and \vv\, the stronger gaussian was associated with the host galaxy ISM while
the blueshifted gaussian represented outflowing gas at several hundred \kms.
For this work, covering a wide range of galaxy star formation rates 
and data quality, a multiple component fit is not feasible.
Therefore, we have used two additional methods to study the strongest
absorbers in our sample.

\subsubsection{Apparent Optical Depth Method}

\begin{figure}
\centering
\leavevmode
\includegraphics[width=4in]{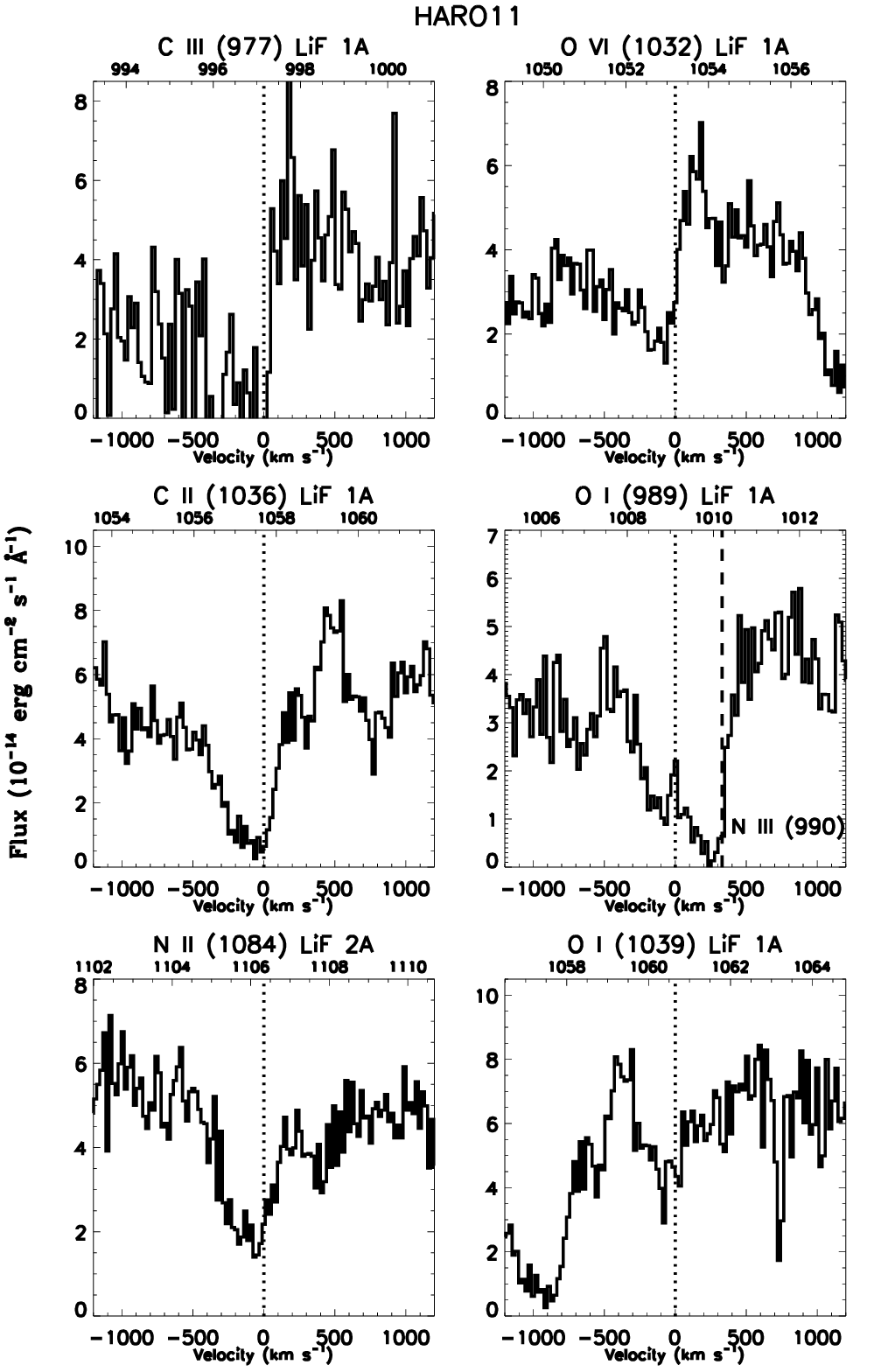}
\caption{Prominent spectral absorption lines plotted for Haro11.
Note: \ion{O}{1}$\lambda$989 is blended with \ion{N}{3}$\lambda$990. 
\label{f:Haro11lines}}
\end{figure}

\begin{figure}
\centering
\leavevmode
\includegraphics[width=4in]{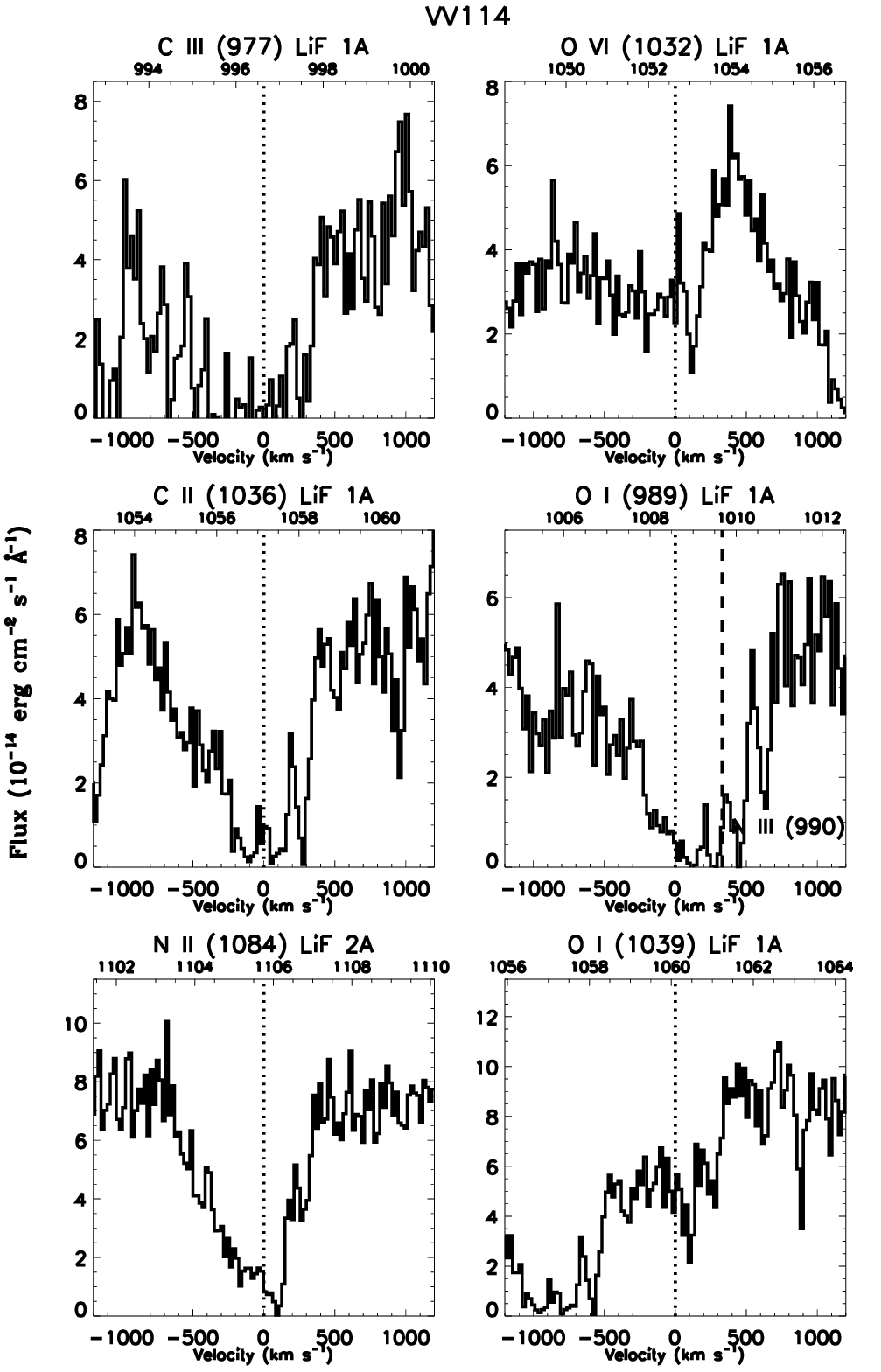}
\caption{Prominent spectral absorption lines plotted for VV114.
Note: \ion{O}{1}$\lambda$989 is blended with \ion{N}{3}$\lambda$990. 
\label{f:VV114lines}}
\end{figure}

For \oviw\, we use the apparent optical depth
method as described in \citet{sem03}.  This method assumes
that the \ovi\, absorption features are not saturated - 
which is consistent with the observed \ovi\, profiles.  
The apparent column density as a function of velocity is given by
\begin{equation}
{\rm N_a}(\nu)=\frac{\tau_{\rm a}(\nu)}{f \lambda}\frac{m_e c}{\pi e^2}=
3.768\times10^{14}\frac{\tau_{\rm a}(\nu)}{f \lambda({\rm \AA})}
\end{equation}
where 
\begin{equation}
\rm{
\tau_a(\nu)=-\ln [I_{obs}(\nu)/I_0(\nu)].}
\end{equation}
$\rm{I_0(\nu)}$ is the continuum intensity while 
$\rm{I_{obs}(\nu)}$ is the observed line intensity.
The \ovi\, line profiles in Figures \ref{f:Haro11lines} - \ref{f:NGC7714lines}
 show that the line is clearly resolved
when present.  Therefore we can
obtain the column density by integrating over velocities
as $\rm{N}=\int\rm{N(\nu)} d\nu=\int\rm{N_a(\nu)} d\nu$.
The apparent optical depth method also allows us to calculate
the velocity centroid and line width using
$\bar{\nu}=\int\nu\tau_{\rm a}(\nu)d\nu/\int\tau_{\rm a}(\nu)d\nu$
and 
$b=[2 \int (\nu - \bar{\nu})^2 \tau_{\rm a}(\nu)d\nu/\int\tau_{\rm a}(\nu)d\nu]^{\onehalf}$
respectively.  Table \ref{t:fuseovi} lists our results.    

In our previous work we have calculated the \ovi\, parameters for 
\vv\, and \hoo\, from gaussian fits.  For \vv\, these two methods
are in excellent agreement as we had found
\novi\, $\sim2.0\times10^{15}\,{\rm cm^{-2}}$ while we now calculate
\novi\, $\sim2.2\times10^{15}\,{\rm cm^{-2}}$ using the apparent 
optical depth measurement.  These results are consistent 
because the \ovi\, profile shape is approximately gaussian for \vv \,(see Figure \ref{f:VV114lines}).
We find however a larger discrepancy for \hoo\, with
\novi\, $\sim4.2\times10^{14}\,{\rm cm^{-2}}$
and \novi\, $\sim8.0\times10^{14}\,{\rm cm^{-2}}$
for the gaussian and apparent optical depth methods respectively.
For \hoo\, (Figure \ref{f:Haro11lines}), a multicomponent profile is observed.
In this case the apparent
optical depth method better includes the extended blue wing.  We also now derive
a larger absorption line width and a more blueshifted
velocity centroid, consistent with a significant
contribution from the  extended blue wing
of the absorption profile.  
The apparent optical depth method 
provides a significant improvement over gaussian profiling
for \ovi\, and is applicable as \ovi\, is not saturated or blended 
significantly with other lines (partly due to our sample selection).
Table \ref{t:fuseovi} does not include
four galaxies. For NGC\,1140, \izw, and \iron\, 
we do not positively detect \ovi\, absorption.  
In \tol\, the \ovi\ is a broad, shallow feature
making it impossible to reliably estimate the velocity limits
over which the above integral should be performed.

\subsubsection{Non-parametric Equivalent Widths\label{s:npeqw}}

Several other strong absorbers exist in the spectra 
of these star-forming galaxies, specifically
\ciiw, \ciiiw, and \niiw.    Unlike \oviw\, however,
these lines are frequently saturated at line center and therefore
we are unable to apply the apparent optical depth method.
We therefore measure the equivalent widths and
velocity centroids of these lines using the \iraf\, \citep{tod96}
tool {\it splot}.  The procedure subtracts a user defined
linear continuum and sums the resulting flux to calculate the
equivalent width.  The comparably strong \oiaw\, and
\niiiw\, lines are frequently blended, so 
these lines have been excluded from the procedure.
Our results are listed in Table \ref{t:fusesplot}.  
These lines are frequently saturated, particularly
\ciii\, and \cii\,, so the equivalent widths are essentially just a measure
of how broad the absorption feature is.

We have also compared the non-parametric equivalent widths, centroids, and widths
with those derived from our previous gaussian fits in Section \ref{s:gaussian}.  There is broad
agreement for the derived values using the \cii\, and \nii\,features.
The principal outliers are the \cii\, centroid velocities for some of the galaxies with
higher SFR ($>$ a few $\rm10^{10}\,L_\odot$) where the 
outflow velocities are found to be higher from the non-parametric fits. 
In the case of \ciii, there is a systematic shift to higher outflow velocities for
the non-parametric values.  This is observed in almost all
of the galaxies in our sample not just in the galaxies with higher SFR.  
These findings are consistent with
our motivation for using a non-parametric method to measure
these absorption features.  Specifically, as line
strengths increase the absorption features are found to be
increasingly non-gaussian and blue-asymmetric.

For the analysis that follows unless otherwise stated, 
we use these non-parametric measurements for the \ciii, \cii, and \nii\, features.
For \ovi\, we use the results derived from the apparent optical depth method and for
the remaining lines we use the gaussian data.

\section{Discussion}

This paper is intended to be a general study of 
the FUV properties of starburst galaxies.  
As stated previously, we will focus on the strongest
ISM absorption features in order to include
the majority of galaxies in our sample.  In particular,
we will focus on the following
lines:  \oviw, \ciiiw, \sthw, \niiw, \ciiw, \oibw.  These lines
are detected and measured throughout most of our sample.
They also represent three different phases of the ISM - coronal (\ovi), HII (\ciii, \sth, \nii) 
and HI (\cii, \oi).

\subsection{Outflow Velocity Distributions\label{s:velvel}}

\begin{figure}
\centering
\leavevmode
\includegraphics[width=5in]{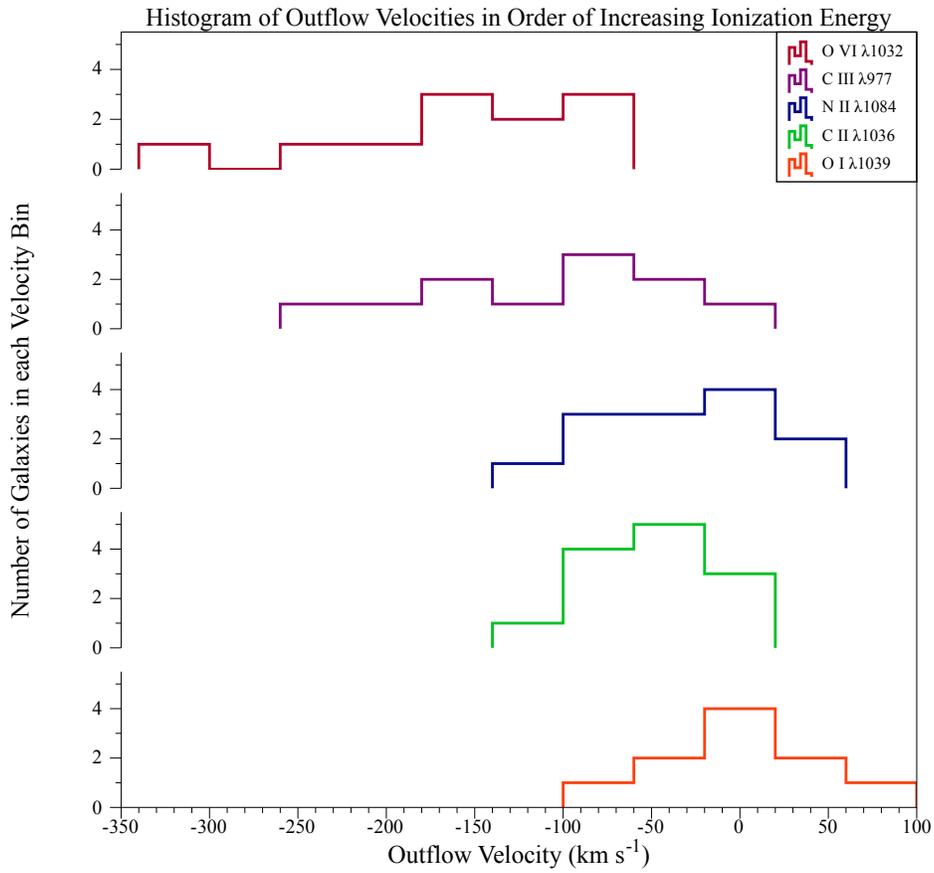}
\caption{
Velocity Outflow Histogram for Prominent Absorption Lines:  Outflows
are observed for all of the absorption features except \oibw.
Observed outflow velocities are significantly higher for
\ovi\, and \ciii.
\label{f:velhist}}
\end{figure}

In Figure \ref{f:velhist} we have plotted histograms of the
relative velocities for the strongest of these absorption features.
These lines are listed in order of increasing ionization
potential.  With the exception of the neutral oxygen ion, the other
absorption features show the presence of outflows
ranging up to a few hundred \kms. 
As the ionization state increases 
the lines generally have
larger outflow velocities.  The sole
exception is \cii\, which is mildly
more blueshifted than \nii.  The \ciii\, and \ovi\, lines however
are significantly more blueshifted than the two 
lower ionization state ions.

\begin{figure}
\centering
\leavevmode
\columnwidth=.5\columnwidth
\includegraphics[width=3.1in]{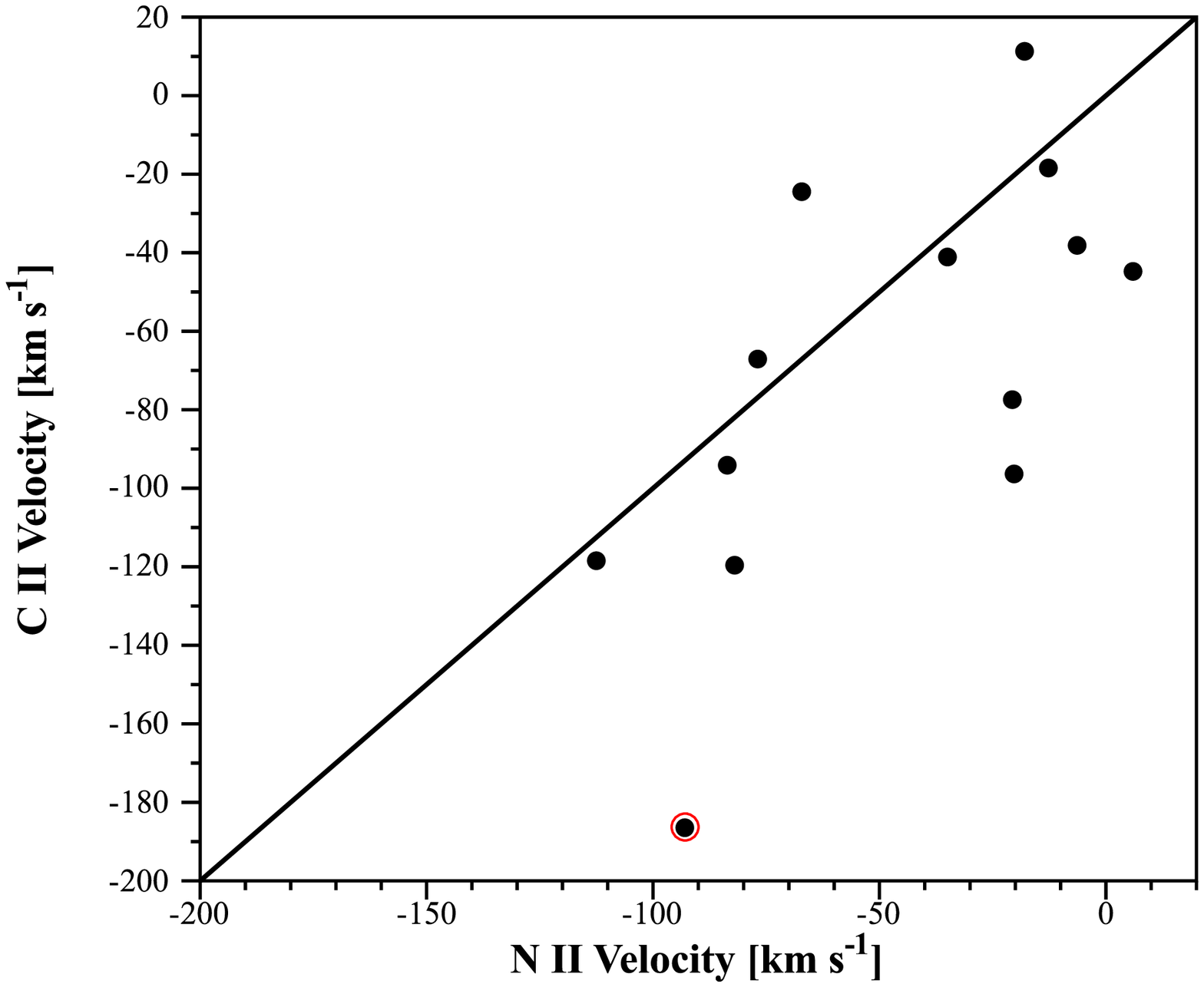}
\includegraphics[width=3.1in]{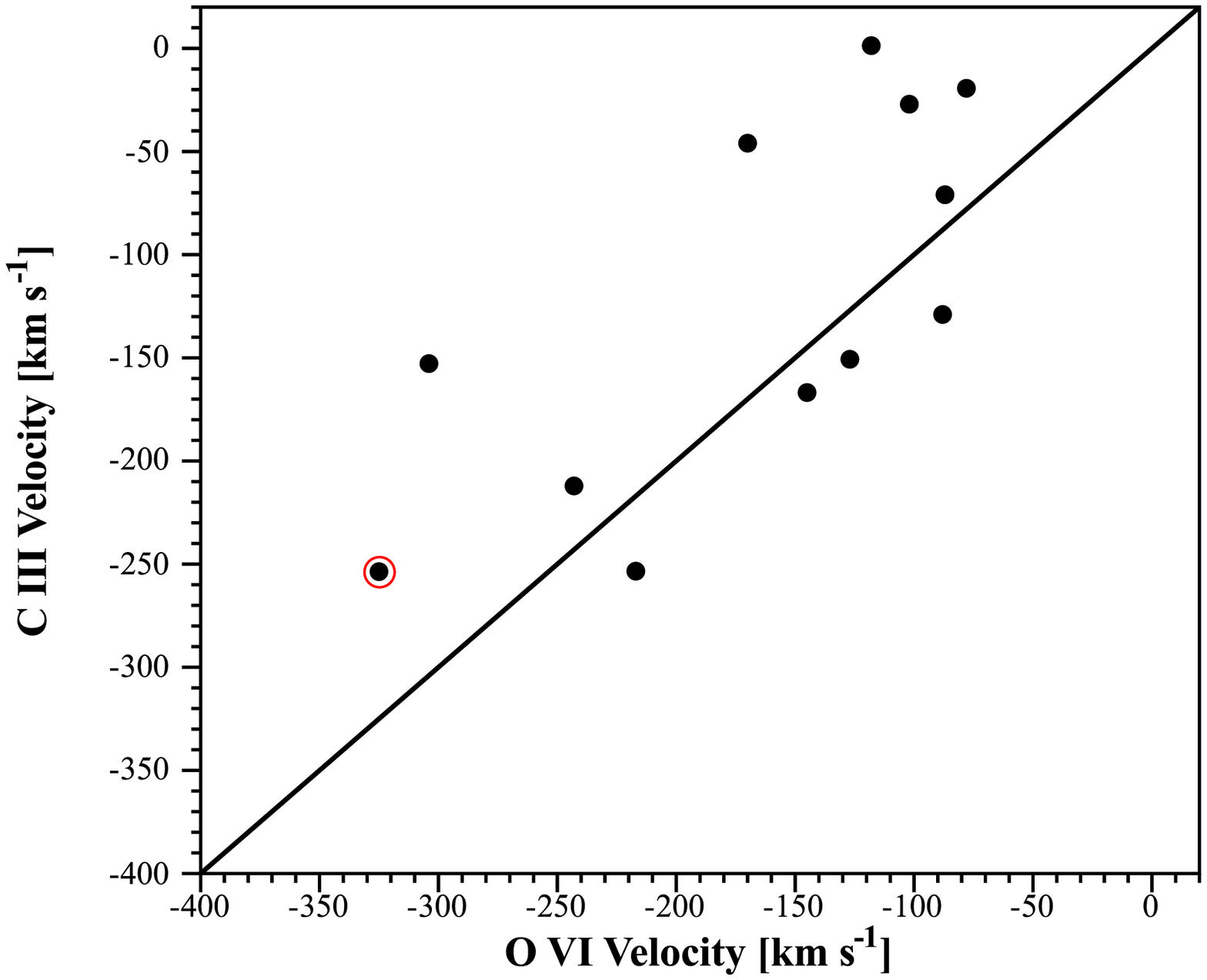}
\caption{
Outflow Velocity Comparison:  The plot on the left shows the
\ciiw\, versus \niiw\, velocities while the right
displays the \ciiiw\, versus \oviw\, velocities.
\cii\, and \ovi\, are generally more blueshifted than
the \nii\, and \ciii\, features respectively.
The lines represents $\rm{v_x = v_y}$ for both plots.
NGC~3310 (which has unusually large outflow velocities for its
mass and star formation rate) is denoted by a red circle.
\label{f:velvel}}
\end{figure}

In Figure \ref{f:velvel} we have directly compared the
velocities of these absorption features.
The left panel shows \cii\, versus \nii\,outflow velocity
while the right panel is \ciii\, versus \ovi.  For the majority
of galaxies in our sample, \cii\, is clearly
more blueshifted than \nii.  Similarly, \ovi\, has an higher outflow 
velocity than \ciii.  This confirms the pattern seen in
Figure \ref{f:velhist}.

As we have previously discussed, the
\ovi\, line (unlike \ciii, \cii, and \nii) 
is not saturated for any of the galaxies in our sample.
Moreover, an examination of the
\ovi\, profile shape suggests a different origin for this feature.
The strongest part
of the \ovi\ profile is blueshifted, and the absorption at or near the
galaxy systemic velocity is weaker. In contrast, the
 \ciii, \cii, and \nii\, lines are broad with extended blueshifted wings,
and have the strongest absorption (the line core)
at or near the galaxy's systemic velocity. Furthermore, the maxiumum outflow speed (defined
by the blueward extent of the absorption-line profile) is similar for \ovi\,
and \ciii. 

Thus, the gas traced by the \ovi\, line
appears to be primarily in the outflow rather than in the quiescent ISM,
while the lower ionization lines are produced by both. 
This difference is not surprising since \ovi\ has a much higher 
ionization potential and is produced in
significantly higher temperature gas.  
We will discuss this further in Section \ref{s:ovi}.

Several previous studies have suggested that the measured outflow 
velocities increase with the ionization potential of the species in starbursts.
\citet{vaz04} used {\sl HST STIS} spectroscopy of 
NGC 1705 (also represented in our \fu\, sample)
and found that several high ionization lines 
had higher outflow speeds.   
\citet{schwa06} similarly used {\sl HST STIS}
to observe starclusters in 17 star-forming galaxies 
(including NGC 3310 and NGC 1741 also in this work).
While \ciibw, \aliiw, and \siiiw\, all trace
the neutral hydrogen phase, 
they found evidence that \ciibw\,
had higher outflow speeds than the two slightly lower
ionization features. 
We also noticed similar behavior previously
when studying the \fu\, observations of Haro 11 
\citep[][ \hoo\, is also in this sample]{grim06}.
\citet{schwa06} suggest that
this could be explained by less dense, hotter
gas escaping the starburst region and
expanding at a faster speed than the 
denser, lower ionization gas.

This picture appears to be different
than what is observed in high redshift
Lyman break galaxies (LBGs).  
\citet{pett02} have examined
high quality spectra of the lensed LBG, MS1512-cB58.
They found the same velocity distributions
for both the low and high ionization features. 
Using a composite LBG spectrum, 
\citet{shap03}  also found 
similar outflow velocities for both
the lower and higher ionized absorption features.

\begin{figure}
\centering
\leavevmode
\includegraphics[width=3.1in]{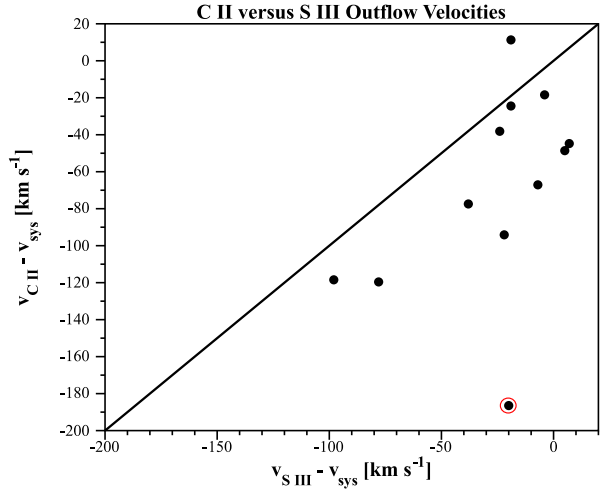}
\caption{
\ciiw\, versus \sthw\,velocities.  While \sth\, has a significantly larger
ionization energy (23.3 eV) than \ciiw\, (11.3 eV), 
higher outflow velocities are observed in \cii.  
The line  represents $\rm{v_{C II} = v_{S III}}$ in both plots.
NGC~3310 (which has unusually large outflow velocities for its
mass and star formation rate) is denoted by a red circle.
\label{f:velsiii}}
\end{figure}

As discussed previously \ciii\,, with an ionization potential of 
24.4 eV, has significantly higher outflow velocities than
\cii\, (11.3 eV), \nii\, (14.5 eV), and \oi.  
This is then consistent with the idea that
higher ionization lines have higher outflow velocities.
On the other hand, the lower ionization \cii\, has a slightly higher
outflow velocity than \nii \,(e.g. left panel of Figure \ref{f:velvel}).
To study this further we have plotted the \ciiw\, versus
\sthw\, velocities (Figure \ref{f:velsiii}).  \sth\, has a much higher
ionization potential (23.3 eV) than \cii\, but is a weaker feature.  It
is not saturated and is well fit by a gaussian model (unlike 
the stronger features).  \cii, \ciii\, and \nii\, all 
have significantly higher
outflow velocities than \sth\,, which is not consistent
with an increase of velocity with ionization energy.

In Tables \ref{t:HARO11dat} - \ref{t:VV114data} 
we listed the absorption
features in order of decreasing values of
$\log(\lambda f N/N_H)$.  The oscillator 
strength ($f$) values are from \citet{mort91} 
and solar relative abundances are assumed for $N/N_H$.  
While the majority of the galaxies in our
sample do not have solar metallicity (Table \ref{t:fuselumins}), 
this is still a useful predictor of the relative
strengths of the absorption features.  
With the exception of \ovi\ (which we have already argued
has a different physical origin than the
other ISM absorption features) 
the values of the equivalent widths
follow decreasing $\log(\lambda f N/N_H)$ (see also Table \ref{t:fusesplot}).
It is particularly striking that
our previous ordering of outflow velocities
follows the same general trend as the line strengths.
Both outflow velocities and line strengths
decrease  from 
\ciii, \cii, \nii,  to \oibw.

An examination of the line profiles
show that the stronger lines have 
extended blue absorption wings.
In fact, these extended absorption wings
were our prime motivation in using 
a non-parametric method for 
measuring the equivalents widths
and centroid velocities (Section \ref{s:npeqw})
of the strongest features.  This was not
necessary for the weaker lines like
\oibw\, and \sthw\,which are
well fit by gaussian modeling.  
As the absorption lines increase in strength,
the extended blue wing
in the absorption-line profile becomes more and more conspicuous.
The simplest explanation for this is that
the high velocity outflowing gas has a lower column density
than the more quiescent gas in the ISM.
We therefore only detect the fastest moving gas in the strongest
absorption features.   
This explanation is consistent with the
observations of the LBG cB58 where the stronger lines
span progressively broader ranges in velocity \citep{pett02}.
It also explains the higher outflow
speeds for \ciibw\, relative to \aliiw\, and \siiiw\, \citep{schwa06}
as \ciibw\, is the strongest feature.
Lastly, in the work by \citet{vaz04}, the
lines with the highest measured equivalent widths
are preferentially those with the highest
ionization energies.  This result
thus appears to be consistent
throughout a wide range of galaxy types and
absorption features.  

This result is particularly relevant to outflow
velocities derived in other wavelength regimes.  
For example, the \naoww\, doublet has 
$\log(\lambda f N/N_H) = - 1.94$, similar to
\niiw.
This suggests that observations of local outflows using \nai\,
absorption are not biased by measuring lower ionic
states than the FUV and UV measurements of
higher ionization absorption lines.  
The derived outflow velocities
should be consistent with those of other similarly strong
absorption features in the neutral and ionized gas.
This could strengthen studies comparing
observations of \nai\,absorption in local galaxies with the
properties of outflows observed in LBGs. Unfortunately, there
is only one galaxy in our sample having published data on the
\nai\, absorption feature: NGC~4214 \citep{schwa04}. In this case
the agreement is good: they find an outflow speed of 23 $\ks$, while we find 
speeds ranging from 0 to 58 $\ks$ depending on the particular line.

\subsection{Origin of \ovi~Absorption\label{s:ovi}}

The \ovi ~line is a particularly interesting one
due to its potential importance in radiative cooling and the relatively
narrow temperature range over which it is a significant (e.g. $\rm T\sim10^5-10^6\,K$).
As we have discussed, the kinematics of the coronal gas as traced by the 
\ovi ~absorption line are different from that seen in the cooler gas.  
In particular, \ovi\, is the most blueshifted ion and (on kinematic grounds) is located {\sl primarily} 
in the outflowing gas and not in
galaxy itself.  

Several previous papers have studied \ovi\ in specific galaxies, 
e.g. NGC 1705 \citep{heck01}, \vv \, \citep{grim06}, and \hoo\, \citep{grim07}.
They attributed the production of  \ovi~
to the intermediate temperature regions created by the 
hydrodynamical interaction between hot outrushing gas
and the cooler denser material seen in H$\alpha$ images. 
Such a situation is predicted to be
created as an overpressured superbubble 
accelerates and then fragments as it expands out of the galaxy
and again as the wind encounters ambient clouds in the galaxy halo.

\begin{figure}
\centering
\leavevmode
\includegraphics[width=5in]{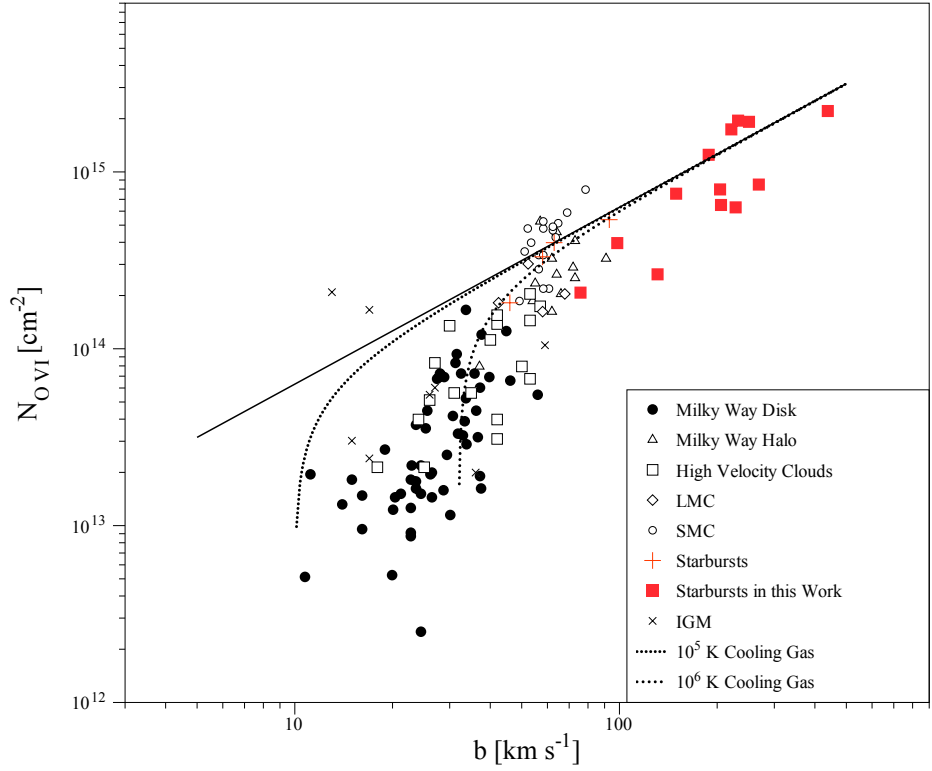}
\caption{
Column density vs. line width for a wide variety of \ovi~absorption line systems including
galactic disk and halo, high velocity clouds, starburst galaxies, and the IGM.
The two dashed lines indicate the predictions  for radiatively cooling gas for assumed
temperatures of $\rm{T_{OVI}=10^5,\, 10^6~K}$.  This plot is originally from \citep{heck02}.
The star-forming galaxies from this work
have been added and extend the relationship to much higher column densities.
\label{f:bovi}}
\end{figure}

\citet{heck02} derived a simple and general relationship between the \ovi~
column density and absorption line width which will hold whenever there 
is a radiatively cooling gas flow passing through the coronal
temperature regime.  They showed that this simple model accounted for the 
properties of \ovi ~ absorption line systems as diverse as clouds in the disk and halo of the
Milky Way, high velocity clouds, the Magellanic Clouds, starburst outflows, 
and some of the clouds in the IGM.  Using the the values found in Table
\ref{t:fuseovi}, in Figure \ref{f:bovi}  we have updated the original plot from \citet{heck02}.
Our results for the \ovi\, measured for the galaxies in our sample are fully consistent with
the model predictions. The fact that starbursts tend to lie at slightly lower column density than the model
predicts is not surprising, given that the model is so generic and idealized (the exact value of the coefficient
relating column density to flow speed depends weakly on the exact nature of the flow, as described
in Heckman et al. (2002).
We conclude that \ovi\, traces radiatively cooling gas produced in an interaction between the
hot outrushing gas seen in X-rays and the cool clouds seen in H$\alpha$ emission, and that this is is a generic feature of starburst
galaxies.

In the case of NGC~1705 \citet{heck01} showed that the overall mass of the coronal phase gas was significantly
smaller than the lower ionization material (by about two orders-of-magnitude). They showed that this is
consistent with the picture in which this gas is rapidly cooling (it is transitory). Following \citet{heck01} we
can make a rough estimate of the gas mass for our sample of galaxies. This mass is simply the product of
the \ovi\, column density, the surface area of the \ovi\ region, and the inverse ratio of \ovi\, to total
column. 
The most physically plausible idealized geometrical 
model is one in which the coronal gas traces the interface between spherically expanding wind and the ambient gas in
the galaxy halo. \citet{grim05} found that the wind radius as traced in soft X-ray emission increased systematically
with both the star formation rate and galaxy stellar mass. Over the range spanned by our sample, the radius would range
from about 0.7 kpc to about 10 kpc. The conversion from \ovi\, to total column density would be factor of about $10^4$
for gas in collisional ionization equilibrium at $T = 3 \times 10^5$ K and with a solar oxygen abundance. The implied
masses of coronal gas range from about $10^6 M_{\odot}$ for the dwarf starbursts like NGC~1705, NGC~4214, and Haro~3 to
about $10^8 M_{\odot}$ for the most powerful starbursts like VV~114 and NGC~3690. 

It is also interesting to note the lack of \ovi\, emission in the sample.
While weak \ovi\, emission is observed in \hoo\, \citep{berg06,grim07}
there is no
convincing evidence for it 
in any of the other galaxies in this sample.  
Very broad (few thousand km s$^{-1}$) \ovi\, emission features appear in a few of the spectra,
consistent with stellar winds in the population of O stars \citep{pell02}.
As \ovi\, emission is the dominant
cooling mechanism for coronal gas, the general lack of observed emission implies that radiative cooling
is not significantly impeding the growth of the outflow.
This is true even in the case of \hoo\
 \citet{grim07}.
They suggested that the \ovi\, emission may not be
from coronal gas cooling but instead could be resonant scattering 
of far-UV continuum photons 
from the starburst off \ovi\, ions in the backside of the galactic wind 
traced by the blueshifted \ovi\ absorption.

\subsection{Dependence of Outflow Velocities on SFR \& Stellar Mass \label{s:velsfr}}

In \citet{grim05} we analyzed
\ch\, data of star-forming galaxies ranging
from dwarf starbursts to ultra-luminous
infrared galaxies (ULIRGs).  We showed 
that the temperature of the hot gas was not
dependent upon the galaxy's SFR.
The thermalized X-ray gas is thought
to be produced in a reverse shock
between the hot out-rushing wind fluid
and the ambient material \citep{strick00},
and its temperature is then determined
by the wind outflow velocity.  
Since a relationship between
X-ray temperature and SFR was not observed,
it follows then that
the wind outflow velocity does not depend
on the SFR. 

In contrast, \citet{mart05b}, \citet{schwa06}, and \citet{rup05} have shown that 
the outflow speed measured in absorption lines does depend on
the SFR.  \citet{mart05b} and \citet{rup05} 
observed outflows in 
luminous infrared galaxies (LIRGs)
and ULIRGs using the NaI D doublet and found that
the outflow velocities increased with the
SFR up to values of SFR of a few tens M$_{\odot}$ yr$^{-1}$.
At higher SFRs (extending up
to several hundred \SFRu),
the relation appears to saturate with  
the outflow velocities no longer increasing.
Similarly, in the STIS starburst sample 
discussed in Section \ref{s:velvel} \citet{schwa06} also 
found that the outflow speeds
increased with the SFR below $\sim$10 \SFRu.

\citet{mart05b} suggests that the simplest explanation
is that there is not enough momentum carried by the wind fluid
in the lower SFR galaxies, so that the winds are unable
to accelerate the entrained absorbing material up to the 
velocity of the wind fluid itself. The relation between
SFR and the outflow speed seen in the absorption-lines 
saturates at about 500 $\ks$ \citep{mart05b} and this is reached at
starburst luminosities $sim$ few $\times 10^{11} L_{\odot}$. If the X-ray emission in galactic
winds arises in wind shocks, the observed temperatures of
$kT_X \sim$ 0.3 to 0.7 keV correspond to wind shock speeds of
$\sim$ 500 to 800 $\ks$ \citep{grim05}. This is then compatible
with the idea that at sufficiently high SFR (high enough wind momentum)
the entrained absorbing gas has been accelerated to near the wind speed.


We can check these results with our data. To that end, we will use the sum of the
IRAS IR and \fu\, UV luminosity as our
tracer of the SFR (see Table \ref{t:fuselumins}).
The UV luminosity is based on the 
continuum flux at 1150 \AA \,and is
corrected for Milky Way extinction.
The combined UV and IR luminosity
provides a reasonable estimate of the
intrinsic UV flux of the galaxy, and thus traces
recent SF.  

\citet{mart05b} also found that
the outflow velocity increased with the
rotational speed of the galaxy.  This suggests that
faster outflows are found in more massive galaxies.
We check this correlation using the near-infrared luminosity of the galaxy
($\rm L_K$) as a proxy for stellar mass.
The K-Band Cousins-Glass-Johnson (CGJ) luminosities are derived
from the 2MASS atlas \citep{jarr03}.  The 2MASS K and J Band
fluxes given in the catalog were converted to K-Band CGJ using the
transformations from \citet{carp01}. We expect the K-band luminosity
to trace the stellar mass to a factor of roughly two (e.g. Leitherer \& Heckman 1995;
Bell \& de Jong 2001; Dasyra et al. 2006).

\begin{figure}
\centering
\leavevmode
\includegraphics[width=6in]{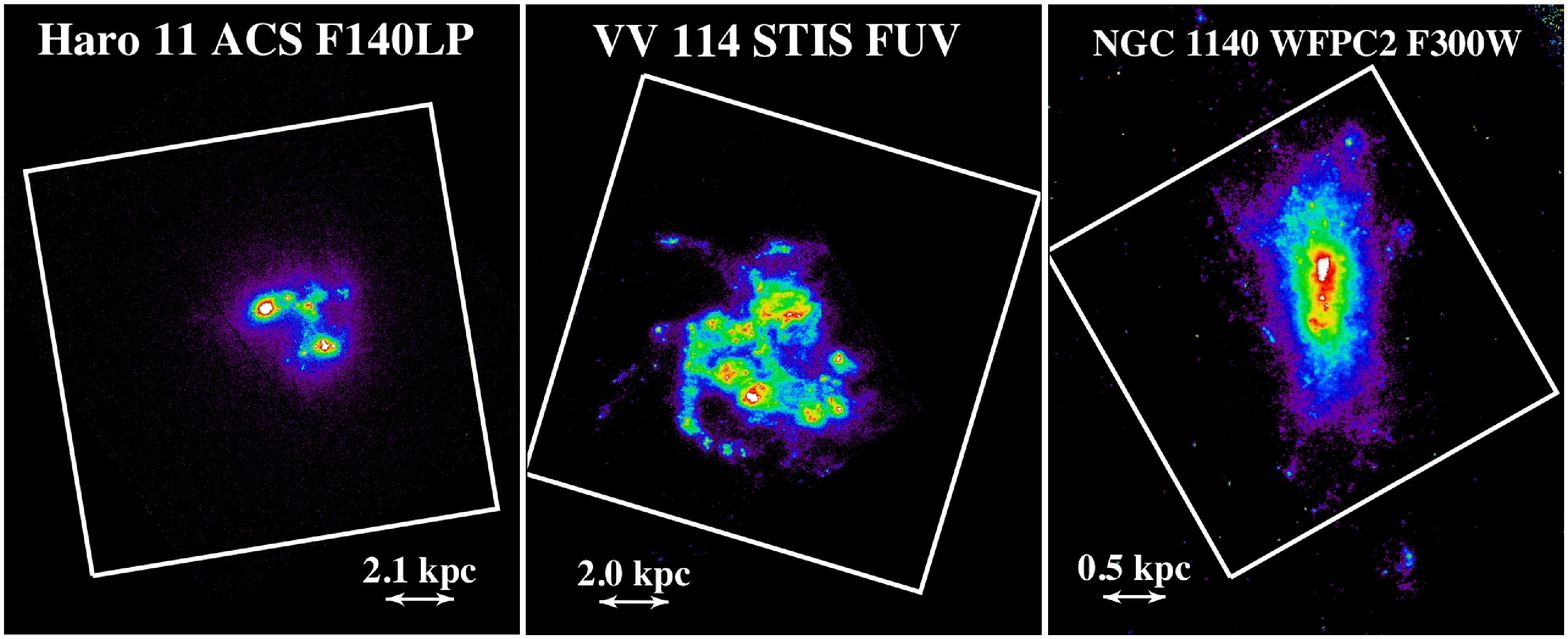}
\includegraphics[width=6in]{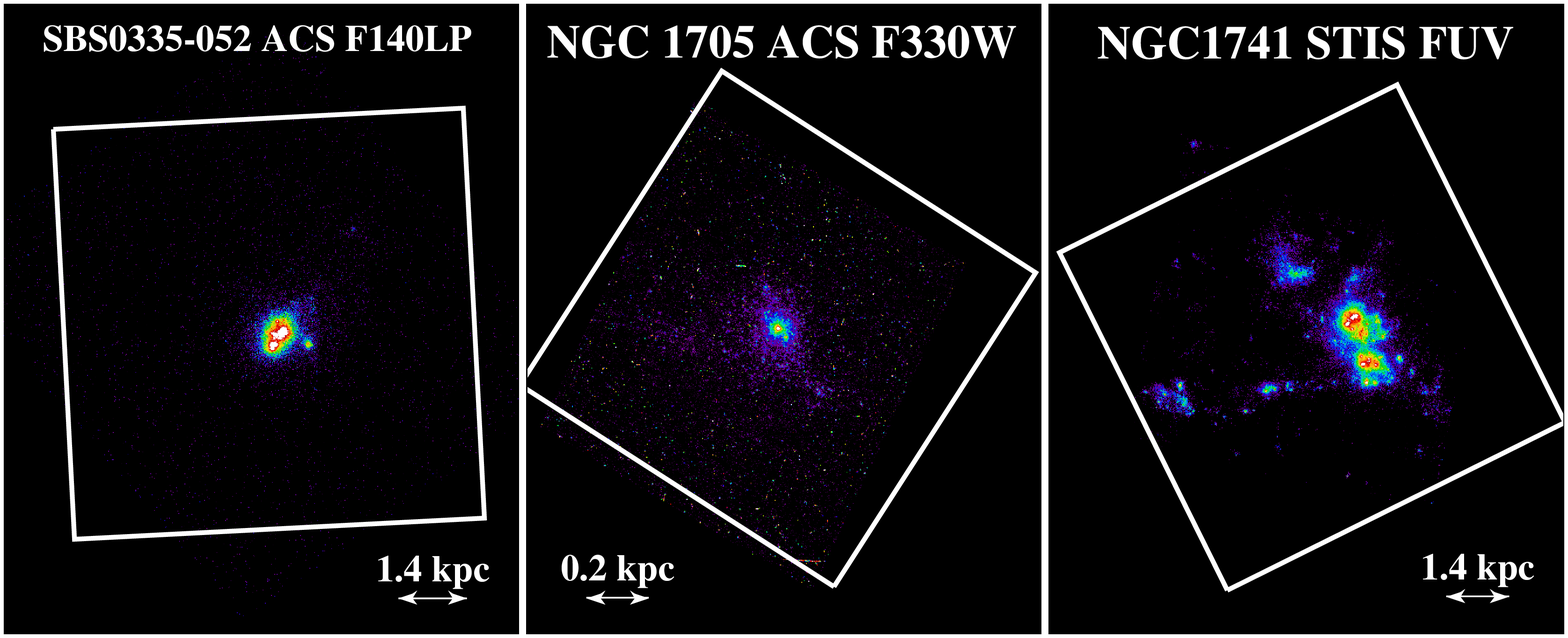}
\includegraphics[width=6in]{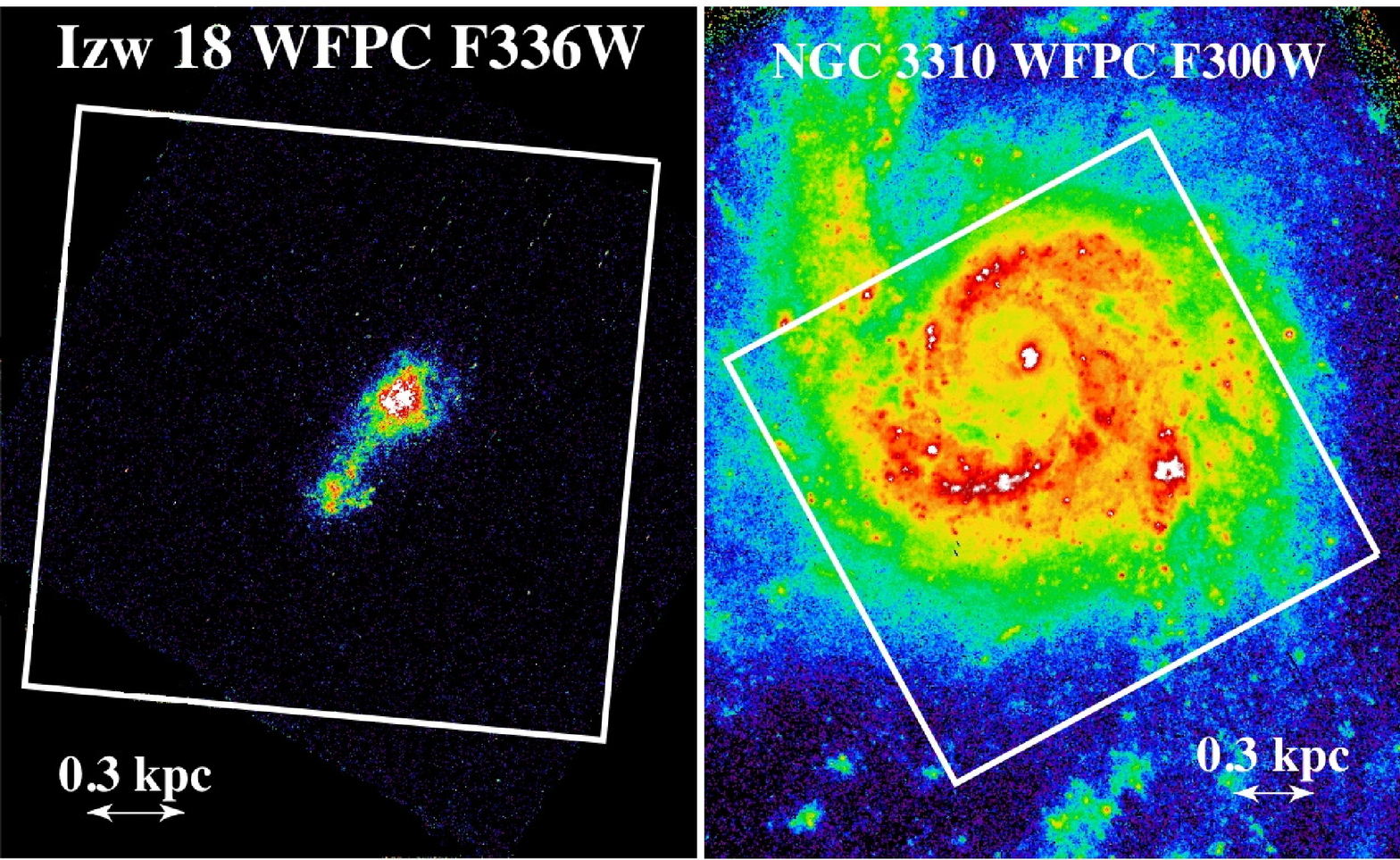}
\caption{
UV or FUV images with \fu\,aperture overlay for \hoo, \vv, NGC\,1140,
\sbs, NGC\,1705, NGC\,1741, \izw, NGC\,3310, and Haro\,3.
The \fu\, LWRS aperture has been overlaid at the nominal
position and orientation for the \fu\,observations.   The image
of Haro\,3 is a {\sl GALEX  } NUV observation at significantly lower
spatial resolution than observed in the other panels.
The overlaid arrows
are 5\arcsec in length.
\label{f:images1}}
\end{figure}

We have made no attempt to correct
for the viewing angle dependence of
the outflow velocity.  This is primarily a practical
decision as we 
do not have a robust way of obtaining the
orientation of the outflow in these galaxies.
While this will increase the scatter in our 
outflow velocity plots,  sample
selection effects probably minimize
the resulting damage.
In particular, an edge-on spiral
galaxy would not have had enough FUV flux to meet our
S/N requirements as the disk heavily absorbs
the nuclear starburst (e.g. M\,82).  
Spiral arm features are apparent for
several of the 
galaxies in our sample (Figures 
\ref{f:images1} and \ref{f:images2}) 
suggesting that we are observing the galaxies 
roughly face-on, and hence roughly along the expected outflow
direction.  Several other of the galaxies
have irregular morphologies.  
In these irregular galaxies it is not
at all apparent how inclination angle would be defined.
As an example, \hoo\, has several
regions of intense SF which high resolution
HST ACS imaging suggests 
could be driving multiple outflows.

\begin{figure}
\centering
\leavevmode
\includegraphics[width=3.1in]{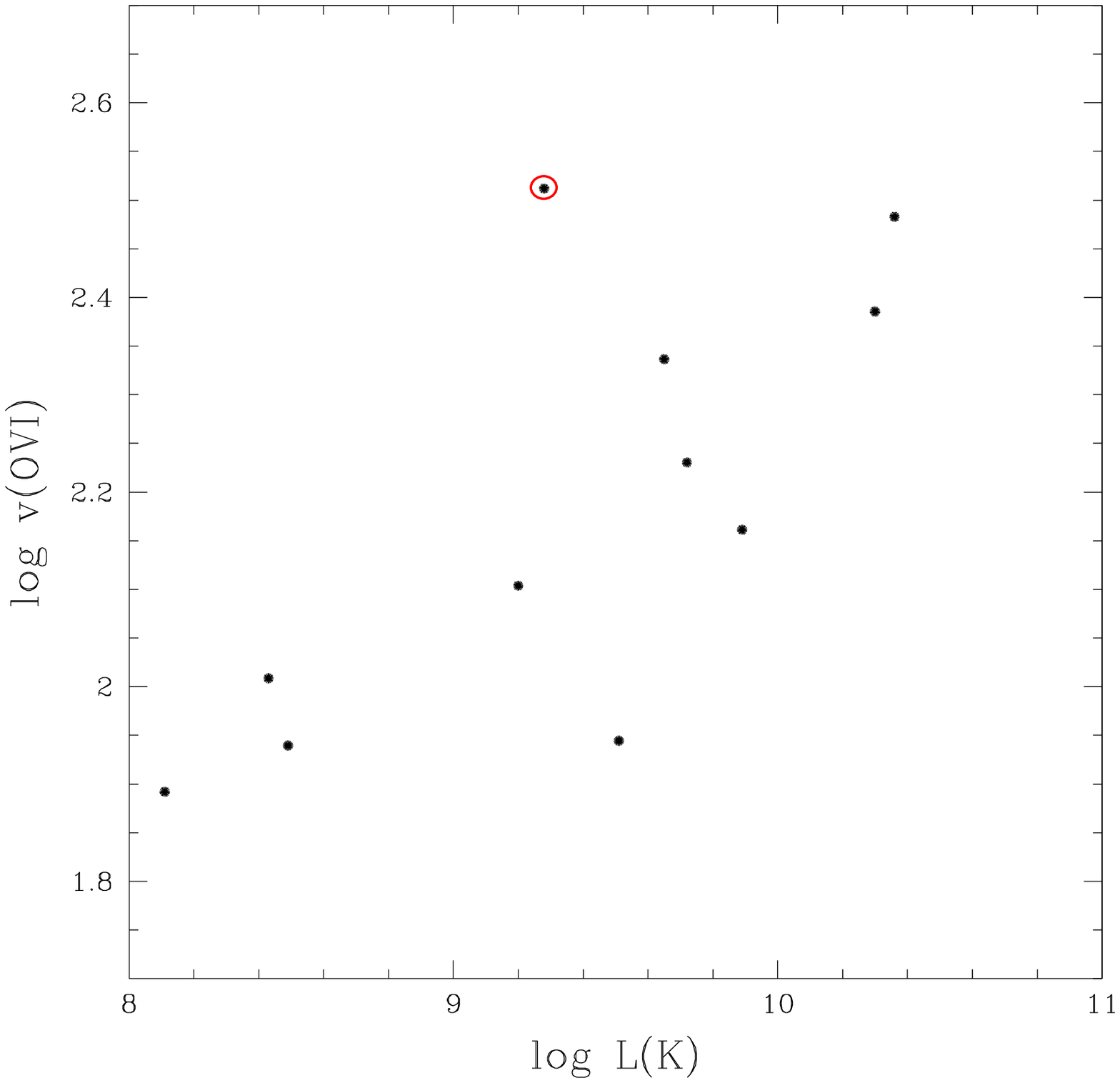}
\includegraphics[width=3.1in]{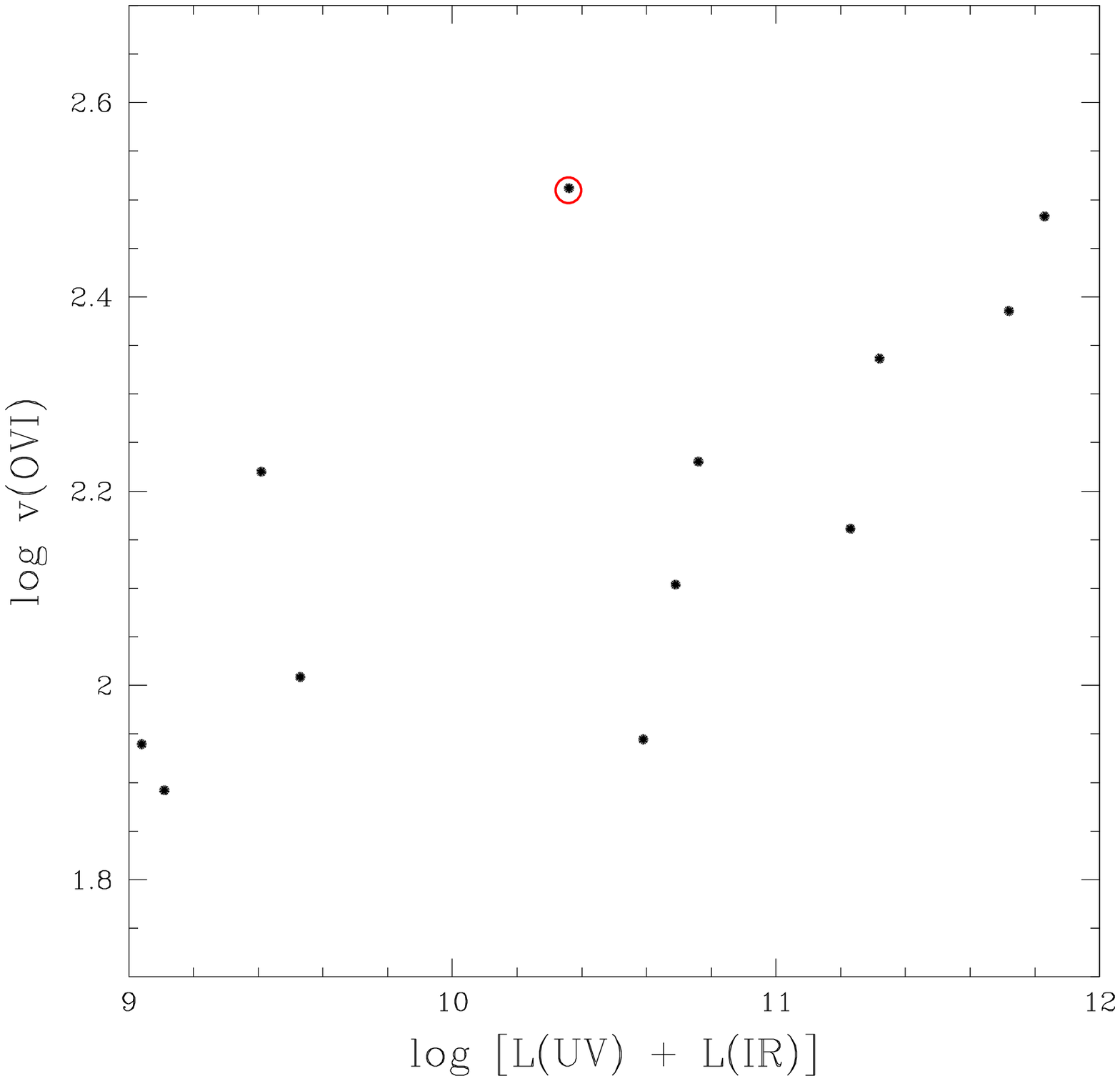}
\caption{
The log of the \ovi\, outflow velocities in  km/s versus the log of the K-Band and 
IR+UV luminosities (in $L_{\odot}$). These are roughly proportional to the stellar mass
and star formation rate respectively. 
NGC~3310 (which has unusually large outflow velocities for its
mass and star formation rate) is denoted by a red circle.
\label{f:vovi}}
\end{figure}

\begin{figure}
\centering
\leavevmode
\includegraphics[width=6in]{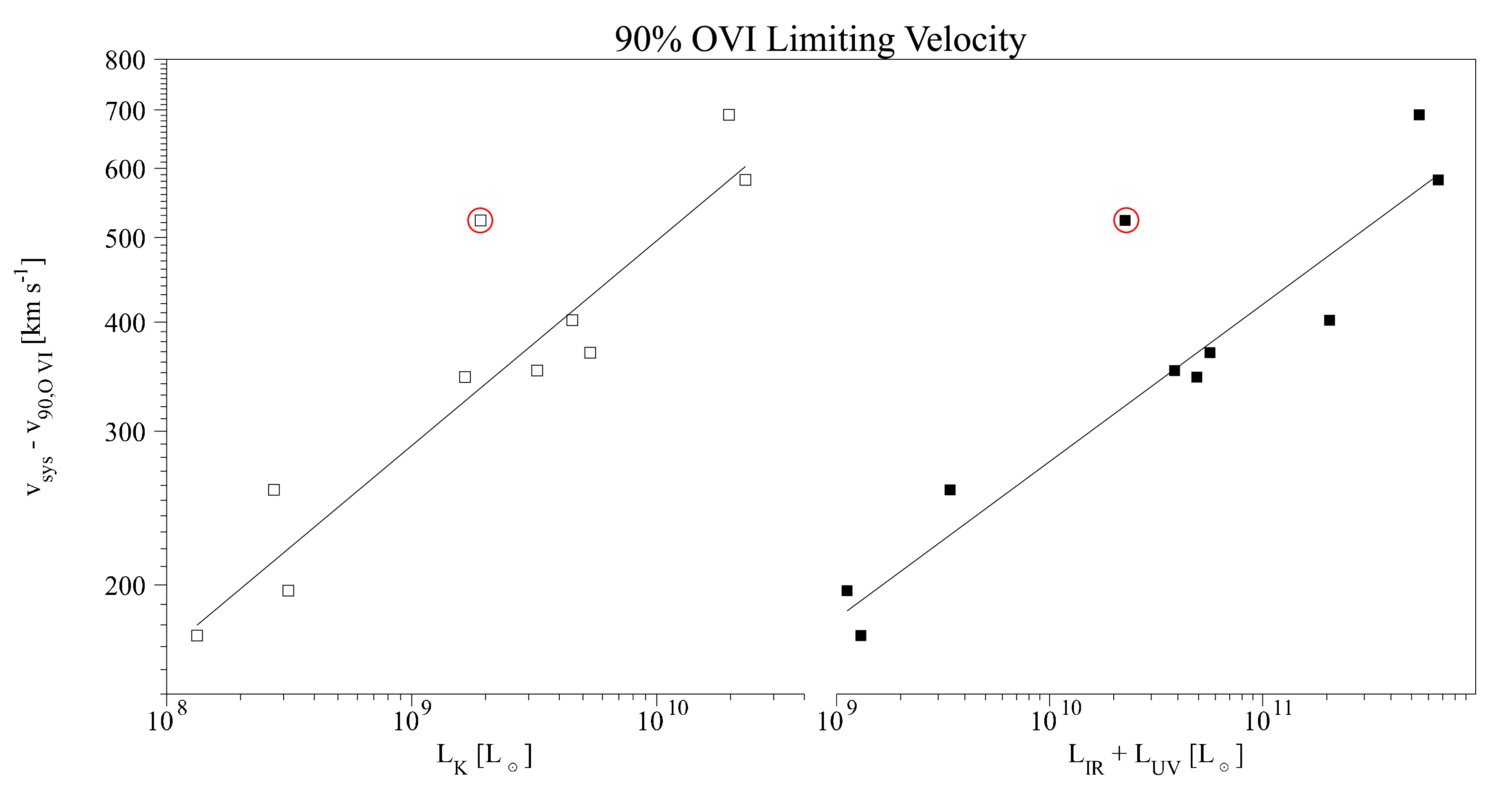}
\caption{
Same as Figure 41, except that here we plot the `maximum' outflow velocity. This is defined so that 90\% (10\%) of 
the absorbing column of gas is to the red (blue) of the velocity. See text for details.
NGC~3310 (which has unusually large outflow velocities for its
mass and star formation rate) is denoted by a red circle.
\label{f:vovi90}}
\end{figure}

\begin{figure}
\centering
\leavevmode
\includegraphics[width=3.1in]{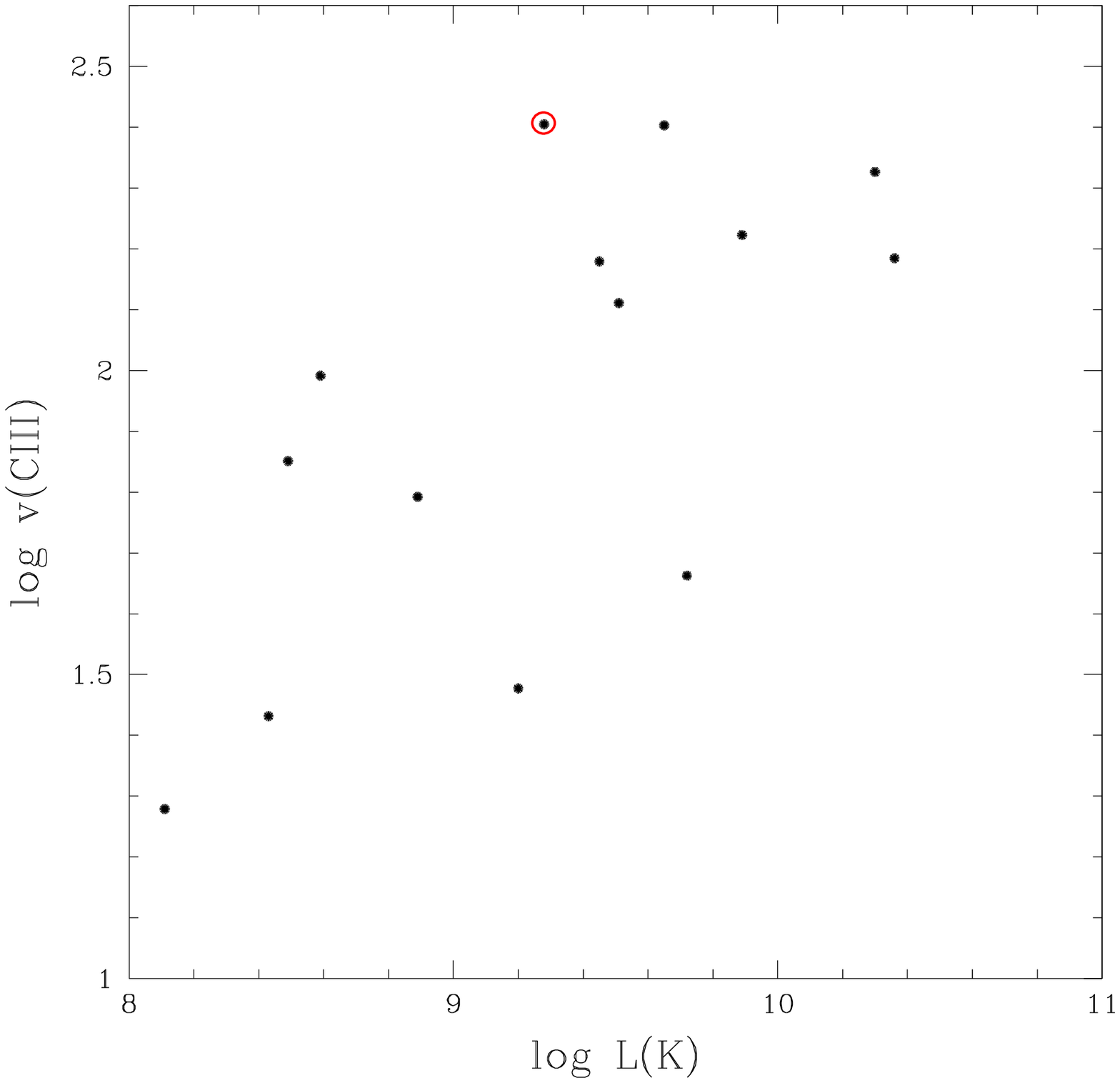}
\includegraphics[width=3.1in]{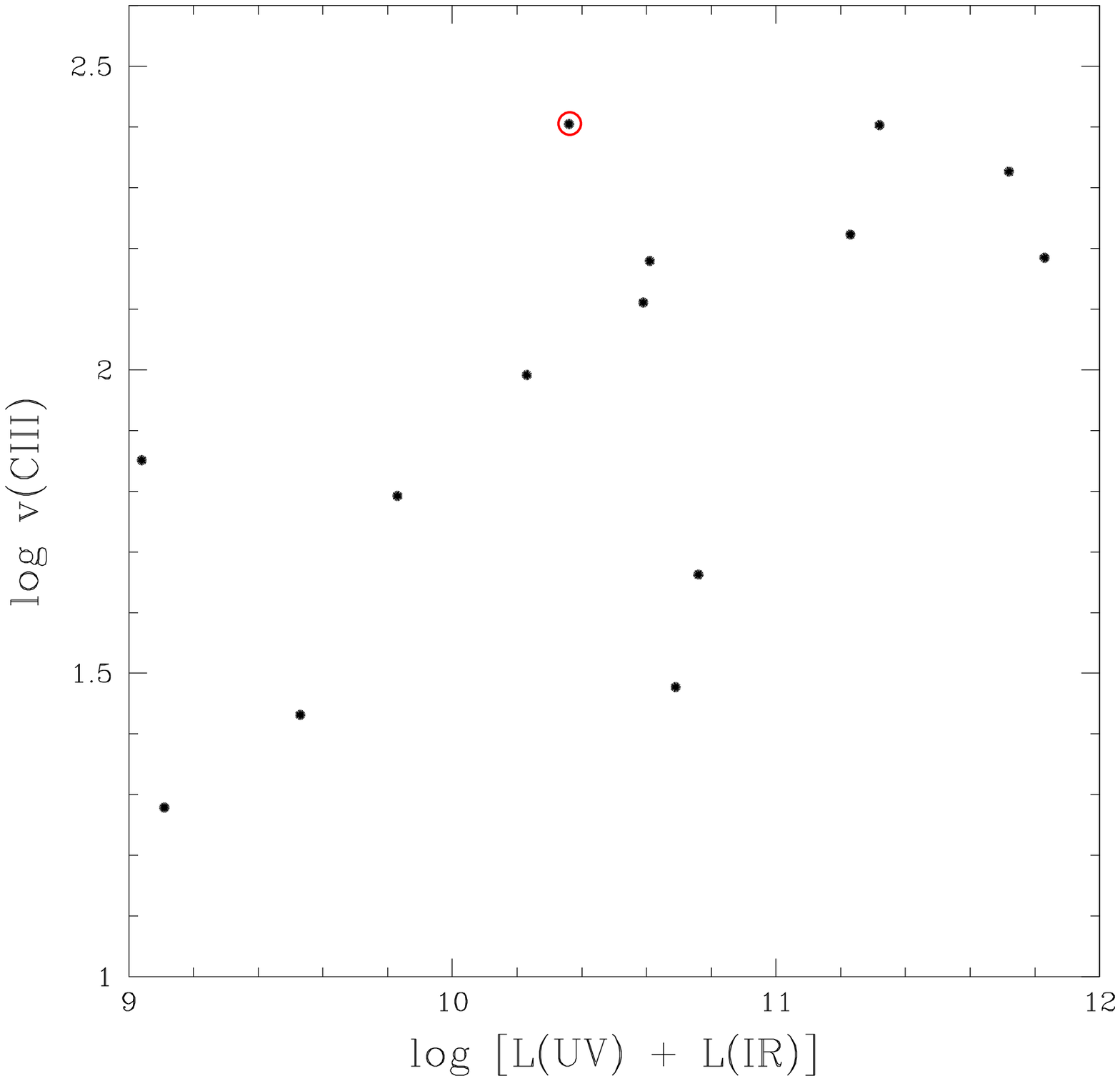}
\caption{
The log of the \ciii\, outflow velocities in  km/s versus the log of the K-Band and 
IR+UV luminosities (in $L_{\odot}$). These are roughly proportional to the stellar mass
and star formation rate respectively.
NGC~3310 (which has unusually large outflow velocities for its
mass and star formation rate) is denoted by a red circle.
\label{f:vciii}}
\end{figure}

In Figures \ref{f:vovi} and \ref{f:vovi90} we
have plotted the \ovi\, outflow velocity
as a function of the near-IR K-band luminosity
and the UV+IR luminosity, using two measurements
of the outflow velocity. The first plot uses the
median velocity defined by the apparent optical
depth method (e.g. half the absorbing column density
lies on either side of this velocity). The second plot
traces the blueward extent of the outflow (the `maximum'
outflow speed). We define this as that velocity for which
90\% (10\%) of the absorbing column density lies to the
red (blue). The maximum outflow velocity correlates better
both the K-band and the UV+IR luminosity that does
the median velocity.
For comparison, in Figure \ref{f:vciii}
we have plotted the \ciii\, outflow
velocities (medians) as a function of 2-Mass K-band
and UV+IR luminosity.


The outflow speed of \ovi\, clearly correlates better with 
with the SFR than does \ciii. The
increase in outflow velocity with increasing SFR is also observed in the other
strong absorption features such as \cii\, and \nii.
We have noted previously that the profile shapes
of these strong lines suggest that there are
two components (higher column density gas at or near the systemic
velocity and lower column density outflowing gas). In contrast,
the \ovi\, absorption seems to originate only in the outflow. This
may explain the stronger correlation ibetween SFR and outflow speed
seen in \ovi.

\begin{figure}
\centering
\leavevmode
\includegraphics[width=3.1in]{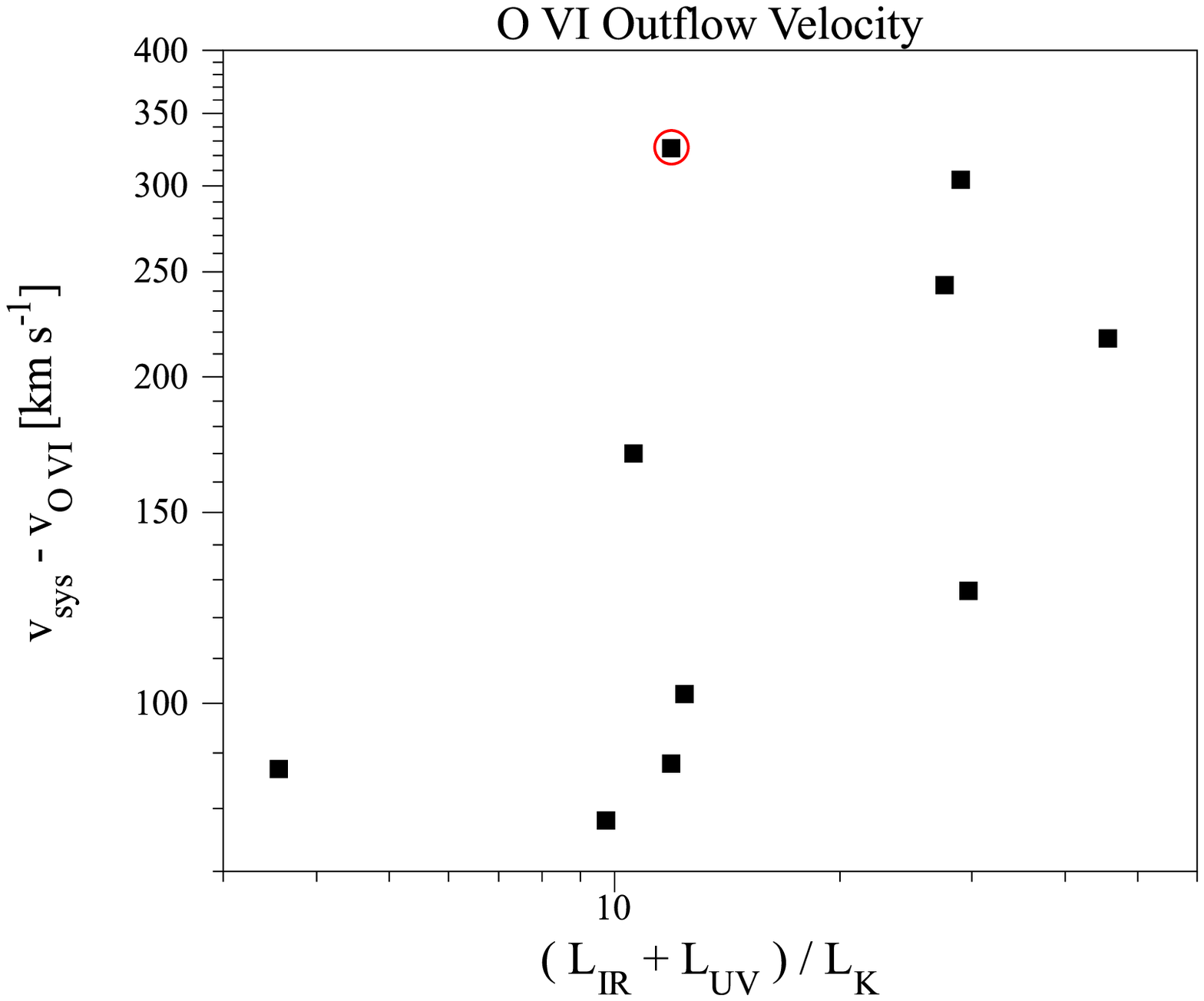}
\includegraphics[width=3.1in]{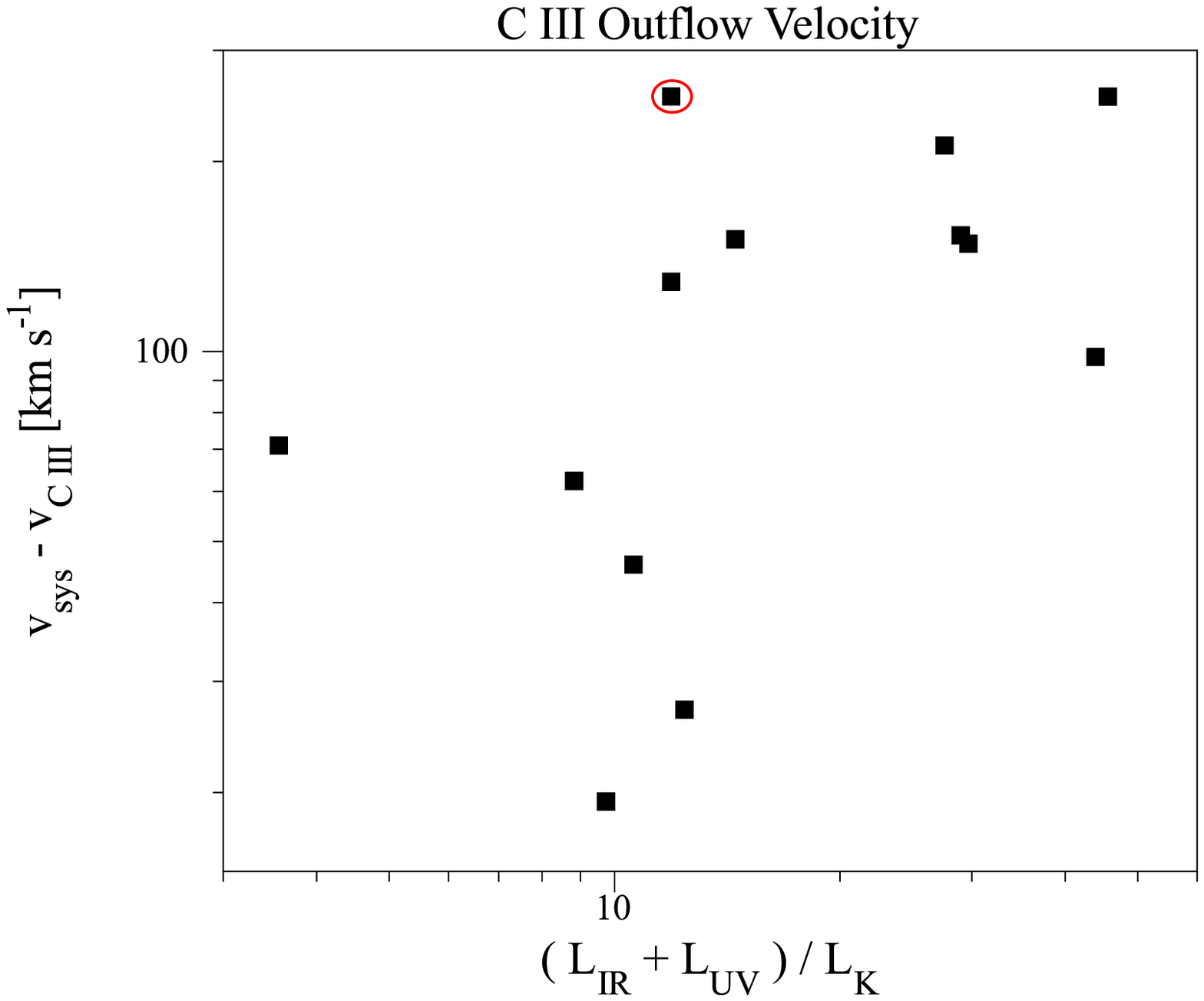}
\caption{
\ovi\, and \ciii\, outflow velocities versus $\rm (L_{IR} + L_{UV})/L_K$.
While these plots have a large degree of scatter, 
the higher SFR/Mass galaxies tend to have higher
outflow velocities. NGC~3310 (which has unusually large outflow velocities for its
mass and star formation rate) is denoted by a red circle.
\label{f:sfrdk}}
\end{figure}

Since the outflow speeds increase with both SFR and stellar mass,
it is instructive to examine the outflow speeds as a function
of the specific SFR (SFR/$M_*$). To do this, we 
have plotted the median \ovi\, and \ciii\,outflow
velocities versus $\rm (L_{IR} + L_{UV})/L_K$ 
in Figure \ref{f:sfrdk}.  
While these plots show a significant amount
of scatter, we see that the
outflow speeds generally increase with the specific SFR.
The \ovi\,outflow velocity correlates
more strongly with $\rm (L_{IR} + L_{UV})/L_K$
than does the \ciii\,velocity.
As previously discussed, the properties of \ovi\,are closely tied
to the wind and thus presumably to the SFR.

The case of NGC\,3310 requires further discussion as it is a significant outlier
in most of our plots, with an outflow speed that is larger than expected for
its SFR or mass.  We note that the very large outflow speeds we see have been
confirmed with STIS far-UV spectra of three individual regions in NGC 3310 (Schwartz et al. 2006).  
Examining the UV morphology of NGC\,3310 in Figure \ref{f:images1},
we see that NGC\,3310 is the only galaxy with significant star formation
spread throughout multiple locations in the \fu\, aperture.
Radio HI velocity maps \citep{kreg01} of the regions observed by \fu\, display 
a spread of almost 200 \kms\, which could increase
the observed width of our absorption features, but should not produce
highly blueshifted absorption.
Another way in which NGC\,3310 stands out
within our sample is that it is the highest metallicity galaxy with 
$\rm\log(O/H)+12=9.0$.  
In principle, higher metallicity 
could allow us to probe gas with lower total column density.
However, as we will show below
the strengths of the strong absorption lines do not depend
significantly on metallicty in our sample. 
Thus, this is unlikely  
to explain the large observed outflow speed in NGC\,3310.

NGC\,3310 is included in the young stellar population study of 16 starburst galaxies
(11 also in our sample) by \citet{pell07}. By fitting the FUV spectra with models
of instantaneous bursts, they
estimated an 
age of $18\pm2$ Myr for the starburst in NGC\,3310 which was 10 Myrs older
than any of the other galaxies in their sample.
As the Type II supernovae that drive the galactic wind persist for
40 Myr, the total amount of kinetic energy injected into the wind
will increase as the starburst ages.
NGC\,3310 appears to be a more mature starburst
and may therefore have a more evolved wind.  

%

\subsection{Absorption-Line Equivalent Widths \label{s:eqw}}

It has known for some time that the strengths of the UV interstellar absorption-lines
correlate with other key properties of starbursts. \citet{heck98} used IUE spectra of 45 local starbursts
to show that the equivalent widths of the strongest lines arising in the HI phase of the ISM correlated
postively with the amount of UV extinction and reddening, the galaxy rotation speed and absolute magnitude,
the metallicity, and SFR. Unfortunately, the poor spectral resolution and modest S/N of these spectra
did not allow them to determine the astrophysics underlying these correlations. Because starbursts
in more massive galaxies are known to have higher metallicity \citep{trem04}, higher SFR, and higher extinction \citep{heck98}
it is especially difficult to sort out the fundamental relations from secondary correlations.

In general, the equivalent width of an interstellar absorption-line will depend on very different
parameters depending on the optical depth of the line. In the limiting case of optically-thick
gas the equivalent width is set purely by the product of velocity width of the line (the Doppler b-parameter)
and the fraction of the UV continuum source covered by the absorber (the covering factor). In the case
of optically thin gas the equivalent width is determined by the product of the ionic column density,
the oscillator strength of the transition, and the covering factor. In our data we have examples
of both optically thick and optically-thin transitions. Because the optically-thick lines are black
(or nearly so) at line center, we know the covering factor is near unity.

\begin{figure}
\centering
\leavevmode
\includegraphics[width=5in]{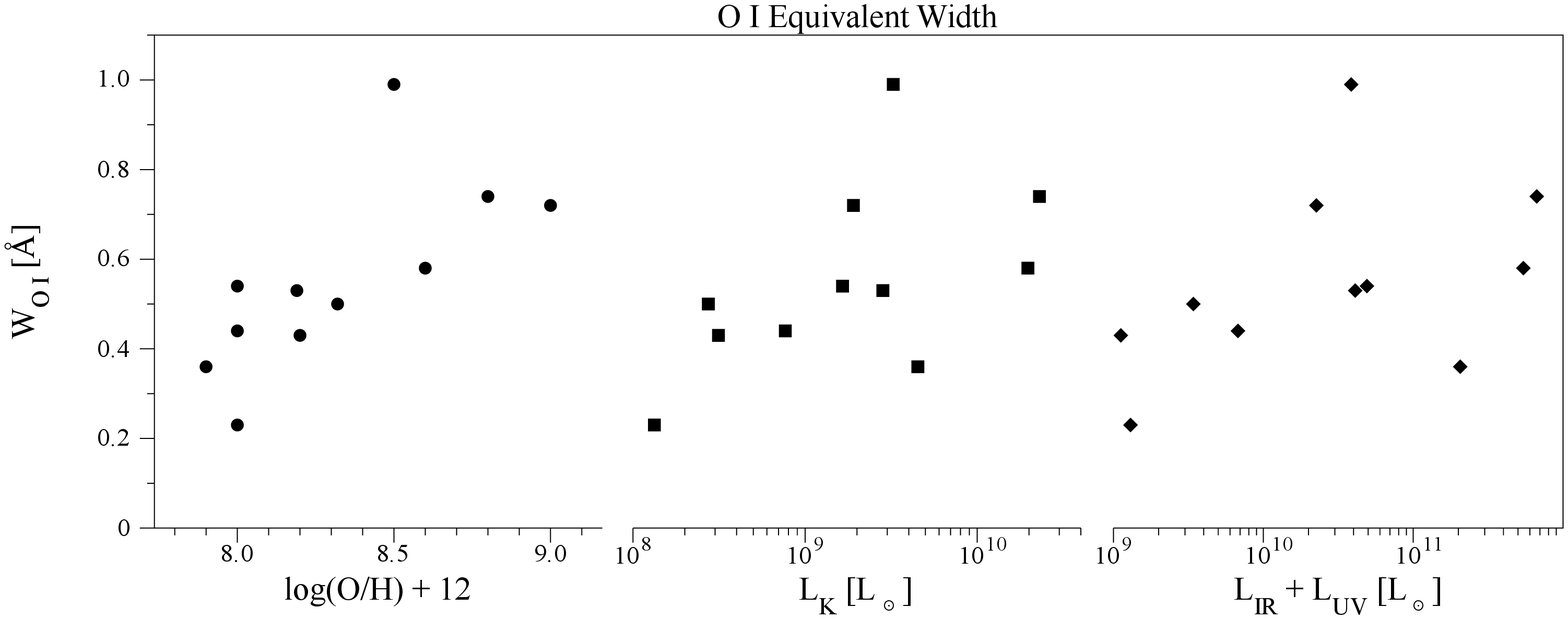}
\includegraphics[width=5in]{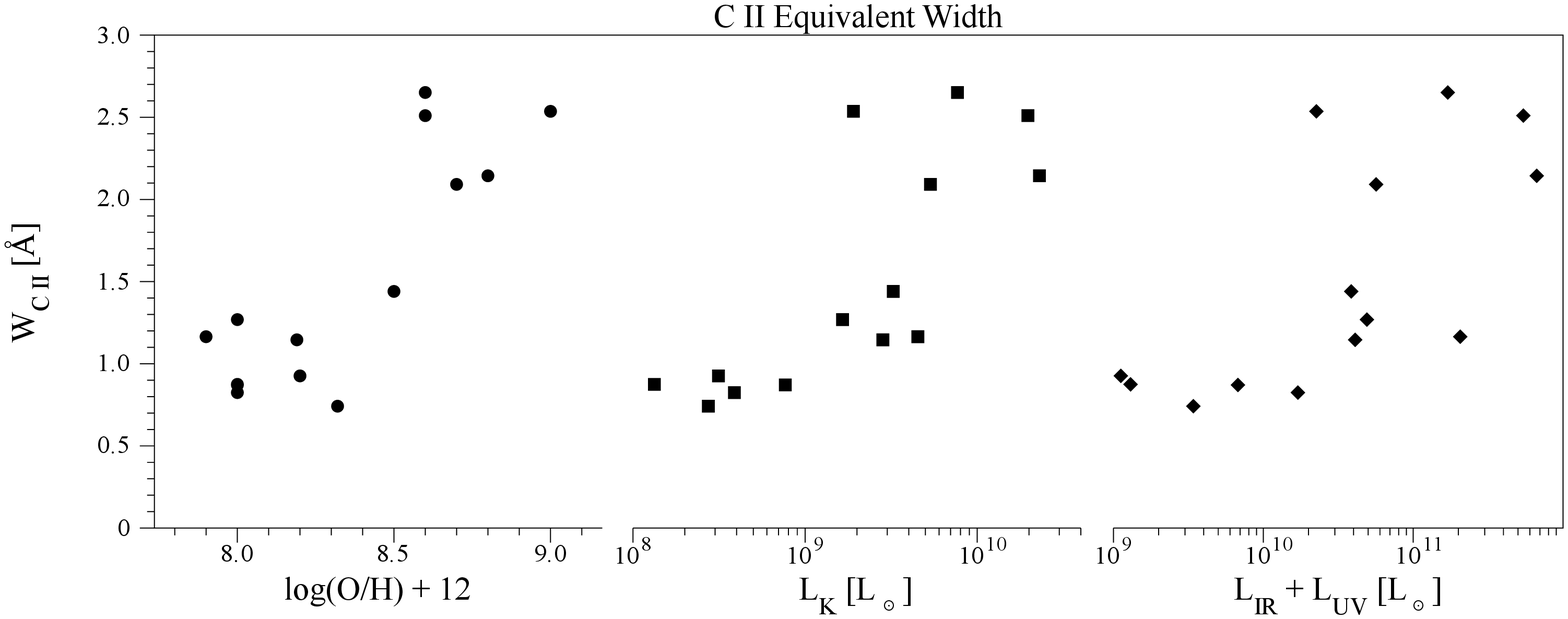}
\includegraphics[width=5in]{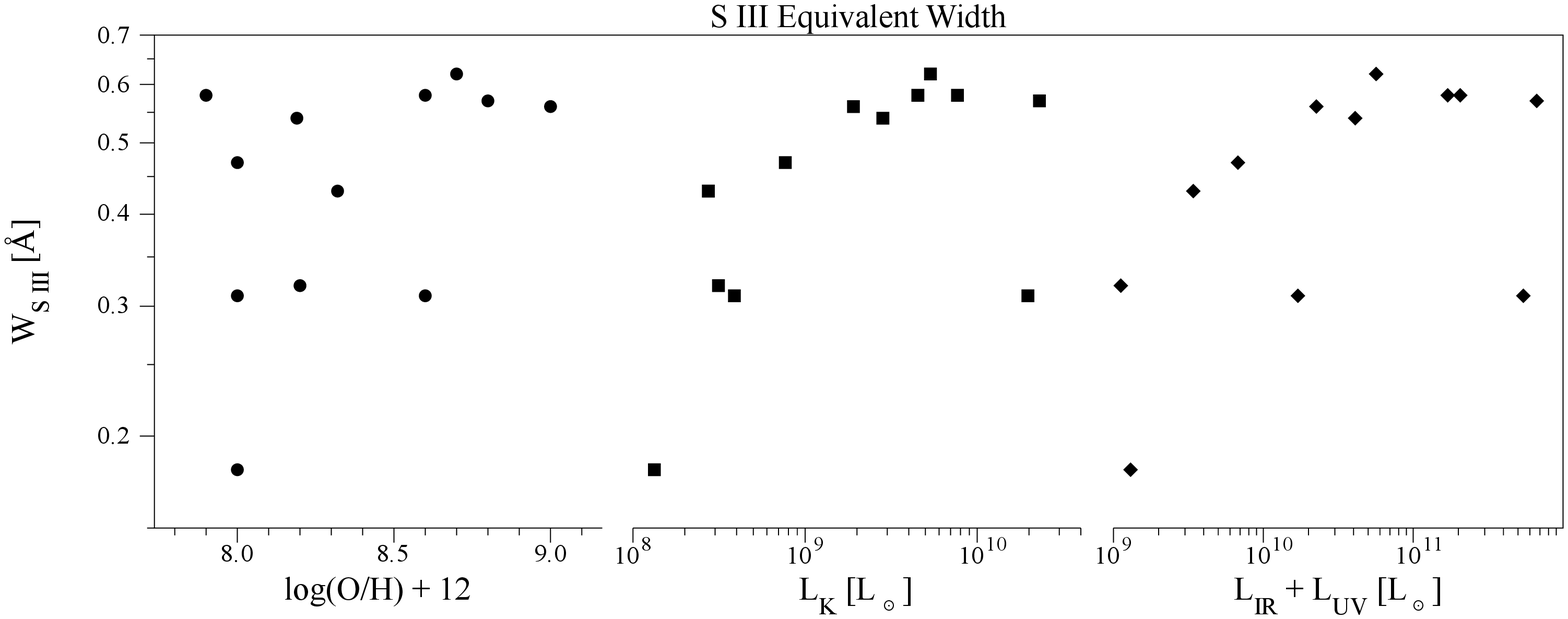}
\includegraphics[width=5in]{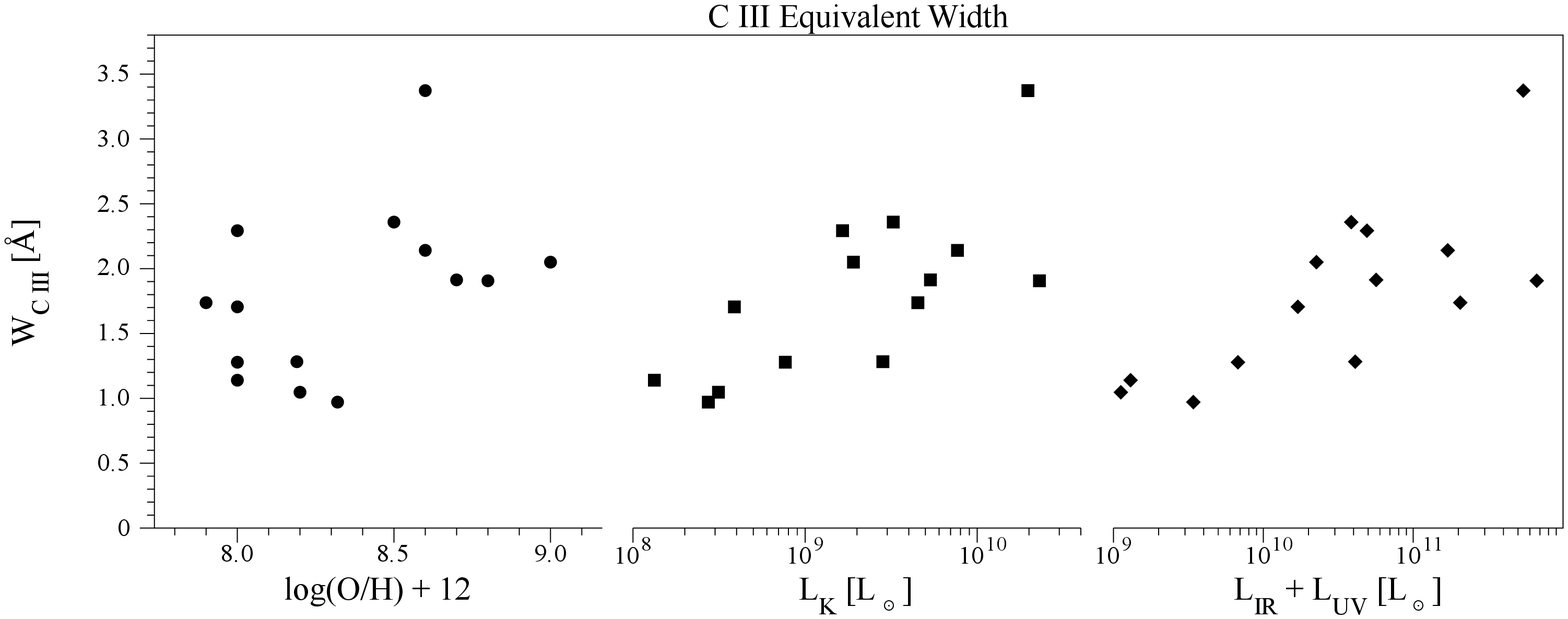}
\caption{
Equivalent widths of \oi, \cii, \sth, and \ciii \,
compared to gas phase metallicity, K-band luminosity,
and IR+UV luminosity.  For the lower ionic states (\oi\, \& \cii)
the gas phase abundance and stellar mass 
track the strength of the lines.  The higher ionization states
however poorly match the metallicity and instead
are correlated with the SFR and stellar mass.  
\label{f:eqw}}
\end{figure}

We also probe gas spanning a range in ionization potential. We have described the properties of the coronal
gas traced by \ovi\ at some length, so here we will confine our attention to the lines tracing the HI phase (ionization potentials less than
a Rydberg) and the HII phase (ionization potentials between 1 and 4 Rydbergs). For each of these two phase we will use one optically-thick
and one optically-thin line. Hence, 
in Figure \ref{f:eqw} we plot the equivalent widths
of \oibw, \ciiw, \sthw, and \ciiiw\, versus 
gas phase metallicity, K-band luminosity,
and IR+UV luminosity.  The first two lines trace the HI phase, with the \oi\, line being optically-thin and \cii\, optically-thick.
The second two lines trace the HII phase, with the \sth\, line being optically-thin and \ciii\, optically-thick.

In fact, the only pattern that emerges is that the poorest correlations are those between
the equivalent widths of the lines tracing the HII phase (the \ciii\, and sthw\, lines) and the metallicity (Kendall $\tau_a=0.38$ and 0.44
respectively). All the other correlations are of comparable strength (Kendall $\tau_a=0.56$ to 0.75). This may reflect the relatively
small sample size and the fact that the different starburst propertiues are interconnected:
galaxies with higher masses will
have higher velocity dispersions, higher SFRs, and higher metallicity.

\subsection{Lyman Continuum Emission \label{s:lyman}}

One of the major puzzles in cosmology today is determining
the source of the UV photons required to reionize the universe 
at  redshifts $\rm{6\lesssim z \lesssim 14}$ \citep{fan06}. 
The known population of AGN do not
 appear to be sufficient by a wide margin, and the leading candidates are star
forming galaxies \citep{sti04,pan05}.  While stars in the early
universe are thought be able to produce enough UV photons, 
it is not well understood what fraction of these photons
can escape through the neutral hydrogen gas of the host galaxy \citep{yan04}
into the IGM.  Superwinds, driven by starbursts, may be able to
enhance the fraction of ionizing radiation by clearing channels
through the surrounding gas \citep{dov00}.

The escape of ionizing continuum radiation from star forming galaxies 
was first observed by \citet{steid01} in a sample of 
$\rm{z\sim 3.4}$ LBGs.  
More recent work by \citet{shap06} focused on deep rest-frame UV spectroscopy 
and found a sample-averaged relative Lyman continuum
escape fraction ($f_{esc}$) of 14\%.  They however detected
Lyman continuum emission in only 2 of the 14 galaxies in
their sample, which suggests that even at high redshifts ($\rm{z\sim3}$)
ionizing radiation is either relatively rare or viewing angle dependent.

Determining the escape fraction in starburst galaxies
is therefore an important step in evaluating 
the importance of starburst galaxies in reionizing the universe.
Local searches for Lyman continuum emission have so far been inconclusive.
 \citet{leith95} and \citet{hurw97} found $f_{esc}\lesssim\,10\%$ in a galaxy sample
 based on {\it HUT} observationsi and Deharveng et al. (2001) obtained
a similar result for Mrk~54 using \fu. 
Most recently \citet{berg06} published an analysis showing 
the detection of Lyman continuum emission in \hoo.  
We however re-analyzed the same data in
\citet{grim07} and did not find any
convincing evidence of Lyman leakage.  
We showed that uncertainties in the background
subtraction and geocoronal contamination
were larger than any detected signal.
We therefore placed a limit on the escaping Lyman continuum
emission in \hoo\, of $f_{esc}\lesssim\,2\%$.

\subsubsection{Below the Lyman Limit\label{s:belowlyman} in Other Starbursts}

\begin{figure}
\centering
\leavevmode
\includegraphics[width=5in,angle=90]{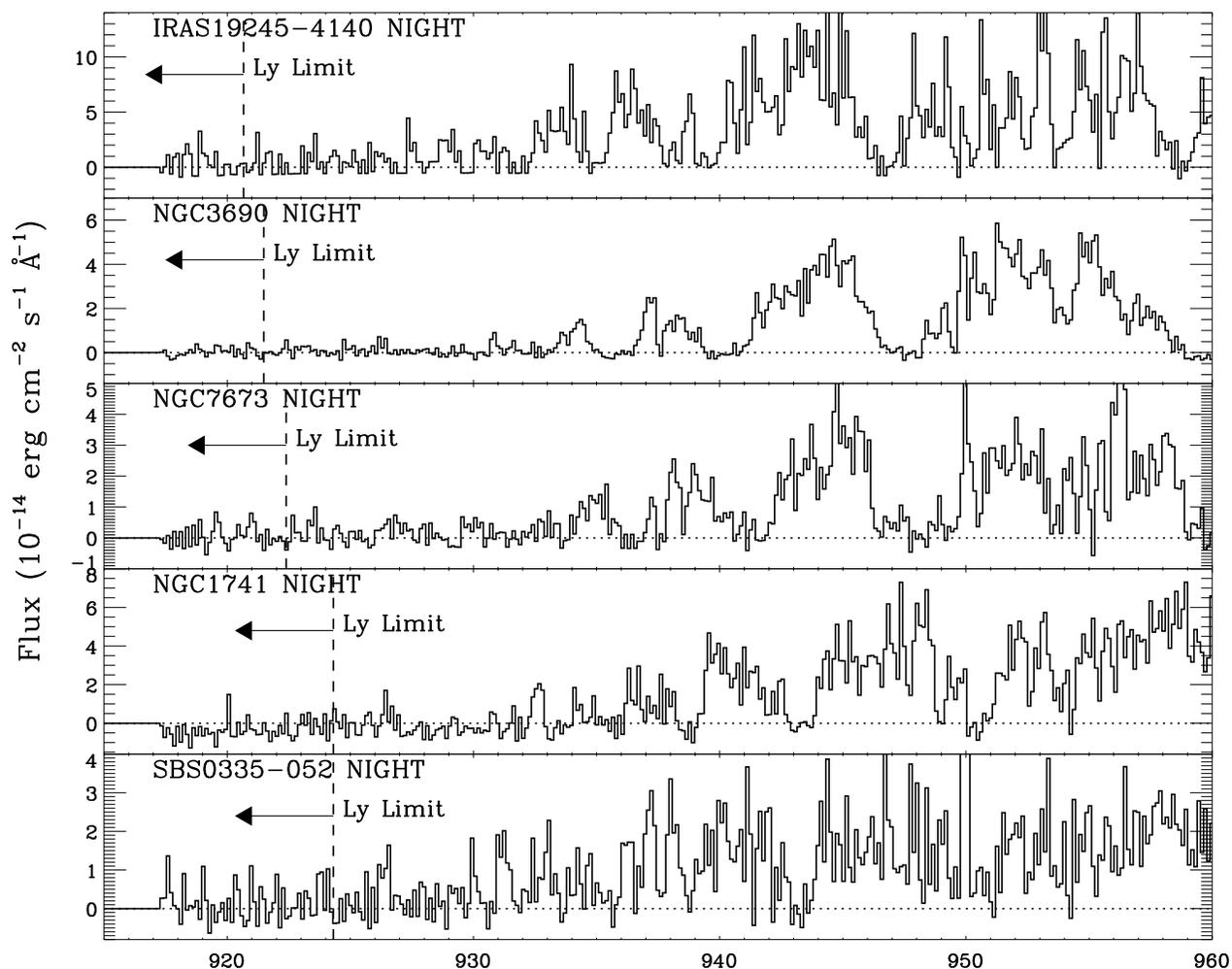}
\caption[SiC 2A Spectra of Lyman Continuum Regions]{
This figure shows the Lyman continuum regions of five galaxies in order of
increasing redshift.  We have plotted only the night data in order to minimize
contamination from airglow and earth limb emission.  There is 
no convincing evidence of Lyman continuum emission in 
 these spectra.  The spectra are generally very noisy in this wavelength region
and any possible signal is dominated by the background uncertainties and 
airglow emission.  These plots do show earth limb and airglow 
contamination from \ion{H}{1}, \ion{O}{1}, \ion{N}{1}, and $\rm{N_2}$ (for a better
example see Figure \ref{f:flyc2}).  
\label{f:flyc1}}
\end{figure}

\begin{figure}
\centering
\leavevmode
\includegraphics[width=5in,angle=90]{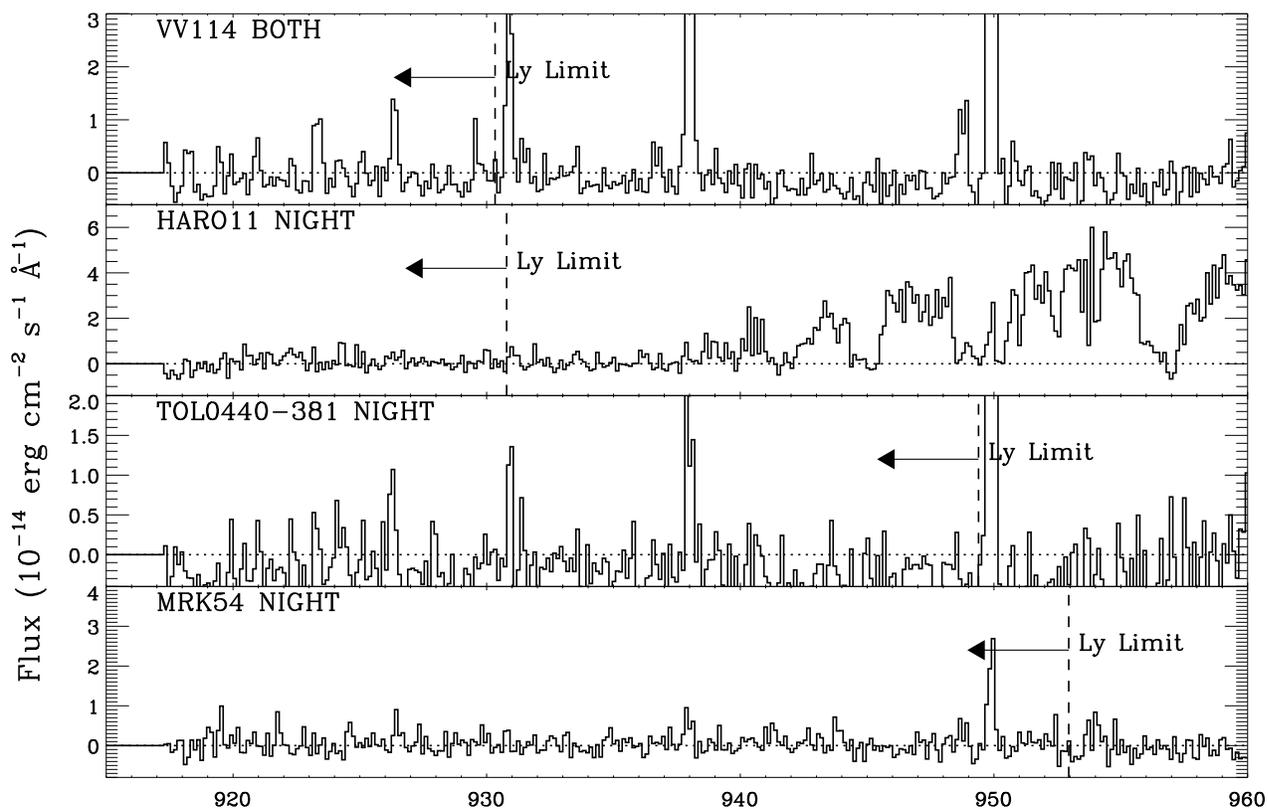}
\caption[SiC 2A Spectra of Lyman Continuum Regions (continued)]{
This is identical to Figure \ref{f:flyc1} but for an additional four galaxies.  
No convincing evidence of Lyman continuum emission is observed in 
 these galaxies.  Earth limb and airglow emission however are
present from \ion{H}{1}, \ion{O}{1}, \ion{N}{1}, and $\rm{N_2}$.  This is 
easily confirmed by comparing the specta, particularly 
\vv\, and \tol\, which both have strong contamination from airglow
in the Lyman continuum regions.  These features however are 
present in all eight spectra in the two figures.
We have included both the night and day time data for \vv\, due to the small
percentage of night time data available for that target.
\label{f:flyc2}}
\end{figure}

We have looked directly for Lyman continuum emission in 
other galaxies observed by \fu.  In order
to move the Lyman limit farther from the
detector edge, we have chosen the nine highest redshift
galaxies in our sample.  
Figures \ref{f:flyc1} and  \ref{f:flyc2} show the
SiC 2A spectra for the nine galaxies plotted
in order of increasing redshift.  While the SiC 1B
spectra also cover this wavelength regime, 
background subtraction problems are even more
severe for the SiC 1B channel. 
In order to minimize 
contamination by dayglow we have plotted the
night only spectra for all of the galaxies with the
exception of \vv.  In the case of \vv, there was 
not enough night time exposure  to exclude the
day observations.

\begin{figure}
\centering
\leavevmode
\includegraphics[width=5in,angle=90]{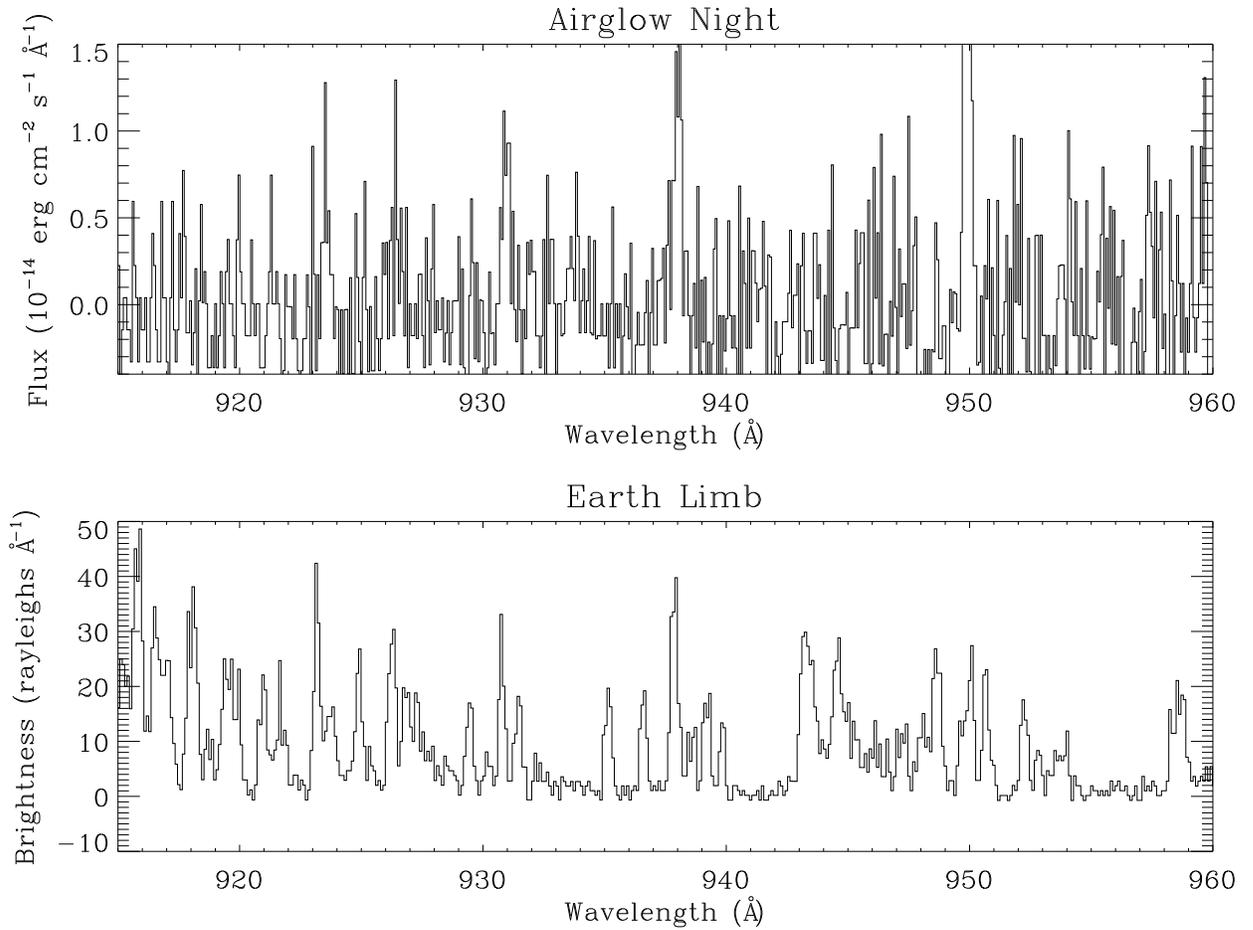}
\caption[]{
Plots of LWRS SiC 1B night airglow (top) and  earth limb (bottom) spectra. 
Data originally from \citet{feld01}.  Though the night airglow spectrum is very noisy
the two spectra show the wealth of geo-coronal
emission features that impact data analysis in these wavelength regions.
\label{f:airglow}}
\end{figure}

We have plotted the spectra in the observed frame
so that airglow features may be more easily identified.  The
intrinsic Lyman continuum region for each galaxy is also 
clearly marked.  Several of the  spectra are very noisy and 
systematic flux offsets caused by background over-subtraction
in the CALFUSE pipeline are observed.  As would
be expected, the background
subtraction problems are much more prevalent in the 
lower S/N spectra , specifically \sbs, \tol, \iron, and NGC\,7673.
A large number of apparent emission features can be observed,
particularly in the wavelength region between 920-930 \AA.
These are easily observed in the day/night spectrum of
\vv\, but also in the night only spectra of \tol.  
These features, while varying widely in
relative strength, are found at the same wavelengths in
most of the spectra.  They are properly identified 
as geocoronal emission features, mostly
 \ion{H}{1}, \ion{O}{1}, \ion{N}{1}, and $\rm{N_2}$.
For comparison, LWRS spectra (SiC 1B) of airglow
and the earth limb are found in Figure \ref{f:airglow}.
The large number of variable  emission
features highlights the difficulty in studying
Lyman continuum emission in this wavelength regime.

No convincing evidence of Lyman continuum emission is seen in any of the 
nine spectra.  All of the galaxies 
have flat spectra in the Lyman continuum regions with continuum
levels at or below zero.
The spectra with continuum
levels below zero are indicative of  background over-subtraction by
the CALFUSE pipeline.  This is commonly observed in low S/N \fu\,
spectra on the SiC 1B and SiC 2A detectors as discussed previously,
and highlights the difficulty in studying Lyman
continuum emission.  The \fu\, background dominates the signal
in these wavelength regions.
The emission features that are observed
can all be identified as geo-coronal emission, 
with varying strengths in all of the spectra.  
In addition to these extracted spectra,
we have also examined the
2-D detector images.  A visual inspection of the 2-D spectra
confirm our analysis of the one-dimensional spectra as we do not observe
any evidence of Lyman continuum emission.

We have chosen to derive numerical limits for the
five galaxies with S/N$>$5 on the LiF 1A channel.  
We have excluded the lower
S/N galaxies such as \tol\,and \iron\, due to the significant errors
introduced by the background subtraction.  
\iron\, for example,
has a S/N of only 1.3 longward of the Lyman limit on the SiC 2A channel.  
While these errors exist even in the five galaxies with
higher S/N, we can still place meaningful limits on the
Lyman escape fraction.

We follow the work of \citet{shap06} in estimating both
the absolute and relative escape fractions. The absolute
escape fraction describes
the fraction of ionizing photons that escape the host galaxy.  The relative escape fraction
neglects the effects of dust extinction and is defined as the percentage of escaping ionizing 
photons relative to escaping non-ionizing FUV photons \citep[1500 \AA, ][]{shap06}.
The relative escape fraction was introduced due to the difficulty in determining the
intrinsic ionizing flux for high-z galaxies.

For these five galaxies we have measured an upper limit
to the flux below the Lyman limit in each galaxy.  We have then
converted these to luminosities with 
$\rm L_{esc}\sim\lambda F_\lambda$ ($\lambda$=900 \AA).
From Starburst99 v5.1 \citep{vaz05} we find an intrinsic ratio of
the bolometric to ionizing radiation of   
$\rm L_{bol}/L_{ioniz}\sim7$.  If we assume that the
IR + UV luminosities dominate the bolometric luminosity
as expected in these star-forming galaxies we derive:

\begin{equation}
\rm f_{esc,abs}=\frac{L_{esc}}{L_{UV}+L_{IR}}\left(\frac{L_{bol}}{L_{ioniz}}\right)_{intrinsic}.
\end{equation}

We slightly modify $\rm f_{esc,rel}$ of \citet{shap06} to use
1150 \AA\, instead of 1500 \AA\, as our reference wavelength.
These values should be roughly comparable as we would 
expect the intrinsic luminosities of starburst galaxies to be 
comparable at 1150 \AA\, and 1500 \AA\, \citep{leith02}.
We estimate $\rm f_{esc,rel}$ from 

\begin{equation}
\rm f_{esc,rel}=\left(\frac{L_{900\, \AA}}{L_{1150\, \AA}}\right)_{obs}
\left(\frac{L_{1150\, \AA}}{L_{900\, \AA}}\right)_{intrinsic}.
\end{equation}

We obtain an intrinsic ratio $\rm L_{1150\, \AA}/L_{900\, \AA}\sim4$ from Starburst99. 
The derived values for $\rm f_{esc,rel}$ and $\rm f_{esc,abs}$
are found in Table \ref{t:lylimit}.  The observed fluxes and luminosities have been
corrected for MW extinction using \citet{card89}.
We find $\rm f_{esc,abs}< 0.2-7\%$ and $\rm f_{esc,rel}<9-26\%$
for the galaxies.  The weakest limit, 
$\rm f_{esc,rel}<26\%$, is from \vv\, which is known to contain
strong intrinsic dust attenuation \citep{gold02}.
Our upper limits for $\rm f_{esc,rel}$ are comparable
to the average value of 14 LBGs with $\rm <f_{esc,rel}>\sim14\%$
from \citet{shap06}.  
This average value in the high redshift sample however is dominated by two galaxies
with $\rm f_{esc,rel}>40\%$.  While our low redshift limits are compatible with
the sample averaged values of $\rm f_{esc,rel}$ in LBGs,
our limits show that the extreme Lyman continuum leakage
observed in some LBGs is not observed in our small, local sample.







\subsubsection{Limits from \cii\label{s:ciilimit}}

Given the difficulties in directly observing Lyman continuum emission
on the SiC detectors it is useful to investigate
the Lyman escape fraction using a different method.
Previously, \citet{heck01b} have argued that since the optical depth
at the Lyman edge must be much larger than that of
the optical depth at line center for \cii,
we can then use the 
residual intensity in the core of the  
\cii\, line to place
upper limits on the escape fraction at the Lyman limit.

Assuming that \cii\, is the dominant ionic species of carbon
in the \ion{H}{1} phase, \cite{heck01b} derive
\begin{equation}
\tau_{\rm Ly} = 4\times10^{-16}\, (N_H/N_j)\,(b/f\lambda)\, \tau_{\rm C}=8.6\, Z_{\rm C}^{-1}\, (b/100\,\ks)\, \tau_{\rm C}
\label{e:cii}
\end{equation}
where $\rm Z_{C}$ is the gas-phase carbon abundance 
in solar units and $b$ is the Doppler parameter.
In our earlier gaussian fits to \cii, we find that $b$ varies from $\sim65-470\,\ks$.
Using the smallest Doppler parameter and solar abundances we expect the optical
depth at the Lyman edge would be a factor of $\sim5$ times larger than the
\cii\, optical depth.  As the majority of the galaxies in our sample have 
broader lines and sub-solar
abundances, we would expect the Lyman limit optical depth
to be enhanced by an even greater value.

An examination of the \cii\, feature for the galaxies in our sample 
(e.g. Figures \ref{f:Haro11lines} - \ref{f:NGC7714lines}) shows that it
is nearly black in almost all cases.  Given our estimates of
the optical depth enhancement expected at the 
Lyman limit, this suggests that none of the galaxies in our sample have
significant amounts of escaping Lyman continuum emission.

The most likely scenario for the escape of Lyman continuum emission
is the picket fence model.  Instead of a uniform slab of hydrogen surrounding
a galaxy, holes in the gas surrounding the galaxy would allow ionizing photons
to escape into the IGM.  Starburst driven outflows could, for example,
create ``tunnels'' in the surrounding neutral gas. 
The picket fence model would explain the variability
seen in the center of some of the \cii\, features.  For example,
in Figure \ref{f:VV114lines} of \vv's \cii\, profile a small fraction 
of photons are escaping at 0 \kms\, while the line is completely black
at both $\sim\pm100\,\ks$.  The few galaxies with
 a significant
\cii\, escape fraction stand out due to their low metallicity, 
suggesting that their Lyman continuum
escape fraction would still be very low.

We have conservatively estimated the continuum level
and calculated a residual relative intensity
for the \cii\,line center.  We calculate an upper limit to $\rm f_{esc,C II}$ by
finding the upper bound to the normalized residual flux
near line center. 
Only our highest redshift galaxy Mrk 54 has been excluded,
as the \cii\, feature falls in the gap between the LiF channels.
We also estimate upper limits for Lyman continuum emission using 
\flyabs$=\rm f_{esc,C II}\,L_{UV}/(L_{UV}+L_{IR})$ following
the methods of the previous section.  For all but one of the galaxies
in our sample we find \flyabs$\lesssim2\%$.  

Using Equation \ref{e:cii} we have also estimated the optical
depth of the Lyman limit using our estimate 
of the \cii\,escape fraction and assuming the slab model
of hydrogen surrounding a galaxy.  The Doppler parameters are found from our
Gaussian fits in Tables \ref{t:HARO11dat} - \ref{t:VV114data}.  As
many of these lines are non-gaussian these must be considered 
only rough estimates.  We have used the gas phase oxygen
metallicities in Table \ref{t:fuselumins} 
for the carbon metallicity and solar oxygen abundances
from \citet{asp05}.  The results can be found in 
Table \ref{t:ciilimit}.  
Our lowest derived optical depth is $\rm \tau_{Ly}\sim20$
found for NGC\,3310 but the majority of galaxies
have $\rm \tau_{Ly}\gtrsim100$.
While these optical depths are only 
rough estimates,  they suggest that the
Lyman limit optical depth is well above the threshold needed to allow
Lyman photons to escape the galaxy in the case of the simple slab model.

\section{Conclusions}

Starburst driven outflows are a cosmologically
important example of galactic feedback.
These outflows drive gas, metals, and energy out
of galaxy centers and possibly into the 
IGM.  Outflows in the early universe 
may also blow out ``tunnels''
allowing Lyman continuum photons from stars
to escape the galaxy and reionize the universe.
Studies of these superwinds
are therefore vital to expanding our understanding of galaxy 
formation and evolution.

The extreme scale and energetics of starburst driven
outflows complicate studies of their properties.  
Originating as a hot (T $>10^8\,$K) wind fluid,
the outflows expand out into ambient gas of the galaxy and possibly 
beyond.  Shocks, other hydrodynamical interactions, and
entrained ambient gas and dust 
produce a complicated, multiphase outflow
that requires  panchromatic observations to properly characterize it.

Observations in the FUV play an important role in understanding
outflows as they probe the neutral, ionized, and coronal gas.
This gas traces the interaction between the hot outflowing wind fluid
and the denser ambient ISM.
Unfortunately, due to observational constraints at both
low redshifts (due to UV atmospheric absorption) and high redshifts (due to contamination by the Ly$\alpha$ forest), 
studies at these wavelengths were severely limited until the launch of
the \fu\,satellite in 1999.  As the number of star-forming galaxies observed by \fu\,
has grown, it now possible to study the global properties of these
outflows in the FUV.

In this work we have analyzed FUV  spectra of a sample of 16 local, star-forming galaxies
covering almost three orders of magnitude in SFR and over two orders of magnitude in stellar mass.
Interstellar medium absorption and stellar photospheric 
features are observed throughout the
sample of spectra.  The strongest ISM absorption lines are generally blueshifted
by $\sim$ 80 to 300 $\ks$ relative to thge galaxy systemic velocity, implying that galactic outflows are the norm
in starbursts.  

Coronal gas is an important coolant for gas at temperatures
T$\sim10^{5.5}\,$K and is only directly observable
in the FUV through observations of the \oviww\, doublet. 
\ovi\, emission is only detected in 1 of the 16 galaxies in our sample.
This suggests that radiative cooling of the coronal gas is not 
dynamically significant in the wind and does not impede its outflow.
In contrast, \ovi\, absorption is nearly ubiquitous in our sample
and is found to trace systematically higher outflow speeds than the
photoionized and neutral gas.  
We find a strong relationship between \ovi\ column density and the width
of the line that is consistent with the 
production of \ovi\, in a radiatively cooling gas flow passing through the coronal
temperature regime.  This is expected to occur
at the interface between the outrushing hot wind seen in X-rays and the 
cool dense material in its path. 

Previous studies \citep{vaz04,schwa04,grim06} have suggested that the observed
outflow velocities in a galaxy increase with the ionization potential of the
species observed. While this is clearly the case for the coronal phase gas,
it is not the case for the neutral and photoionized material. Our new results show that the
derived outflow velocities in thius gas in a given galaxy increase with line strength (equivalent width)
and not ionization state.
This could be explained if the
observations of the stronger lines probe lower column density
gas than the weaker absorption features.  The lower column density
gas extends to greater outflow velocities and is not readily observable
in weaker absorption features.  We conclude that the extensive studies
of outflows in the local star forming galaxies using the
\naiww\ doublet in the optical
should be able to reliably measure the outflow velocities
that would be seen in the strong FUV absorption-lines. This is because the
\naiww\  line should be sufficiently strong to probe the low column density
outflowing gas.

In agreement with previous UV \citep{schwa06} and optical \citep{mart05b,rup05} studies 
we find that the outflow velocities of the strongest absorption lines
increase with increasing SFR and increasing specific SFR (SFR/M$_*$). We speculate that 
starbursts with higher SFR generate more
wind momentum which can accelerate the absorbing material (clouds) entrained in the wind to 
higher outflow velocities. Our sample does not probe the regime of very high SFR (of-order $10^2$
M$_{\odot}$ per year) where
the relationship appears to flatten \citep{mart05b}.

We find that the equivalent width of the optically thin lines arising
in the HI phase correlates most strongly with the gas-phase metallicity (as expected
since the equivalent width should be proportional to the column density of the metal species).
In contrast, the equivalent widths of the optically-thick lines in the HI and HII phase
correlate better with the galaxy mass and the SFR. This
is consistent with the fact that the equivalent width for these optically-thick lines
is determined by the velocity dispersion in the ISM. More massive galaxies and
galaxies undergoing elevated SFR will have higher velocity dispersions due to
both gravity and the kinetic energy input from massive stars.

Lastly, we have looked for escaping Lyman continuum emission
in our sample.  We find no convincing evidence of escaping
ionizing radiation in any of the observed galaxies.
In a subsample of five of our galaxies with suitable redshifts and 
data quality we find that the upper limit to the absolute fraction of escaping Lyman continuum photons
ranges from \flyabs$< 0.2 - 6.3 \%$.  Following 
\citet{shap06} we also derive upper limits to the realtive escape fractioni (the ratio of the fraction
of escaping Lyman continuum and non-ionizing FUV continuum). These range from \flyrel$< 9 - 26\%$ for this subsample.
While our upper limits are compatible with 
observations of high redshift Lyman Break Galaxies, we do not observe any
evidence of Lyman continuum emission as observed in some
Lyman Break Galaxies.

Using the escape fraction of light at the center of
the \cii\, feature we also place limits on the
escape fraction of Lyman continuum emission in 15 of the
galaxies included in our sample.
In a simple model where a galaxy is surrounded
by a slab of neutral hydrogen we find significant
optical depths at the Lyman limit of 
$\tau_{\rm Ly}>20$.  
In a more realistic model with holes
in the neutral gas surrounding a galaxy, we
find \flyabs$<2\%$ for almost all of the galaxies.
This suggests that local starburst galaxies are not allowing the escape of 
significant amounts of hydrogen ionizing photons, even in galaxies
with strong winds.

This work provides a unique view of the 
properties of outflows in star-forming galaxies.
The wealth of absorption features over a variety
of gas phases enables direct comparisons difficult
in other wavelength regimes.  The projected physical size of \fu's 
large spectroscopic aperture for these local starbursts is particularly well matched to high redshift
spectrographic observations of Lyman Break Galaxies and other star forming galaxies  - both
probe the outflows on the galactic scale.  Thus, this study
should provide an excellent complement
to the high redshift observations.

\acknowledgements


\clearpage
\begin{deluxetable}{lrrlllllcc}
\rotate
\tabletypesize{\scriptsize}
\tablecolumns{10}
\tablewidth{0pc}
\tablecaption{\fu~Galaxy Properties\label{t:fuselumins}}
\tablehead{
\colhead{Galaxy} &  \colhead{$\rm{v_{sys}}$} & \colhead{Distance} & \colhead{MW E(B-V)}  &
	\colhead{$\rm{L_{FUV}}$\tablenotemark{a}} & \colhead{$\rm{L_{K}}$}  &
	 \colhead{$\rm{L_{IR}}$} &  \colhead{$\rm{L_{UV}+L_{IR}}$} &
	\colhead{$\rm{F_{60\mu m}/F_{100\mu m}}$} & \colhead{Z} \\
	& \colhead{(\kms)} & \colhead{(Mpc)} & \colhead{(mag)} & \colhead{$\rm{(L_\odot)}$} & 
	\colhead{$\rm{(L_\odot)}$} &
	\colhead{$\rm{(L_\odot)}$} & \colhead{$\rm{(L_\odot)}$} &
	& \colhead{(log[O/H]+12)} }
\startdata
HARO11  & 6175 &  88 & 0.011 & $1.6\times 10^{10}$ & $4.5\times 10^{9}$ & $1.9\times 10^{11}$ & $2.1\times 10^{11}$ & 1.37 & 7.9 \\ 
VV114  & 6016 &  86 & 0.016 & $2.4\times 10^{10}$ & $2.0\times 10^{10}$ & $5.2\times 10^{11}$ & $5.4\times 10^{11}$ & 0.73 & 8.6 \\ 
NGC1140  & 1501 &  21 & 0.038 & $2.4\times 10^{9}$ & $7.7\times 10^{8}$ & $4.4\times 10^{9}$ & $6.8\times 10^{9}$ & 0.68 & 8.0 \\ 
SBS0335-052  & 4043 &  57 & 0.047 & $2.6\times 10^{9}$ &  \nodata & \nodata & \nodata & \nodata & 7.3 \\ 
TOL0440-381  & 12291 & 178 & 0.015 & $1.7\times 10^{10}$ & $2.8\times 10^{9}$ & $2.4\times 10^{10}$ & $4.1\times 10^{10}$ & \nodata  & 8.2 \\ 
NGC1705  & 632 &   8 & 0.008 & $1.0\times 10^{9}$ & $1.3\times 10^{8}$ & $2.6\times 10^{8}$ & $1.3\times 10^{9}$ & 0.38 & 8.0 \\ 
NGC1741  & 4037 &  57 & 0.051 & $1.1\times 10^{10}$ & $1.6\times 10^{9}$ & $3.8\times 10^{10}$ & $4.9\times 10^{10}$ & 0.71 & 8.0 \\ 
IZW18  & 751 &  10 & 0.032 & $1.3\times 10^{8}$ &  \nodata & \nodata & \nodata & \nodata & 7.2 \\
\enddata
\tablecomments{$\rm{v_{sys}}$, MW  E(B-V), IRAS, and K-Band fluxes were obtained from NED where available. 
IRAS \citep{soif89,mosh93}  and 2Mass \citep{jarr03} luminosities are 
transformed using methods of \citet{sand96} and \citet{carp01} respectively.
The references for the gas-phase nebular metallicity measurements are as follows, 
\hoo\, \citep{berg02}, \vv\,\citep{kim95,char01}, 
NGC\,1140, NGC\,1705, NGC\,3310, NGC\,3690, NGC\,4214, NGC\,7673, NGC\,7714 \citep{heck98},
\sbs\,\citep{izo97}, \tol\,\citep{terl91,pell07}, NGC\,1741 \citep{cont96}, 
Haro\,3 \citep{izo04}, \izw\,\citep{izo99},  \iron\,\citep{ost07}, and Mrk\,54 \citep[spectrum from 
A. Boselli with conversions from][]{char01}.
}
\tablenotetext{a}{FUV luminosity derived from \fu\, LiF 2A continuum flux at 1150\,\ang.  Values are corrected for MW extinction using
the method of \citet{card89} and the E(B-V) values given in this table.}
\end{deluxetable}

\clearpage

\begin{deluxetable}{lrrcclllrrr}
\tabletypesize{\scriptsize}
\rotate
\tablecolumns{11}
\tablewidth{0pc}
\tablecaption{\fu~Observation Data\label{t:fusesample}}
\tablehead{\colhead{Galaxy} & \multicolumn{2}{c}{\fu\, Pointing} & 
	\colhead{Aperture} & \colhead{Region\tablenotemark{a}} 
	& \colhead{DAY/NIGHT} &
	\colhead{Obs. Date} & \colhead{PI} &
	\colhead{Obs. Time\tablenotemark{b}} & \colhead{Exposures} & \colhead{S/N\tablenotemark{c}}\\ 
	& \multicolumn{2}{c}{(J2000)} & & & & \colhead{(UT)} & & \colhead{(ks)} & }
\startdata
HARO 11  & $00^{\rm h}36^{\rm m}52\fs5$ & $-33\degr33\arcmin19\arcsec$ & LWRS & PC & NIGHT & 	
	2001-10-12 & Bergvall &	11.9 & 6 & 7.2\\ 
VV 114  & $01^{\rm h}07^{\rm m}46\fs6$ & $-17\degr30\arcmin24\arcsec$ & LWRS & PC & BOTH 
	& 2003-07-26 & Heckman & 11.2 & 4 & 6.6\\ 
NGC 1140  & $02^{\rm h}54^{\rm m}33\fs6$ & $-10\degr01\arcmin39\arcsec$ & LWRS & PC & BOTH &
	2001-11-28 & Buat & 3.9 & 2 & 4.8\\ 
SBS 0335-052  & $03^{\rm h}37^{\rm m}43\fs1$ & $-05\degr02\arcmin39\arcsec$ & LWRS & PC & BOTH &
	2001-09-26 & Thuan & 24.4 & 8 & 4.1\\ 
TOL 0440-381  & $04^{\rm h}42^{\rm m}08\fs0$ & $-38\degr01\arcmin10\arcsec$ & LWRS & PC & BOTH &
	2000-12-13 & Deharveng & 31.2 & 11 & 4.0\\ 
NGC 1705  & $04^{\rm h}54^{\rm m}13\fs5$ & $-53\degr21\arcmin39\arcsec$ & LWRS & EC & BOTH &
	2000-02-04 & Heckman & 24.0 & 13 & 27.2\\ 
NGC 1741  & $05^{\rm h}01^{\rm m}37\fs7$ & $-04\degr15\arcmin30\arcsec$ & LWRS & PC & NIGHT &
	2003-12-23 & Heckman & 18.5 & 10 & 8.4\\ 
IZW 18  & $09^{\rm h}34^{\rm m}02\fs3$ & $+55\degr14\arcmin25\arcsec$ & LWRS & PC & BOTH &
	2001-02-11 & Moos & 58.2 & 37 & 9.7\\
\enddata
\tablenotetext{a}{Detector region used for spectra extraction.  Although all targets are extended (EC) the majority have
flux falling entirely within the smaller point source (PC) extraction region.  Using the smaller point source region
minimizes background subtraction errors which are very important in the lower S/N data.}
\tablenotetext{b}{Exposure time in LiF 1A channel.}
\tablenotetext{c}{S/N in 0.078\,\ang\,bins in LiF 1A channel.}
\end{deluxetable}

			\begin{deluxetable}{lcccccc}
			\tabletypesize{\scriptsize}
			\tablecolumns{7}
			\tablewidth{0pc}
			\tablecaption{       HARO11         ISM Absorption Line Fit Data
			\label{t:HARO11dat}}
			\tablehead{
			\colhead{Ion} & \colhead{$\lambda_{0}$} & \colhead{$log(\lambda f N/N_H)$} &
			                 \colhead{Instrument} & \colhead{$\rm{W}_\lambda$} &
			                 \colhead{$\rm{v_{c}}$} & \colhead{FWHM} \\
			                  & \colhead{\AA} & & & \colhead{\AA} & \colhead{\kms} &
			                  \colhead{\kms}}
			\startdata
\ion{C}{3} & 977.02 & -0.61 & 2asic & 1.83$\pm$0.57 & 5998$\pm$6 & 269$\pm$10\\ 
\ion{O}{6} & 1031.93 & -1.13 & 1alif & 0.38$\pm$0.06 & 6039$\pm$16 & 193$\pm$34\\ 
 & & & 2blif & 0.39$\pm$0.07 & 6050$\pm$14 & 159$\pm$34\\ 
\ion{C}{2} & 1036.34 & -1.37 & 1alif & 1.29$\pm$0.05 & 6063$\pm$6 & 381$\pm$18\\ 
 & & & 2blif & 1.24$\pm$0.13 & 6074$\pm$8 & 344$\pm$36\\ 
\ion{O}{1} & 988.73 & -1.56 & 1alif & 0.72$\pm$0.05 & 6092$\pm$14 & 276$\pm$14\\ 
 & & & 2blif & 0.66$\pm$0.09 & 6079$\pm$23 & 304$\pm$26\\ 
\ion{N}{2} & 1083.99 & -2.00 & 1blif & 0.92$\pm$0.07 & 6081$\pm$10 & 353$\pm$28\\
\enddata
	\tablenotetext{a}{Using $\rm{v_{sys}}=$      6175       \kms}
	\tablenotetext{b}{Errors are undetermined for this value.}
	\end{deluxetable}

\clearpage
			\begin{deluxetable}{lcccccc}
			\tabletypesize{\scriptsize}
			\tablecolumns{7}
			\tablewidth{0pc}
			\tablecaption{        HARO3         ISM Absorption Line Fit Data
			\label{t:HARO3data}}
			\tablehead{
			\colhead{Ion} & \colhead{$\lambda_{0}$} & \colhead{$log(\lambda f N/N_H)$} &
			                 \colhead{Instrument} & \colhead{$\rm{W}_\lambda$} &
			                 \colhead{$\rm{v_{c}}$} & \colhead{FWHM} \\
			                  & \colhead{\AA} & & & \colhead{\AA} & \colhead{\kms} &
			                  \colhead{\kms}}
			\startdata
\ion{C}{3} & 977.02 & -0.61 & 2asic & 1.16\tablenotemark{b} & 913$\pm$26 & 202$\pm$16\\ 
\ion{O}{6} & 1031.93 & -1.13 & 1alif & 0.71$\pm$0.08 & 787$\pm$26 & 324\tablenotemark{b}\\ 
 & & & 2blif & 0.85$\pm$0.12 & 796$\pm$21 & 324$\pm$53\\ 
\ion{C}{2} & 1036.34 & -1.37 & 1alif & 0.97$\pm$0.07 & 868$\pm$3 & 229$\pm$26\\ 
 & & & 2blif & 1.11$\pm$0.13 & 876$\pm$10 & 259$\pm$54\\ 
\ion{O}{1} & 988.73 & -1.56 & 1alif & 0.97\tablenotemark{b} & 912$\pm$11 & 150\tablenotemark{b}\\ 
 & & & 2blif & 0.46$\pm$0.06 & 906$\pm$4 & 88$\pm$14\\ 
\ion{N}{3} & 989.80 & -2.03 & 1alif & 0.87\tablenotemark{b} & 914$\pm$6 & 127$\pm$2\\
\enddata
	\tablenotetext{a}{Using $\rm{v_{sys}}=$      963        \kms}
	\tablenotetext{b}{Errors are undetermined for this value.}
	\end{deluxetable}

\clearpage
			\begin{deluxetable}{lcccccc}
			\tabletypesize{\scriptsize}
			\tablecolumns{7}
			\tablewidth{0pc}
			\tablecaption{   IRAS19245-4140     ISM Absorption Line Fit Data
			\label{t:IRAS19245}}
			\tablehead{
			\colhead{Ion} & \colhead{$\lambda_{0}$} & \colhead{$log(\lambda f N/N_H)$} &
			                 \colhead{Instrument} & \colhead{$\rm{W}_\lambda$} &
			                 \colhead{$\rm{v_{c}}$} & \colhead{FWHM} \\
			                  & \colhead{\AA} & & & \colhead{\AA} & \colhead{\kms} &
			                  \colhead{\kms}}
			\startdata
\ion{C}{3} & 977.02 & -0.61 & 2asic & 2.50$\pm$0.74 & 2748$\pm$29 & 663$\pm$235\\ 
\ion{C}{2} & 1036.34 & -1.37 & 1alif & 0.86$\pm$0.09 & 2746$\pm$11 & 272$\pm$30\\ 
 & & & 2blif & 0.75$\pm$0.25 & 2761$\pm$31 & 256$\pm$96\\ 
\ion{O}{1} & 988.73 & -1.56 & 1alif & 0.31$\pm$0.03 & 2774$\pm$8 & 81$\pm$13\\ 
\ion{N}{2} & 1083.99 & -2.00 & 2alif & 0.17\tablenotemark{b} & 2792$\pm$14 & 72\tablenotemark{b}\\ 
\ion{N}{3} & 989.80 & -2.03 & 1alif & 0.32\tablenotemark{b} & 2708$\pm$7 & 89\tablenotemark{b}\\ 
\ion{N}{1} & 1134.41 & -2.57 & 1blif & 0.38\tablenotemark{b} & 2782$\pm$1 & 50$\pm$11\\ 
 & & & 2alif & 0.30\tablenotemark{b} & 2794$\pm$0 & 37\tablenotemark{b}\\
\enddata
	\tablenotetext{a}{Using $\rm{v_{sys}}=$      2832       \kms}
	\tablenotetext{b}{Errors are undetermined for this value.}
	\end{deluxetable}

\clearpage
			\begin{deluxetable}{lcccccc}
			\tabletypesize{\scriptsize}
			\tablecolumns{7}
			\tablewidth{0pc}
			\tablecaption{        IZW18         ISM Absorption Line Fit Data
			\label{t:IZW18data}}
			\tablehead{
			\colhead{Ion} & \colhead{$\lambda_{0}$} & \colhead{$log(\lambda f N/N_H)$} &
			                 \colhead{Instrument} & \colhead{$\rm{W}_\lambda$} &
			                 \colhead{$\rm{v_{c}}$} & \colhead{FWHM} \\
			                  & \colhead{\AA} & & & \colhead{\AA} & \colhead{\kms} &
			                  \colhead{\kms}}
			\startdata
\ion{C}{2} & 1036.34 & -1.37 & 1alif & 0.63$\pm$0.05 & 748$\pm$7 & 180$\pm$16\\ 
 & & & 2blif & 0.65$\pm$0.06 & 757$\pm$8 & 190$\pm$18\\ 
\ion{O}{1} & 988.73 & -1.56 & 1alif & 0.64$\pm$0.23 & 784$\pm$4 & 106$\pm$12\\ 
 & & & 2asic & 0.54\tablenotemark{b} & 816$\pm$5 & 84$\pm$15\\ 
\ion{N}{3} & 989.80 & -2.03 & 1alif & 0.51$\pm$0.10 & 711$\pm$9 & 122$\pm$28\\ 
 & & & 2asic & 0.69$\pm$0.15 & 754$\pm$11 & 180$\pm$68\\ 
\ion{O}{1} & 1039.23 & -2.29 & 1alif & 0.18$\pm$0.02 & 759$\pm$5 & 76$\pm$9\\ 
 & & & 2blif & 0.23$\pm$0.03 & 770$\pm$5 & 79$\pm$13\\
\enddata
	\tablenotetext{a}{Using $\rm{v_{sys}}=$      751        \kms}
	\tablenotetext{b}{Errors are undetermined for this value.}
	\end{deluxetable}

\clearpage
			\begin{deluxetable}{lcccccc}
			\tabletypesize{\scriptsize}
			\tablecolumns{7}
			\tablewidth{0pc}
			\tablecaption{        MRK54         ISM Absorption Line Fit Data
			\label{t:MRK54data}}
			\tablehead{
			\colhead{Ion} & \colhead{$\lambda_{0}$} & \colhead{$log(\lambda f N/N_H)$} &
			                 \colhead{Instrument} & \colhead{$\rm{W}_\lambda$} &
			                 \colhead{$\rm{v_{c}}$} & \colhead{FWHM} \\
			                  & \colhead{\AA} & & & \colhead{\AA} & \colhead{\kms} &
			                  \colhead{\kms}}
			\startdata
\ion{C}{3} & 977.02 & -0.61 & 1alif & 1.76$\pm$0.12 & 13382$\pm$12 & 434$\pm$46\\ 
 & & & 2blif & 1.48$\pm$0.09 & 13402$\pm$6 & 368$\pm$28\\ 
\ion{O}{6} & 1031.93 & -1.13 & 1alif & 0.77$\pm$0.09 & 13321$\pm$14 & 287$\pm$38\\ 
\ion{O}{1} & 988.73 & -1.56 & 1alif & 1.18$\pm$0.34 & 13541$\pm$46 & 386$\pm$112\\ 
 & & & 2blif & 0.82$\pm$0.19 & 13526$\pm$30 & 337$\pm$86\\ 
\ion{N}{2} & 1083.99 & -2.00 & 1blif & 1.41$\pm$0.12 & 13466$\pm$9 & 409$\pm$39\\ 
 & & & 2alif & 1.60$\pm$0.13 & 13469$\pm$10 & 497$\pm$44\\ 
\ion{N}{3} & 989.80 & -2.03 & 1alif & 0.67$\pm$0.11 & 13481$\pm$13 & 177$\pm$45\\
\enddata
	\tablenotetext{a}{Using $\rm{v_{sys}}=$     13450       \kms}
	\tablenotetext{b}{Errors are undetermined for this value.}
	\end{deluxetable}

\clearpage
			\begin{deluxetable}{lcccccc}
			\tabletypesize{\scriptsize}
			\tablecolumns{7}
			\tablewidth{0pc}
			\tablecaption{       NGC1140        ISM Absorption Line Fit Data
			\label{t:NGC1140da}}
			\tablehead{
			\colhead{Ion} & \colhead{$\lambda_{0}$} & \colhead{$log(\lambda f N/N_H)$} &
			                 \colhead{Instrument} & \colhead{$\rm{W}_\lambda$} &
			                 \colhead{$\rm{v_{c}}$} & \colhead{FWHM} \\
			                  & \colhead{\AA} & & & \colhead{\AA} & \colhead{\kms} &
			                  \colhead{\kms}}
			\startdata
\ion{C}{3} & 977.02 & -0.61 & 2asic & 1.32$\pm$0.18 & 1438$\pm$11 & 268$\pm$46\\ 
\ion{C}{2} & 1036.34 & -1.37 & 1alif & 0.93$\pm$0.08 & 1481$\pm$6 & 250$\pm$21\\ 
 & & & 2blif & 1.33$\pm$0.27 & 1487$\pm$27 & 358$\pm$76\\ 
\ion{O}{1} & 988.73 & -1.56 & 2asic & 0.51$\pm$0.12 & 1542$\pm$5 & 100\tablenotemark{b}\\ 
 & & & 2blif & 0.75\tablenotemark{b} & 1548$\pm$6 & 100\tablenotemark{b}\\ 
\ion{N}{2} & 1083.99 & -2.00 & 2alif & 0.63$\pm$0.10 & 1517$\pm$7 & 177$\pm$28\\ 
\ion{N}{3} & 989.80 & -2.03 & 2asic & 0.99\tablenotemark{b} & 1502$\pm$5 & 138$\pm$39\\ 
 & & & 2blif & 0.69$\pm$0.10 & 1517$\pm$7 & 130\tablenotemark{b}\\
\enddata
	\tablenotetext{a}{Using $\rm{v_{sys}}=$      1501       \kms}
	\tablenotetext{b}{Errors are undetermined for this value.}
	\end{deluxetable}

\clearpage
			\begin{deluxetable}{lcccccc}
			\tabletypesize{\scriptsize}
			\tablecolumns{7}
			\tablewidth{0pc}
			\tablecaption{       NGC1705        ISM Absorption Line Fit Data
			\label{t:NGC1705da}}
			\tablehead{
			\colhead{Ion} & \colhead{$\lambda_{0}$} & \colhead{$log(\lambda f N/N_H)$} &
			                 \colhead{Instrument} & \colhead{$\rm{W}_\lambda$} &
			                 \colhead{$\rm{v_{c}}$} & \colhead{FWHM} \\
			                  & \colhead{\AA} & & & \colhead{\AA} & \colhead{\kms} &
			                  \colhead{\kms}}
			\startdata
\ion{C}{3} & 977.02 & -0.61 & 1bsic & 1.21$\pm$0.29 & 613$\pm$20 & 310\tablenotemark{b}\\ 
 & & & 2asic & 1.23\tablenotemark{b} & 651$\pm$28 & 325\tablenotemark{b}\\ 
\ion{O}{6} & 1031.93 & -1.13 & 1alif & 0.18$\pm$0.01 & 554$\pm$3 & 111$\pm$9\\ 
 & & & 2blif & 0.20$\pm$0.01 & 559$\pm$3 & 125$\pm$9\\ 
\ion{C}{2} & 1036.34 & -1.37 & 1alif & 1.01\tablenotemark{b} & 600$\pm$37 & 262\tablenotemark{b}\\ 
 & & & 2blif & 1.02$\pm$0.26 & 597$\pm$24 & 270$\pm$84\\ 
\ion{O}{1} & 988.73 & -1.56 & 1alif & 0.58$\pm$0.04 & 600$\pm$3 & 162$\pm$21\\ 
 & & & 2blif & 0.53$\pm$0.04 & 596$\pm$4 & 153$\pm$14\\
\enddata
	\tablenotetext{a}{Using $\rm{v_{sys}}=$      632        \kms}
	\tablenotetext{b}{Errors are undetermined for this value.}
	\end{deluxetable}

\clearpage
			\begin{deluxetable}{lcccccc}
			\tabletypesize{\scriptsize}
			\tablecolumns{7}
			\tablewidth{0pc}
			\tablecaption{       NGC1741        ISM Absorption Line Fit Data
			\label{t:NGC1741da}}
			\tablehead{
			\colhead{Ion} & \colhead{$\lambda_{0}$} & \colhead{$log(\lambda f N/N_H)$} &
			                 \colhead{Instrument} & \colhead{$\rm{W}_\lambda$} &
			                 \colhead{$\rm{v_{c}}$} & \colhead{FWHM} \\
			                  & \colhead{\AA} & & & \colhead{\AA} & \colhead{\kms} &
			                  \colhead{\kms}}
			\startdata
\ion{C}{3} & 977.02 & -0.61 & 2asic & 1.08\tablenotemark{b} & 4007$\pm$20 & 164$\pm$15\\ 
\ion{O}{6} & 1031.93 & -1.13 & 1alif & 0.58$\pm$0.08 & 3833$\pm$32 & 400\tablenotemark{b}\\ 
 & & & 2blif & 0.58$\pm$0.20 & 3921$\pm$92 & 400\tablenotemark{b}\\ 
\ion{C}{2} & 1036.34 & -1.37 & 1alif & 1.38$\pm$0.07 & 3985$\pm$4 & 300$\pm$28\\ 
 & & & 2blif & 1.39$\pm$0.19 & 3986$\pm$4 & 270$\pm$29\\ 
\ion{O}{1} & 988.73 & -1.56 & 1alif & 1.16$\pm$0.25 & 4024$\pm$10 & 234$\pm$56\\ 
 & & & 2blif & 0.98\tablenotemark{b} & 4046$\pm$11 & 182$\pm$50\\ 
\ion{N}{2} & 1083.99 & -2.00 & 1blif & 0.93$\pm$0.19 & 3957$\pm$9 & 277$\pm$57\\
\enddata
	\tablenotetext{a}{Using $\rm{v_{sys}}=$      4037       \kms}
	\tablenotetext{b}{Errors are undetermined for this value.}
	\end{deluxetable}

\clearpage
			\begin{deluxetable}{lcccccc}
			\tabletypesize{\scriptsize}
			\tablecolumns{7}
			\tablewidth{0pc}
			\tablecaption{       NGC3310        ISM Absorption Line Fit Data
			\label{t:NGC3310da}}
			\tablehead{
			\colhead{Ion} & \colhead{$\lambda_{0}$} & \colhead{$log(\lambda f N/N_H)$} &
			                 \colhead{Instrument} & \colhead{$\rm{W}_\lambda$} &
			                 \colhead{$\rm{v_{c}}$} & \colhead{FWHM} \\
			                  & \colhead{\AA} & & & \colhead{\AA} & \colhead{\kms} &
			                  \colhead{\kms}}
			\startdata
\ion{C}{3} & 977.02 & -0.61 & 1bsic & 2.44$\pm$0.05 & 701$\pm$5 & 583$\pm$19\\ 
 & & & 2asic & 2.10$\pm$0.61 & 705$\pm$19 & 450$\pm$169\\ 
\ion{O}{6} & 1031.93 & -1.13 & 1alif & 1.67$\pm$0.04 & 616$\pm$5 & 556$\pm$15\\ 
 & & & 2blif & 1.71$\pm$0.04 & 620$\pm$5 & 554$\pm$15\\ 
\ion{C}{2} & 1036.34 & -1.37 & 1alif & 3.31$\pm$0.07 & 749$\pm$6 & 786$\pm$31\\ 
 & & & 2blif & 3.41$\pm$0.07 & 752$\pm$6 & 803$\pm$35\\ 
\ion{O}{1} & 988.73 & -1.56 & 1alif & 1.57\tablenotemark{b} & 877$\pm$10 & 270$\pm$50\\ 
 & & & 2blif & 1.52\tablenotemark{b} & 859$\pm$6 & 259$\pm$42\\
\enddata
	\tablenotetext{a}{Using $\rm{v_{sys}}=$      983        \kms}
	\tablenotetext{b}{Errors are undetermined for this value.}
	\end{deluxetable}

\clearpage
			\begin{deluxetable}{lcccccc}
			\tabletypesize{\scriptsize}
			\tablecolumns{7}
			\tablewidth{0pc}
			\tablecaption{       NGC3690        ISM Absorption Line Fit Data
			\label{t:NGC3690da}}
			\tablehead{
			\colhead{Ion} & \colhead{$\lambda_{0}$} & \colhead{$log(\lambda f N/N_H)$} &
			                 \colhead{Instrument} & \colhead{$\rm{W}_\lambda$} &
			                 \colhead{$\rm{v_{c}}$} & \colhead{FWHM} \\
			                  & \colhead{\AA} & & & \colhead{\AA} & \colhead{\kms} &
			                  \colhead{\kms}}
			\startdata
\ion{C}{3} & 977.02 & -0.61 & 1bsic & 1.69$\pm$0.10 & 3033$\pm$84 & 385\tablenotemark{b}\\ 
 & & & 2asic & 1.53$\pm$0.20 & 3111$\pm$9 & 364$\pm$104\\ 
\ion{O}{6} & 1031.93 & -1.13 & 1alif & 0.78\tablenotemark{b} & 2866$\pm$37 & 429\tablenotemark{b}\\ 
 & & & 2blif & 0.98$\pm$0.16 & 2842$\pm$24 & 564$\pm$89\\ 
\ion{C}{2} & 1036.34 & -1.37 & 1alif & 2.40$\pm$0.10 & 3051$\pm$6 & 580$\pm$39\\ 
 & & & 2blif & 2.20$\pm$0.08 & 3052$\pm$6 & 501$\pm$36\\ 
\ion{N}{2} & 1083.99 & -2.00 & 2alif & 2.27$\pm$0.08 & 3075$\pm$5 & 589$\pm$22\\ 
\ion{O}{1} & 1039.23 & -2.29 & 1alif & 0.63$\pm$0.07 & 3169$\pm$4 & 172$\pm$20\\
\enddata
	\tablenotetext{a}{Using $\rm{v_{sys}}=$      3121       \kms}
	\tablenotetext{b}{Errors are undetermined for this value.}
	\end{deluxetable}

\clearpage
			\begin{deluxetable}{lcccccc}
			\tabletypesize{\scriptsize}
			\tablecolumns{7}
			\tablewidth{0pc}
			\tablecaption{       NGC4214        ISM Absorption Line Fit Data
			\label{t:NGC4214da}}
			\tablehead{
			\colhead{Ion} & \colhead{$\lambda_{0}$} & \colhead{$log(\lambda f N/N_H)$} &
			                 \colhead{Instrument} & \colhead{$\rm{W}_\lambda$} &
			                 \colhead{$\rm{v_{c}}$} & \colhead{FWHM} \\
			                  & \colhead{\AA} & & & \colhead{\AA} & \colhead{\kms} &
			                  \colhead{\kms}}
			\startdata
\ion{C}{3} & 977.02 & -0.61 & 2asic & 1.04$\pm$0.33 & 233$\pm$4 & 194$\pm$28\\ 
\ion{O}{6} & 1031.93 & -1.13 & 1alif & 0.29$\pm$0.04 & 212$\pm$5 & 124$\pm$16\\ 
 & & & 2blif & 0.35$\pm$0.03 & 225$\pm$5 & 152$\pm$15\\ 
\ion{C}{2} & 1036.34 & -1.37 & 1alif & 1.04$\pm$0.04 & 273$\pm$2 & 263$\pm$13\\ 
 & & & 2blif & 1.06$\pm$0.03 & 279$\pm$2 & 265$\pm$12\\ 
\ion{O}{1} & 988.73 & -1.56 & 1alif & 0.47$\pm$0.14 & 283$\pm$12 & 110\tablenotemark{b}\\ 
 & & & 2blif & 0.50$\pm$0.15 & 285$\pm$6 & 110\tablenotemark{b}\\ 
\ion{N}{2} & 1083.99 & -2.00 & 1asic & 0.63$\pm$0.03 & 289$\pm$4 & 162$\pm$8\\
\enddata
	\tablenotetext{a}{Using $\rm{v_{sys}}=$      291        \kms}
	\tablenotetext{b}{Errors are undetermined for this value.}
	\end{deluxetable}

\clearpage
			\begin{deluxetable}{lcccccc}
			\tabletypesize{\scriptsize}
			\tablecolumns{7}
			\tablewidth{0pc}
			\tablecaption{       NGC7673        ISM Absorption Line Fit Data
			\label{t:NGC7673da}}
			\tablehead{
			\colhead{Ion} & \colhead{$\lambda_{0}$} & \colhead{$log(\lambda f N/N_H)$} &
			                 \colhead{Instrument} & \colhead{$\rm{W}_\lambda$} &
			                 \colhead{$\rm{v_{c}}$} & \colhead{FWHM} \\
			                  & \colhead{\AA} & & & \colhead{\AA} & \colhead{\kms} &
			                  \colhead{\kms}}
			\startdata
\ion{O}{6} & 1031.93 & -1.13 & 1alif & 0.95$\pm$0.17 & 3323$\pm$28 & 386$\pm$66\\ 
 & & & 2blif & 1.07$\pm$0.15 & 3323\tablenotemark{b} & 386\tablenotemark{b}\\ 
\ion{C}{2} & 1036.34 & -1.37 & 1alif & 2.06$\pm$0.34 & 3355$\pm$30 & 469$\pm$166\\ 
 & & & 2blif & 2.07$\pm$0.58 & 3382$\pm$14 & 350$\pm$11\\ 
\ion{N}{2} & 1083.99 & -2.00 & 1blif & 1.52$\pm$0.12 & 3401$\pm$15 & 440$\pm$40\\ 
 & & & 2alif & 1.45$\pm$0.12 & 3405$\pm$14 & 411$\pm$32\\ 
\ion{O}{1} & 1039.23 & -2.29 & 1alif & 1.02$\pm$0.12 & 3435$\pm$8 & 283$\pm$29\\ 
 & & & 2blif & 0.92$\pm$0.17 & 3419$\pm$10 & 254$\pm$53\\ 
\ion{N}{1} & 1134.41 & -2.57 & 2alif & 0.47$\pm$0.06 & 3404$\pm$26 & 381$\pm$53\\ 
\ion{P}{5} s & 1117.98 &   & 2alif & 0.45$\pm$0.05 & 3417$\pm$19 & 250\tablenotemark{b}\\ 
\ion{Si}{4} s & 1128.33 &   & 1blif & 0.32$\pm$0.09 & 3402$\pm$33 & 250\tablenotemark{b}\\ 
 & & & 2alif & 0.57$\pm$0.10 & 3455$\pm$19 & 259$\pm$50\\ 
	\enddata
	\tablenotetext{a}{Using $\rm{v_{sys}}=$      3392       \kms}
	\tablenotetext{b}{Errors are undetermined for this value.}
	\end{deluxetable}

\clearpage
			\begin{deluxetable}{lcccccc}
			\tabletypesize{\scriptsize}
			\tablecolumns{7}
			\tablewidth{0pc}
			\tablecaption{       NGC7714        ISM Absorption Line Fit Data
			\label{t:NGC7714da}}
			\tablehead{
			\colhead{Ion} & \colhead{$\lambda_{0}$} & \colhead{$log(\lambda f N/N_H)$} &
			                 \colhead{Instrument} & \colhead{$\rm{W}_\lambda$} &
			                 \colhead{$\rm{v_{c}}$} & \colhead{FWHM} \\
			                  & \colhead{\AA} & & & \colhead{\AA} & \colhead{\kms} &
			                  \colhead{\kms}}
			\startdata
\ion{O}{6} & 1031.93 & -1.13 & 1alif & 1.54$\pm$0.01 & 2589$\pm$3 & 363$\pm$6\\ 
 & & & 2blif & 1.56$\pm$0.14 & 2594$\pm$16 & 522$\pm$51\\ 
\ion{C}{2} & 1036.34 & -1.37 & 1alif & 1.94$\pm$0.35 & 2739$\pm$42 & 532$\pm$110\\ 
 & & & 2blif & 2.76$\pm$0.10 & 2726$\pm$9 & 709$\pm$30\\ 
\ion{N}{2} & 1083.99 & -2.00 & 2alif & 2.36$\pm$0.11 & 2753$\pm$8 & 628$\pm$33\\ 
\ion{Fe}{2} & 1096.88 & -2.96 & 1blif & 0.49$\pm$0.09 & 2725$\pm$11 & 220$\pm$48\\ 
 & & & 2alif & 0.46$\pm$0.08 & 2757$\pm$14 & 349$\pm$55\\ 
\ion{Fe}{2} & 1121.97 & -3.14 & 2alif & 0.50$\pm$0.10 & 2727$\pm$19 & 258$\pm$36\\ 
\ion{S}{3} & 1012.50 & -3.24 & 1alif & 0.58\tablenotemark{b} & 2750$\pm$61 & 188\tablenotemark{b}\\ 
 & & & 2blif & 0.62$\pm$0.04 & 2751$\pm$9 & 200\tablenotemark{b}\\ 
\ion{Si}{4} s & 1122.48 &   & 2alif & 0.80$\pm$0.08 & 2766$\pm$9 & 255$\pm$22\\ 
	\enddata
	\tablenotetext{a}{Using $\rm{v_{sys}}=$      2803       \kms}
	\tablenotetext{b}{Errors are undetermined for this value.}
	\end{deluxetable}

\clearpage
			\begin{deluxetable}{lcccccc}
			\tabletypesize{\scriptsize}
			\tablecolumns{7}
			\tablewidth{0pc}
			\tablecaption{     SBS0335-052      ISM Absorption Line Fit Data
			\label{t:SBS0335- }}
			\tablehead{
			\colhead{Ion} & \colhead{$\lambda_{0}$} & \colhead{$log(\lambda f N/N_H)$} &
			                 \colhead{Instrument} & \colhead{$\rm{W}_\lambda$} &
			                 \colhead{$\rm{v_{c}}$} & \colhead{FWHM} \\
			                  & \colhead{\AA} & & & \colhead{\AA} & \colhead{\kms} &
			                  \colhead{\kms}}
			\startdata
\ion{O}{6} & 1031.93 & -1.13 & 1alif & 0.16\tablenotemark{b} & 3863$\pm$105 & 155\tablenotemark{b}\\ 
 & & & 2blif & 0.13\tablenotemark{b} & 3919$\pm$82 & 132\tablenotemark{b}\\ 
\ion{C}{2} & 1036.34 & -1.37 & 1alif & 0.47$\pm$0.08 & 4040$\pm$10 & 138$\pm$26\\ 
 & & & 2blif & 0.38$\pm$0.07 & 4051$\pm$7 & 94$\pm$19\\ 
\ion{O}{1} & 988.73 & -1.56 & 1alif & 0.40\tablenotemark{b} & 4043$\pm$12 & 76\tablenotemark{b}\\ 
 & & & 2blif & 0.39\tablenotemark{b} & 4075$\pm$10 & 56\tablenotemark{b}\\ 
\ion{N}{2} & 1083.99 & -2.00 & 2alif & 0.23\tablenotemark{b} & 4095$\pm$45 & 176\tablenotemark{b}\\ 
\ion{N}{3} & 989.80 & -2.03 & 1alif & 0.46$\pm$0.10 & 4058$\pm$11 & 115$\pm$24\\ 
 & & & 2blif & 0.51$\pm$0.18 & 4093$\pm$7 & 86$\pm$12\\ 
\ion{O}{1} & 1039.23 & -2.29 & 1alif & 0.23\tablenotemark{b} & 4066$\pm$3 & 35\tablenotemark{b}\\ 
 & & & 2blif & 0.21$\pm$0.02 & 4081$\pm$3 & 54$\pm$3\\ 
\ion{N}{1} & 1134.98 & -2.39 & 2alif & 0.10$\pm$0.03 & 4069$\pm$4 & 31\tablenotemark{b}\\ 
\ion{N}{1} & 1134.41 & -2.57 & 2alif & 0.11$\pm$0.03 & 4073$\pm$5 & 31$\pm$9\\ 
	\enddata
	\tablenotetext{a}{Using $\rm{v_{sys}}=$      4043       \kms}
	\tablenotetext{b}{Errors are undetermined for this value.}
	\end{deluxetable}

\clearpage
			\begin{deluxetable}{lcccccc}
			\tabletypesize{\scriptsize}
			\tablecolumns{7}
			\tablewidth{0pc}
			\tablecaption{     TOL0440-381      ISM Absorption Line Fit Data
			\label{t:TOL0440- }}
			\tablehead{
			\colhead{Ion} & \colhead{$\lambda_{0}$} & \colhead{$log(\lambda f N/N_H)$} &
			                 \colhead{Instrument} & \colhead{$\rm{W}_\lambda$} &
			                 \colhead{$\rm{v_{c}}$} & \colhead{FWHM} \\
			                  & \colhead{\AA} & & & \colhead{\AA} & \colhead{\kms} &
			                  \colhead{\kms}}
			\startdata
\ion{C}{3} & 977.02 & -0.61 & 1alif & 1.23$\pm$0.22 & 12063$\pm$16 & 369$\pm$76\\ 
 & & & 2blif & 1.23\tablenotemark{b} & 12063$\pm$7 & 369\tablenotemark{b}\\ 
\ion{C}{2} & 1036.34 & -1.37 & 1alif & 1.25$\pm$0.42 & 12171$\pm$9 & 222$\pm$31\\ 
\ion{N}{2} & 1083.99 & -2.00 & 2alif & 0.60$\pm$0.07 & 12194$\pm$11 & 206$\pm$25\\ 
\ion{S}{3} & 1012.50 & -3.24 & 1alif & 0.52$\pm$0.05 & 12197$\pm$12 & 252$\pm$26\\ 
 & & & 2blif & 0.58$\pm$0.06 & 12205$\pm$9 & 192$\pm$22\\ 
\ion{S}{4} s & 1062.66 &   & 2alif & 0.55$\pm$0.11 & 12254$\pm$32 & 376$\pm$73\\ 
\ion{Si}{4} s & 1128.33 &   & 2alif & 0.82$\pm$0.13 & 12298$\pm$28 & 351$\pm$54\\ 
	\enddata
	\tablenotetext{a}{Using $\rm{v_{sys}}=$     12291       \kms}
	\tablenotetext{b}{Errors are undetermined for this value.}
	\end{deluxetable}

\clearpage
			\begin{deluxetable}{lcccccc}
			\tabletypesize{\scriptsize}
			\tablecolumns{7}
			\tablewidth{0pc}
			\tablecaption{        VV114 ISM Absorption Line Fit Data
			\label{t:VV114data}}
			\tablehead{
			\colhead{Ion} & \colhead{$\lambda_{0}$} & \colhead{$log(\lambda f N/N_H)$} &
			                 \colhead{Instrument} & \colhead{$\rm{W}_\lambda$} &
			                 \colhead{$\rm{v_{c}}$} & \colhead{FWHM} \\
			                  & \colhead{\AA} & & & \colhead{\AA} & \colhead{\kms} &
			                  \colhead{\kms}}
			\startdata
\ion{C}{3} & 977.02 & -0.61 & 1alif & 2.55$\pm$0.81 & 5971$\pm$8 & 350\tablenotemark{b}\\ 
 & & & 2blif & 2.39\tablenotemark{b} & 5990$\pm$5 & 304$\pm$5\\ 
\ion{O}{6} & 1031.93 & -1.13 & 1alif & 1.52$\pm$0.10 & 5684$\pm$25 & 750\tablenotemark{b}\\ 
 & & & 2blif & 1.51$\pm$0.47 & 5759$\pm$28 & 783$\pm$196\\ 
\ion{C}{2} & 1036.34 & -1.37 & 1alif & 2.70$\pm$0.48 & 5964$\pm$22 & 746$\pm$134\\ 
 & & & 2blif & 2.54$\pm$0.17 & 5962$\pm$17 & 635$\pm$66\\ 
\ion{O}{1} & 988.73 & -1.56 & 1alif & 1.59\tablenotemark{b} & 6084$\pm$312 & 350\tablenotemark{b}\\ 
 & & & 2blif & 2.32$\pm$0.80 & 6114$\pm$40 & 448$\pm$37\\
\enddata
	\tablenotetext{a}{Using $\rm{v_{sys}}=$      6016       \kms}
	\tablenotetext{b}{Errors are undetermined for this value.}
	\end{deluxetable}

\clearpage

\begin{deluxetable}{lrrr}
\tabletypesize{\scriptsize}
\tablecolumns{4}
\tablewidth{0pc}
\tablecaption{\ovi~Data: Apparent Optical Depth Method\label{t:fuseovi}}
\tablehead{\colhead{Galaxy} &  \colhead{$\rm{N_{O\,VI}}$} & \colhead{b} & 
	\colhead{$\rm v_{O\,VI} - v_{sys}$} \\
	& \colhead{(\cms)} & \colhead{(\kms)} & \colhead{(\kms)}}
\startdata
Haro 11	& $8.0\times10^{14}$ &	204 &	-217 \\
VV 114	& $2.2\times10^{15}$ &	438 &	-243 \\
SBS 0335-052	& $2.6\times10^{14}$ &	131 &	-166 \\
NGC 1705	 & $2.1\times10^{14}$ &	76 &	-78 \\
NGC 1741	& $6.3\times10^{14}$ &	228 &	-127 \\
NGC 3310	 & $1.9\times10^{15}$ &	232 &	-325 \\
Haro 3	& $7.5\times10^{14}$ &	149 &	-102	 \\		
NGC 3690	 & $8.5\times10^{14}$ &	269 &	-304 \\
NGC 4214	 & $4.0\times10^{14}$ &	98 &	-87 \\
Mrk 54	& $1.3\times10^{15}$ &	188 &	-145 \\
NGC 7673	 & $1.9\times10^{15}$ &	251 &	-88 \\
NGC 7714	 & $1.7\times10^{15}$ &	221 &	-170 
\enddata
\end{deluxetable}

\begin{deluxetable}{lrrrrrr}
\tabletypesize{\scriptsize}
\tablecolumns{7}
\tablewidth{0pc}
\tablecaption{Line Data: Non-parametric Measurement\label{t:fusesplot}}
\tablehead{ & \multicolumn{2}{c}{\ion{C}{3}} & \multicolumn{2}{c}{\ion{C}{2}}  & \multicolumn{2}{c}{\ion{N}{2}} \\
	\cline{2-3} \cline{4-5} \cline{6-7}
	\colhead{Galaxy} &  \colhead{$\rm{W_{C III}}$} & \colhead{$\rm{v_{C III}}$} &
	\colhead{$\rm{W_{C II}}$} & \colhead{$\rm{v_{C II}}$} &
	\colhead{$\rm{W_{N II}}$} & \colhead{$\rm{v_{N II}}$} \\
	& \colhead{(\ang)} & \colhead{(\kms)} & \colhead{(\ang)} & \colhead{(\kms)} & \colhead{(\ang)} & \colhead{(\kms)}} 
\startdata
Haro\,11 & 1.74 & -253 & 1.16 & -118 & 0.92 & -113 \\
VV114 & 3.37 & -212 & 2.51 & -94 & 2.11 & -84 \\
NGC\,1140 & 1.28 & -62 & 0.87 & -45 & 0.66 & 6 \\
SBS\,0335-052 &  \nodata & \nodata & 0.47 & -9 & 0.28 & 34 \\
Tol\,0440-381 & 1.28 & -151 & 1.15 & -120 & 0.71 & -82 \\
NGC\,1705 & 1.14 & -19 & 0.87 & -38 & 0.56 & -6 \\
NGC\,1741 & \nodata & \nodata & 1.27 & -41 & 0.94 & -35 \\
IZW\,18 & 0.61 & -137 &  \nodata &  \nodata &  \nodata & \nodata \\
NGC\,3310 & 2.05 & -254 & 2.54 & -186 & 1.86 & -93 \\
Haro\,3 & 0.97 & -27 & 0.74 & -49 & 0.74 & 49 \\
NGC3690 & 1.91 & -153 & 2.14 & -67 & 1.80 & -77 \\
NGC\,4214 & 1.05 & -71 & 0.93 & -18 & 0.60 & -13 \\
Mrk\,54 & 2.15 & -167 & 2.65 & 11 & 1.42 & -18 \\
IRAS\,19245-4140 & 1.71 & -98 & 0.82 & -77 & 0.20 & -21 \\
NGC\,7673 & 2.36 & -129 & 1.44 & -96 & 1.29 & -20 \\
NGC\,7714 & 1.91 & -46 & 2.09 & -24 & 2.69 & -67 \\
\enddata
\tablecomments{These measurements were made using the IRAF splot command ``e''.}
\end{deluxetable}

\clearpage

\begin{deluxetable}{llrr}
\tabletypesize{\scriptsize}
\tablecolumns{4}
\tablewidth{0pc}
\tablecaption{Correlations of Outflow Speeds with Galaxy Properties\label{t:voutcorr}}
\tablehead{
        \colhead{Galaxy Parameter} & \colhead{Outflow Measure} & \colhead{$\tau_a$} & \colhead{$P_{null}$}}
\startdata
$L_{IR}+L_{UV}$ & $v_{sys}-v_{OVI}$ &  -0.636 & 0.00644 \\
$L_K$           & $v_{sys}-v_{OVI}$ &  -0.564 & 0.0158 \\
$L_{IR}+L_{UV}$ & $v_{sys}-v_{CIII}$ & -0.516 & 0.0101 \\
$L_K$           & $v_{sys}-v_{CIII}$ & -0.495 & 0.0138 \\
\enddata
\tablecomments{Col 3. measures the strength of the correlation between the parameters in the first two columns, using the
value $\tau_a$ from the Kendall rank test. Col.  4 specifies the probability of the null hypothesis that the
two parameters are uncorrelated.}
\end{deluxetable}

\begin{deluxetable}{llrr}
\tabletypesize{\scriptsize}
\tablecolumns{4}
\tablewidth{0pc}
\tablecaption{Correlations of Equivalent Widths with Galaxy Properties\label{t:eqwcorr}}
\tablehead{
        \colhead{Galaxy Parameter} & \colhead{Line} & \colhead{$\tau_a$} & \colhead{$P_{null}$}}
\startdata
$log(O/H)+12$ & EQW$_{OI}$ & 0.748 & 0.000710 \\
$L_K$ & EQW$_{OI}$ & 0.689 & 0.00556 \\
$L_{IR}+L_{UV}$ & EQW$_{OI}$ & 0.600 & 0.0157 \\
$log(O/H)+12$ & EQW$_{CII}$ & 0.694 & 0.000960 \\
$L_K$ & EQW$_{CII}$ & 0.718 & 0.000634 \\
$L_{IR}+L_{UV}$ & EQW$_{CII}$ & 0.564 & 0.00727\\
$log(O/H)+12$ & EQW$_{SIII}$ & 0.442 & 0.0753 \\
$L_K$ & EQW$_{SIII}$ & 0.697 & 0.00283 \\
$L_{IR}+L_{UV}$ & EQW$_{SIII}$ & 0.624 & 0.00756 \\
$log(O/H)+12$ & EQW$_{CIII}$ & 0.379 & 0.0712 \\
$L_K$ & EQW$_{CIII}$ & 0.615 & 0.00341 \\
$L_{IR}+L_{UV}$ & EQW$_{CIII}$ & 0.590 & 0.00510 \\
\enddata
\tablecomments{Col 3. measures the strength of the correlation between the parameters in the first two columns, using the 
value $\tau_a$ from the Kendall rank test. Col.  4 specifies the probability of the null hypothesis that the
two parameters are uncorrelated.}
\end{deluxetable}

\begin{deluxetable}{lrrrrrrrrr}
\tabletypesize{\scriptsize}
\tablecolumns{10}
\tablewidth{0pc}
\tablecaption{Lyman Limit Escape Fraction\label{t:lylimit}}
\tablehead{ 
	\colhead{Galaxy} &  \colhead{MW E(B-V)} & \colhead{$\rm F_{1150 \AA}$} &
	\colhead{$\rm F_{1150 \AA,corr}$} &  \colhead{$\rm F_{900 \AA}$} &
	\colhead{$\rm F_{900 \AA,corr}$} &  \colhead{$\rm L_{1150 \AA} + L_{IR}$} &
	\colhead{$\rm L_{900 \AA}$} & \colhead{$\rm f_{esc,abs}$} & \colhead{$\rm f_{esc,rel}$} \\
	 & \colhead{(mag)} & \multicolumn{2}{c}{(\ergcmsa)}  &   \multicolumn{2}{c}{(\ergcmsa)} & 
	 \colhead{\Lsun} & \colhead{\Lsun}
	& \colhead{(\%)} & \colhead{(\%)} \\ }
\startdata
Haro\,11 & 0.011 & $5.2\times10^{-14}$ & $5.9\times10^{-14}$ & $<1.3\times10^{-15}$ & 
	$<1.6\times10^{-15}$ & $2.1\times10^{11}$ & $4.6\times10^{8}$ & $<$1.6 & $<$11 \\
Mrk\,54 & 0.015 & $3.2\times10^{-14}$ & $3.8\times10^{-14}$ & $<6.5\times10^{-16}$ & 
	$<8.9\times10^{-16}$ & $1.7\times10^{11}$ & $1.2\times10^{9}$ & $<$5.0 & $<$9 \\
NGC\,1741 & 0.051 & $5.2\times10^{-14}$ & $9.2\times10^{-14}$ & $<1.3\times10^{-15}$ & 
	$<3.7\times10^{-15}$ & $4.9\times10^{10}$ & $4.4\times10^{8}$ & $<$6.3 & $<$16 \\
NGC\,3690 & 0.017 & $7.4\times10^{-14}$ & $8.9\times10^{-14}$ & $<1.6\times10^{-15}$ & 
	$<2.3\times10^{-15}$ & $6.7\times10^{11}$ & $1.6\times10^{8}$ & $<$0.2 & $<$10 \\
VV\,114 & 0.016 & $7.6\times10^{-14}$ & $9.1\times10^{-14}$ & $<4.2\times10^{-15}$ & 
	$<5.8\times10^{-15}$ & $5.4\times10^{11}$ & $1.6\times10^{9}$ & $<$2.0 & $<$26 \\
\enddata
\tablecomments{We have estimated an upper limit to the flux below the Lyman limit for the galaxies
with S/N$>$5 in the LiF 1A channel.  See Section \ref{s:belowlyman} for a discussion of the derived
values in this table.  We find $f_{esc}<7\%$ for all five galaxies.}
\end{deluxetable}

\begin{deluxetable}{lrrrrrr}
\tabletypesize{\scriptsize}
\tablecolumns{7}
\tablewidth{0pc}
\tablecaption{Lyman Limit: C II Limits on the Escape Fraction\label{t:ciilimit}}
\tablehead{ & & \colhead{P.F.} & & \multicolumn{3}{c}{Slab}  \\
	\cline{3-3}  \cline{5-7} 
	\colhead{Galaxy} &  \colhead{$\rm f_{esc,C\,II}$} & \colhead{$\rm f_{esc,abs}$} & &
	\colhead{$\rm Z$\tablenotemark{a}} & \colhead{$b$\tablenotemark{b}} & 
	\colhead{$\tau_{\rm Ly}$} \\
	& \colhead{(\%)} & \colhead{(\%)} & & \colhead{($\rm Z_\odot$)} & \colhead{(\kms)} & }
\startdata
Haro\,11 & $\leq$8 & $\leq$0.6 & & 0.17 & 225 & 281.2 \\
Haro\,3 & $<0.5$ & $<0.1$ & & 0.46 & 141  & $\infty$\\
IRAS\,19245-4140 & $\leq$21 & $\leq$11 & & 0.22 & 163  & 100.0 \\
IZw\,18 & $\leq$10 &  \nodata & & 0.03 & 111 &  633.9 \\
NGC\,1140 & $<$4 & $<$1.4 & & 0.22 & 181  & 229.0 \\
NGC\,1705 & $<$1 & $<$0.8 & & 0.22 & 155  & 280.6 \\
NGC\,1741 & $<$1 & $<$0.2 & & 0.22 & 161  & 291.5 \\
NGC\,3310 & $\leq$5 & $\leq$0.5 &  & 2.19 & 172  & 20.3\\
NGC\,3690 & $<1$ & $<0.1$ & & 1.38 & 477   & $\infty$\\
NGC\,7673 & $\leq10$ & $\leq1.7$ & & 0.69 & 229   & $\infty$\\
NGC\,7714 & $<$3 & $<$0.1 & & 1.10 & 323  & 88.8 \\
SBS\,0335-052 & $<$18 &  \nodata & & 0.05 & 66 & 208.1\\
Tol\,0440-381 & $<$4 & $<$1.6 & & 0.34 & 211  & 172.4\\
VV\,114 & $<$2 & $<$0.1 & & 0.87 & 419 &  161.9\\
NGC\,4214 & $<$1 & $<$0.2 & & 0.35 & 66  & 75.4\\
\enddata
\tablecomments{We have conservatively estimated the 
continuum level and found the lowest convincing escape fraction in
C\,II at line center.  We then estimate $\rm f_{esc,abs}$ using the
picket fence (P.F.) model for escaping radiation.  We also
find extremely high optical depths ($\geq20$) below
the Lyman limit assuming the slab model. }
\tablenotetext{a}{Derived using the gas phase oxygen abundances in Table \ref{t:fuselumins} and a solar C/O abundance ratio from \citet{asp05}.}
\tablenotetext{b}{From gaussian fits in Tables \ref{t:HARO11dat} - \ref{t:VV114data}.}
\end{deluxetable}

\clearpage

\clearpage

\clearpage

\clearpage

\clearpage

\clearpage

\clearpage

\clearpage

\clearpage

\clearpage

\clearpage

\clearpage

\clearpage

\clearpage

\clearpage

\clearpage

\clearpage

\clearpage

\clearpage

\clearpage

\clearpage

\clearpage

\clearpage

\clearpage

\clearpage

\clearpage

\clearpage

\clearpage

\clearpage

\clearpage

\clearpage

\clearpage

\end{document}